\documentclass{jfm}
\usepackage{etoolbox}
\usepackage{subfiles}
\usepackage{cite}

\usepackage{xcolor}
\usepackage{amsmath}
\usepackage{amsfonts}   % if you want the fonts
\usepackage{amssymb}    % if you want extra symbols
\usepackage{graphicx}
\usepackage{subfigure} % subfig/subfloat causes error with fancy-preview
\usepackage{lipsum}

\newcommand\nc{\newcommand}

\nc\cB{{\mathcal{B}}}
\nc\cC{{\mathcal{C}}}
\nc\cN{{\mathcal{N}}}
\nc\tcN{\tilde{{\cal{N}}}}
\nc\cG{{\mathcal{G}}}
\nc\cD{{\mathcal{D}}}
\nc\cU{{\mathcal{U}}}
\nc\cL{{\mathcal{L}}}
\nc\cT{{\mathcal{T}}}
\nc\cF{{\mathcal{F}}}
\nc\cK{{\mathcal{K}}}
\nc\Grad{\nabla}
\nc\Lapcian{\nabla^2}
\nc\hGrad{\hat{\nabla}}
\nc\tGrad{\tilde{\nabla}}
\nc\Div{\nabla \cdot}
\nc{\curl}[1]{ \nabla \times { #1 } }

\nc\Pieff{{\Pi_{\textrm{S}}}}

\nc{\ep}{\epsilon}
\nc{\pa}{\partial}
\nc{\pad}[2]{\frac{\partial #1}{\partial #2}}
\nc{\padn}[3]{\frac{\partial^{#3} #1}{\partial #2^{#3}}}

\nc{\std}[2]{\frac{d #1}{d #2}}
\nc{\stdn}[3]{\frac{d^{#3} #1}{d #2^{#3}}}

\nc{\vad}[2]{\frac{\delta #1}{\delta #2}}
\nc{\myTitle}{My Title}

\graphicspath {{figures/}} 
\usepackage[acronym]{glossaries}

\newglossaryentry{NLC}
{
  name={NLC},
  description={nematic liquid crystal},
  first={\glsentrydesc{NLC} (\glsentrytext{NLC})},
  plural={NLCs},
  descriptionplural={nematic liquid crystals},
  firstplural={\glsentrydescplural{NLC} (\glsentryplural{NLC})}
} 
\newacronym{LSA}{LSA}{linear stability analysis}
\newacronym{ODE}{ODE}{ordinary differential equation}
\newacronym{PDE}{PDE}{partial differential equation}
\newacronym{HAN}{HAN}{hybrid aligned nematic}

\newacronym{nCB}{nCB}{cyanobiphenyl}
\newacronym{5CB}{5CB}{4-Cyano-4'-pentylbiphenyl}
\newacronym{6CB}{6CB}{4-cyano-4'-hexylbiphenyl}
\newacronym{7CB}{7CB}{4-Cyano-4'-heptylbiphenyl}
\newacronym{8CB}{8CB}{4-Octyl-4’-cyanobiphenyl}
\newacronym{5AB}{5AB}{tris(trimethylsiloxy)silane-ethoxycyanobiphenyl}

\usepackage[inline]{trackchanges}

\addeditor{ML}
\addeditor{LC}
\addeditor{LK}

\makeglossaries
\title[Stability of thin fluid films with complex structural/effective disjoining pressure]
      {Stability of thin fluid films characterised by a complex form of structural/effective disjoining pressure}
\author[M.-A. Y.-H. Lam, L. J. Cummings and L. Kondic]{Michael-Angelo Y.-H. Lam, Linda J. Cummings, Lou Kondic}
\affiliation{Department of Mathematical Sciences, New Jersey Institute of Technology, Newark, NJ, 07102, USA}
\begin{document}

\maketitle
%\tableofcontents
%\newpage

\begin{abstract}
We discuss instabilities of fluid films of nanoscale thickness, with a particular focus on 
films where the destabilising mechanism allows for linear instability, metastability, 
and absolute stability.   Our study is motivated by nematic liquid crystal films; however we note 
that similar instability mechanisms, and forms of the effective disjoining pressure, appear
in other contexts, such as the well-studied problem of polymeric films on two-layered substrates.  
The analysis is carried out within the framework of the long-wave approximation, which leads
to a fourth order nonlinear partial different equation for the film thickness.  Within the considered
formulation, the nematic character of the film leads to an additional contribution to the 
disjoining pressure, changing its functional form.  This effective disjoining pressure is 
characterised by the presence of a local maximum for non-vanishing film thickness.  Such 
a form leads to complicated instability evolution that we study by analytical means,
including application of marginal stability criteria, and by extensive numerical simulations 
that help us develop a better understanding of instability evolution in the nonlinear regime.   
This combination of analytical and computational techniques allows us to reach novel 
understanding of relevant instability mechanisms, and of their influence on transient and
fully developed fluid film morphologies. 
\end{abstract}

\section{Introduction} \label{sec:Intro}

Thin films, characterised by a small ratio of film thickness to typical lateral lengthscale, arise in many applications on scales that range from nano to macro.  Recently, 
there has been much interest in instabilities of thin films on the nanoscale, in particular for polymeric
films \citep[see][]{seeman_prl01,Sharma2004,Jacobs2008}; nematic
liquid crystal (NLC) films \citep[see][]{Herminghaus1998,Cazabat2011,Effenterre2003,Schlagowski2002,Ziherl2000};
and metal films \citep[see][]{favazza_apl07,fowlkes_nano11}, 
to name just a few examples. In many of the considered cases, the instabilities that develop
are complicated by interplay of a number of different effects.  On the nanoscale, these necessarily 
include fluid/solid interaction, often modelled by including in the governing equations 
a disjoining pressure term that plays an important role in determining the film stability. The relevant
forms of the disjoining pressure may be rather complicated, reflecting, for example, complex
Si/Si0$_2$ substrates in the case of polymer films \citep[see][]{Jacobs2008}, or complex physics of the fluid/solid interaction in the case of metal films, see~\citet{ajaev_pof03,wu_lang10} and the references therein. 
Often, it is convenient to include additional physical effects to form a so-called ``effective disjoining pressure''; these may include additional
forces of electromagnetic origin, such as for ferrofluids in electric or magnetic fields~\citep[see][]{Seric2014}. 
For NLC films, such effective disjoining pressure is often called a structural disjoining pressure, since it includes
the elastic forces that are due to the local structure of the director field (molecular orientation).  Most work on
this type of problem is carried out within the framework of long wave theory, which allows for simplifications of otherwise
extremely complex free surface problems; we adopt this approach also.  

In the context of NLC films, there are 
several experimental studies of dewetting on the nanoscale, 
e.g.,~\citet{Cazabat2011,Effenterre2003,Schlagowski2002,Ziherl2000}. 
For thin \gls{NLC} films deposited on SiO$_2$ wafer, the films were found to be unstable (dewetted) for a certain range of thicknesses,
referred to as the ``forbidden range'' by \citet{Cazabat2011}. Several different physical instability mechanisms 
have been proposed, and the discussion regarding the relevance of each is far from complete.  Possibly relevant physical effects discussed in the 
literature include thermal effects \citep[see][]{Schlagowski2002,Effenterre2003} and the
pseudo-Casimir effect \citep[see][]{Ziherl2000}, as well as the interplay of elastic and disjoining 
pressure already discussed above \citep[see][]{Vandenbrouck1999}.

The functional form of the (effective) disjoining pressure turns out to be similar for many of the setups 
discussed above; for example, for NLC films, the elastic effects lead to a disjoining pressure qualitatively similar 
to that found for polymer films on Si/SiO$_2$ substrates, or even for Newtonian films under inverted
gravity conditions (hanging films); see \citet{Lin2013,lin_pof10}.  
Our present work is motivated by NLC films, however as discussed
above, the results are generally applicable for many other thin films with a similar form of effective disjoining pressure 
(see figure~\ref{fig:DisjoiningPressureCompare} later in the text).  

Various instability mechanisms relevant to thin films have been discussed in the literature. For example,
in addition to the standard linear stability analysis, which focuses on instabilities due to the presence 
of random infinitesimal perturbations (often called spinodal dewetting), there are instabilities due to 
the presence of nucleation centres 
% analysis, which describes spinodal decomposition type instabilities,
% there are nucleation type instabilities
\citep[]{Seemann2001b,Sharma2004,Thiele2001,Diez2007}. \citet{Seemann2001b}
further separate nucleation instabilities into two types: heterogeneous nucleation,
characterised by nucleation holes with random (Poisson) distribution that emerge within a sharp time window;
and homogeneous nucleation, possibly of thermal origin in experiments, where breakup is continuous 
and more and more nucleation holes form.
In a similar direction, \citet{Thiele2001} and \citet{Diez2007} considered the metastable and absolutely stable regimes (stability
of films with respect to perturbations of finite size). In this work, we explore these various 
stability regimes in the context of our NLC model and demonstrate parallels between stability properties
of our model and the work of \citet{Seemann2001b}; \citet{Sharma2004}; \citet{Thiele2001} and \citet{Diez2007}, as well
as similarities and differences between the referenced works.

The remainder of this paper is organised as follows.   In \S~\ref{sec:GovEqn} we outline 
the derivation of the governing equation in the context of NLC films, highlighting differences 
between our model and others in the literature, especially with regard to the form of the effective 
(structural) disjoining pressure. In addition, appropriate scalings and physical parameters are discussed,
and linear stability analysis (LSA) of a flat film is presented.  The issues of particular relevance to 
NLC films are mostly considered in this section.  We then proceed with a general discussion of film 
instability.  In \S~\ref{sec:RandomPert} we discuss so-called spinodal instabilities, due to 
the presence of global (random) perturbations, intended to model the noise
that is always present in physical experiments.   Among other topics, we here discuss the formation of 
secondary (satellite) drops that are found only for a specific range of film thicknesses, and the distance between the 
emerging drops, whose dependence on the film thickness shows unexpected and rich structure.
In \S~\ref{sec:LocalizedPert} we consider the influence
of localised perturbations (nucleation centres) on the film stability.  In this section we also apply the Marginal 
Stability Criterion to help us develop additional analytical insight regarding instability development; this 
approach (to the best of our knowledge) has not yet been discussed in the present context.   We discuss both linearly
stable and metastable regimes of instability.  \S~\ref{sec:conclusions} summarises our
findings; of particular interest to the reader may be Fig.~\ref{fig:CompleteDiagram}, which provides a summary of
various instability mechanisms found and discussed throughout the paper. 

\section{Problem Formulation} \label{sec:GovEqn}

\begin{figure}
\centering
  \subfigure[]{
   \includegraphics[width=0.45\textwidth]{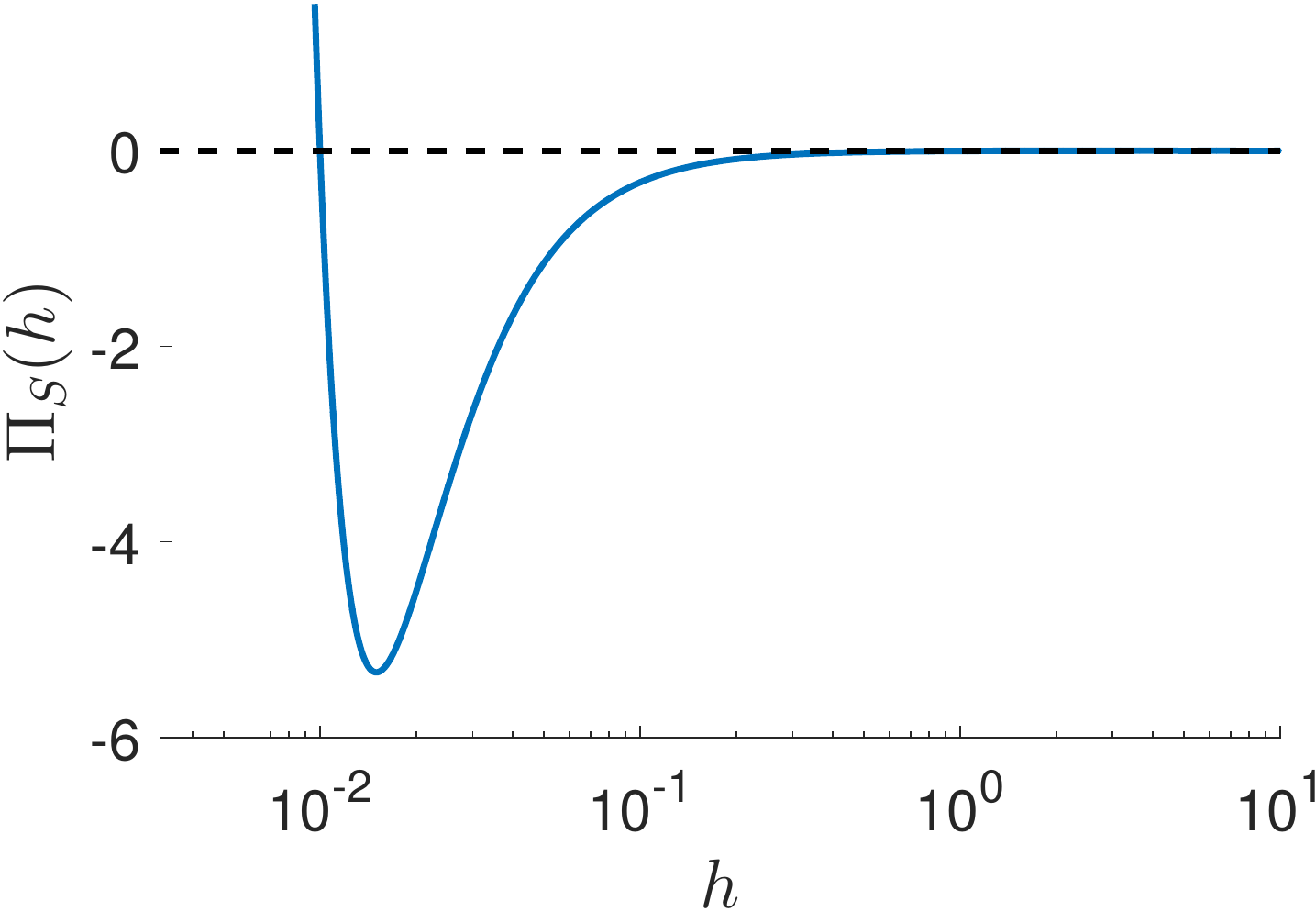}
  \label{fig:vdwDisjoiningPressure} }
  \subfigure[]{
   \includegraphics[width=0.45\textwidth]{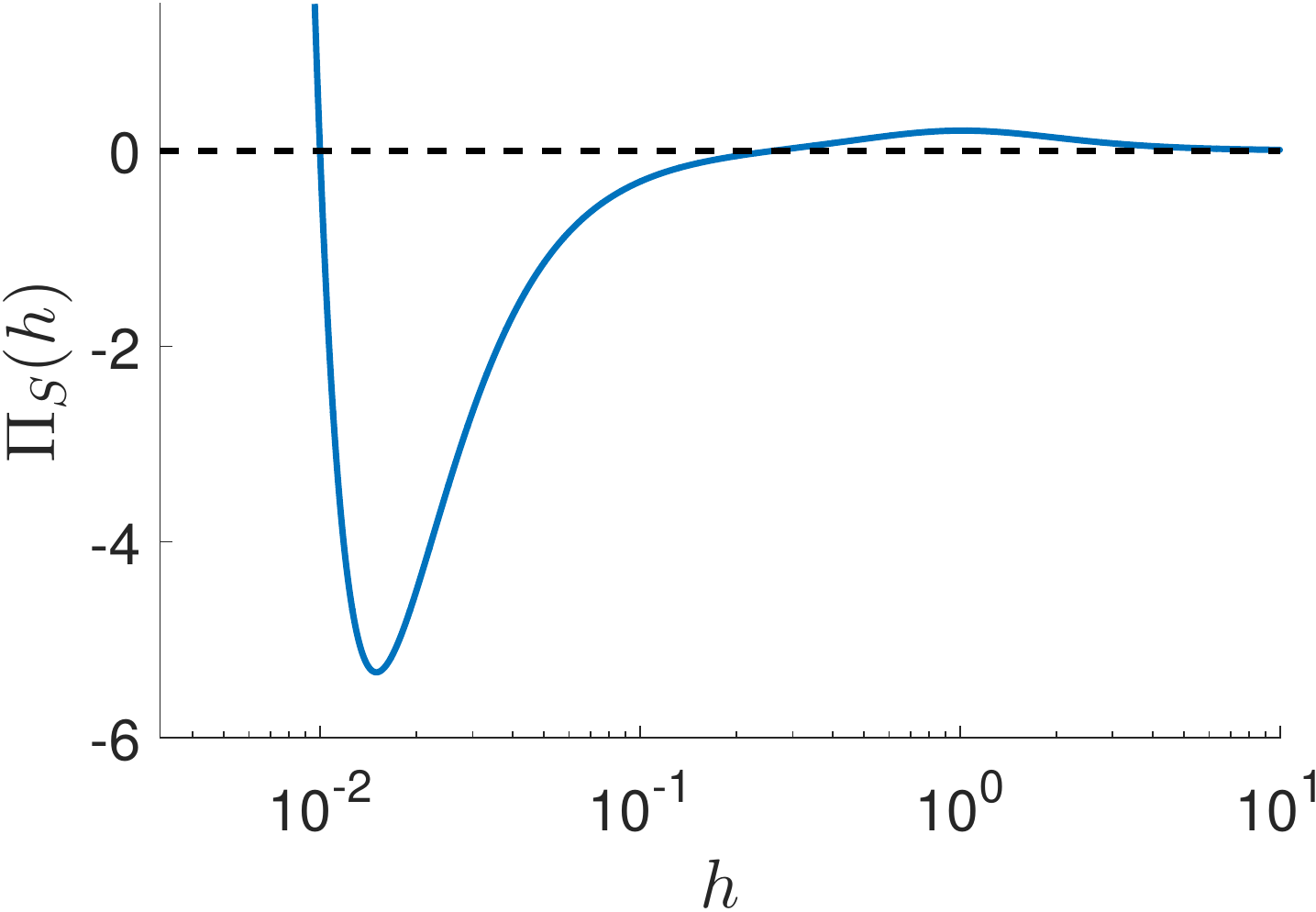}
  \label{fig:StructDisjoiningPressure} }
  \caption[]{  a) The disjoining pressure due to Van der Waals forces only (\ref{eq:DisjoingPressure}) ($\cN=0$)
and b) the total structural disjoining pressure, specified by (\ref{eq:EffDisjoingPressure}); $h$ is the film thickness.}
  \label{fig:DisjoiningPressureCompare}
\end{figure}

We first outline our model for \gls{NLC} films of thickness on the nano/micro scale, but emphasise that our results are of relevance to other
types of thin films (such as polymeric films), which may be described by long-wave equations similar to ours 
(equation (\ref{eq:NanoGov-PDE}) later) and which involve an effective disjoining pressure qualitatively similar to that sketched in figure~\ref{fig:StructDisjoiningPressure}. 
For the thin films that we consider, the effects of gravity are negligible; however, van der
Waals' interactions, neglected in our previous work on thin film dynamics of \glspl{NLC}
\citep[see][]{Lam2015,Lin2013},
become important. Our model derives from the Leslie-Ericksen equations~\citep[][]{Leslie1979},
which comprise a fluid momentum equation plus incompressibility, analogous to Navier-Stokes;
and an additional equation modelling conservation of energy (linear translational energy of the
\gls{NLC} molecules, and angular energy about each \gls{NLC} molecule's centre of mass). 
The fluid momentum equation differs from Navier-Stokes in two key ways: 1) by introducing a new anisotropic stress tensor (linear momentum), and 2)
by an additional term modelling angular momentum.
For brevity, the full derivation of the long wave model from the Leslie-Ericksen equations is skipped in the present paper; 
we present just a brief overview, highlighting where the modelling differs from our earlier work~\citep[][]{Lin2013}. 

There are three crucial quantities in
our long wave model: $h$, the free surface height; and the angles $\theta$, $\phi$ that characterise the director field, $\mathbf{n}$, a unit vector representing the 
local average orientation of the long axis of the rod-like \gls{NLC} molecules.
It is convenient to express the director
field in polar coordinates, $\mathbf{n}=(\sin\theta \cos\phi , \sin\theta \sin \phi, \cos \theta )$,
and consider the variables: $\theta$, the azimuthal angle; and $\phi$, the polar angle,
which vary quasistatically via the film height $h$.
Note that throughout this paper we use hats to distinguish between dimensional and dimensionless variables or parameters, 
hatted quantities being dimensional.

Analogous to Newtonian fluid models we include surface tension with
constant surface energy density $\hat \gamma$, and van der Waals forces 
with volume energy density $\hat{A}$ (proportional to the Hamaker constant).
Unique to \glspl{NLC} (with respect to simple fluids) 
are the isotropic viscosity stress tensor, characterised by 6 coefficients, $\hat\alpha_i$ ($i=1,6$);
and the elastic response, characterised by three constants $\hat {K}_i$ ($i=1,2,3$).
Similar to other authors~\citep[][]{Delabre2009,Rey2008,Effenterre2003,Ziherl2003}, we use the one-constant 
approximation, $ \hat K= \hat K_1= \hat K_2= \hat K_3$. Furthermore,
under the long wave approximation, only three of the six viscosities 
($\hat\alpha_3$, $\hat\alpha_4$, and $\hat\alpha_6$) are relevant.

To nondimensonalise the Leslie-Ericksen equations five scaling factors are defined: $\hat{H}$, the film thickness;
$\hat{L}$, the lengthscale of variations in the plane in the film, $(\hat x, \hat y)$;
$\delta=\hat{H}/\hat{L}$, the small aspect ratio;
$\hat{U}$, a characteristic flow velocity;
$\hat{T}_F =\hat{L}/\hat{U}$, the timescale of fluid flow; and
$\hat{\mu}=\hat \alpha_4 /2 $, a representative viscosity scale analogous to the 
Newtonian viscosity (with $\hat\alpha_i = 0$ for $i\ne4$, the isotropic viscosity stress tensor
reduces to a simple Newtonian stress tensor). 
The choice of the scales is 
based on experimental data, and is
further discussed in $\S$~\ref{sec:Scalings}.

\subsection{Weak Free Surface Anchoring} \label{sec:WeakAnchoring}

In addition to the fluid flow timescale, $\hat T_F$, we may consider
the timescale of elastic reorientation across the layer, $\hat T_R = \hat \mu \hat H^2 / \hat K$.
Assuming the timescale of reorientation is much faster 
than that of fluid flow, a quasistatic approximation (instantaneous elastic equilibrium) may be justified, 
and the energy equation may be solved to give the director field, $\mathbf{n}$, as a function of the 
instantaneous film height, $h$. In our previous work~\citep[see][]{Lin2013} we showed that the spherical director angles $\theta$ and $\phi$ are then given as
 \begin{equation} 
  \theta = a(x,y,t)z + \theta_S(x,y,t) \quad \textrm{and} \quad \phi=\phi_S(x,y,t) \;,
 \end{equation}
where $a(x,y,t)$ depends on the free surface height, $h(x,y,t)$, as discussed below. 
The functions $\theta_S$ and $\phi_S$ depend on the preferred orientation of 
the \gls{NLC} molecules at the interfaces, a phenomenon known as anchoring. 
We assume that the anchoring on the substrate is stronger than that at the free surface, and
therefore that $\theta_S(x,y)$ and $\phi_S(x,y)$ are imposed by the substrate. In line with 
observations for physical systems, we assume $\theta_S$ is constant, while $\phi_S(x,y)$ may vary only spatially. 
This solution for the director feeds into the anisotropic stress tensor. 
The particular state here in which $\theta$ is linear in $z$ is often called the \gls{HAN} state of \glspl{NLC}.

In general, the preferred polar anchoring angles at the free surface, 
$\theta_F$, and at the substrate, $\theta_S$ (both assumed constant), are different (antagonistic anchoring). 
For very thin films, adjusting between these anchoring conditions
incurs a large energy penalty in the bulk of the film.
To reduce the energy penalty for very thin films, we allow the (weakly anchored) free surface director angle
$\theta_\mathcal{F}$ to relax to the same value as the (strongly anchored) substrate director angle $\theta_\mathcal{S}$
for very thin films. Specifically, we take
 \begin{equation} \label{eq:theta}
   \theta =  \theta_\mathcal{S} + (\theta_\mathcal{F}-\theta_\mathcal{S}) \frac{m(h)}{h}z  \;,
 \end{equation}
 \begin{equation} \label{eq:mh}
   m(h) =  g(h) \frac{h^2}{h^2+\beta^2} \;, \quad 
   g(h) = \frac{1}{2} \left[1 + \tanh \left( \frac{h-2b}{w}\right) \right] \;, 
 \end{equation}
where $m(h)$ is a relaxation function (directly related to the anchoring energy at the free surface, see~\citet{Lin2013}) in which $\beta$
is the thickness scale at which the bulk elastic energy
is comparable to the free surface anchoring energy,
and $g(h)$ is a `cut-off' function (controlled
by $w$) that enforces the substrate 
anchoring on the free surface when the film thickness is close to the equilibrium 
film thickness, $b$ (discussed in more detail later in the text).

Note that with this form of the director field, thicker \gls{NLC} films 
are in the \gls{HAN} state, while films close to the equilibrium thickness have a director field nearly 
homogeneous across the film thickness. This transition between the states is in agreement
with several other models, e.g. \citet{Cazabat2011,Delabre2009,Ziherl2000}.
Furthermore, dimensional analysis shows that our lengthscale $\beta$ 
may be related to the extrapolation length discussed in the literature~\citep{Delabre2009,Effenterre2001,Ziherl2003}.

\subsection{Long Wave Model} \label{sec:LongWaveEqn}
To derive a long wave model for nanoscale thin films, we modify our  earlier model, described in~\citet{Lin2013},
by neglecting gravity and including van der Waals' fluid-solid interaction. With this modification,
there are five nondimensional parameters:
$\cC$, the inverse Capillary number, representing the ratio of surface tension  and viscous forces;
$\cK$, the ratio of van der Waals' and viscous forces;
$\cN$, a scaled inverse Ericksen number, representing the ratio of elastic and viscous forces;
$\lambda$ and $\nu$, scaled effective viscosities. 
In terms of physical parameters, %

\begin{equation} \label{eq:nondimensionalParas}
\cC = \frac{\delta^3 \hat{\gamma} }{ \hat{\mu} \hat{U}}  \; , \quad
\cK =\frac{\delta^2 \hat{A} \hat{L}}{\hat{\mu} \hat{U}} \; , \quad
   \cN = (\Delta \theta)^2 \frac{  \hat{K}}{  \hat{\mu} \hat{U} \hat{L}} \;, \quad
   \Delta \theta = \theta_\mathcal{F}-\theta_\mathcal{S} \;,
 \end{equation}
\begin{eqnarray} \label{eq:viscosities} 
 \lambda  =  \frac{2+\eta}{4\left( 1 + \eta \right)} \; , \quad
 \nu =  -\frac{\eta}{4\left( 1 +\eta \right)} \; , \quad
 \eta = \frac{ \hat{\alpha}_3 + \hat{\alpha}_6}{ \hat\mu} \;,
\end{eqnarray}
where, recall, $\hat{\mu}=\hat{\alpha}_4/2$. 
For the remainder of the paper, only two-dimensional geometry will be considered; we leave the three-dimensional case for future work.
The governing equation for
the free surface height, $h(x,t)$, may then be expressed as 
\begin{equation} \label{eq:NanoGov-PDE-Phi}
h_t + \left[ \Phi \left(   \cC h^3  h_{xxx}  + h^3 \Pieff'(h)  h_x \right) \right]_x    =  0\, ,
\end{equation}
\begin{equation} \label{eq:PhiFunction}
\Phi= \left\{ \begin{array}{rlr}
 \Phi_{\parallel} & = \lambda  - \nu  & \phi = 0 \\
 \Phi_{\perp} & = \lambda  +\nu & \phi = \frac{\pi}{2}
\end{array} \right. \;.
\end{equation} 
Here $\Phi_{\parallel}$ denotes substrate polar anchoring parallel to the $x$-axis ($\phi=0$)
and $\Phi_{\perp}$ denotes anchoring perpendicular to the $x$-axis ($\phi=\pi/2$);
 note that the value of $\Phi$ influences only 
only the timescale of fluid flow. Furthermore, if $\eta=0$, the governing equation (\ref{eq:NanoGov-PDE-Phi})
is independent of $\phi$ and our model corresponds to a fluid of
isotropic viscosity.  The structural disjoining pressure is defined as 
 \begin{equation} \label{eq:EffDisjoingPressure}
   \Pieff(h) = \cK f(h) +  \frac{\cN}{2}\left[\frac{m(h)}{h} \right]^2 \; ,
 \end{equation}
with $m(h)$ as defined in (\ref{eq:mh}) above.
Here, the second term on the right hand side is the response
due to the antagonistic anchoring conditions, while the first term corresponds to the van der Waals' disjoining pressure, taken here to be of 
power-law form
\begin{equation} \label{eq:DisjoingPressure}
   \Pi(h)= \cK f(h)  = \cK \left[ \left( \frac{b}{h} \right)^3 - \left( \frac{b}{h} \right)^2 \right] \;.
 \end{equation}
The form of the structural disjoining pressure,  (\ref{eq:EffDisjoingPressure}), is similar
to that considered by~\citet{Vandenbrouck1999}; however, there
are two important differences. First, the van der Waals term~(\ref{eq:DisjoingPressure})
considered here contains the repulsive term stabilising short wavelengths; this stabilisation 
is crucial if one wants to carry out simulations.   
Second, in the model of~\citet{Vandenbrouck1999}, the free surface azimuthal anchoring is
assumed to be determined by the  
average initial thickness of the film and does not evolve with $h$.   The power-law form of disjoining
pressure (\ref{eq:DisjoingPressure})
is used extensively in the literature; for a review see \cite{cm_rmp09} and for further background
regarding accounting for solid-fluid interactions, see~\cite{Israelachvili}.  

In the case (assumed henceforth) that $\phi$ is constant, 
a simple rescaling of time in (\ref{eq:NanoGov-PDE-Phi}) yields
\begin{equation} \label{eq:NanoGov-PDE}
h_t + \left[  \cC h^3 h_{xxx}  + h^3 \Pieff'(h)  h_x \right]_x    =  0 \; ,
\end{equation}
with $\Pieff$ as defined in (\ref{eq:EffDisjoingPressure}).
This equation will be considered in the rest of this paper.   Note that in this model nematic
effects enter only through modification of the disjoining pressure, by adding the structural component. 
Figure~\ref{fig:DisjoiningPressureCompare} shows typical forms of the disjoining pressure obtained using
the parameter values discussed in the following.  The structural disjoining pressure proposed in 
(\ref{eq:EffDisjoingPressure}) is similar to that expected for many other film problems, in particular the 
well studied polymer films on Si/SiO$_2$ substrates, see, e.g.~\citet{seeman_prl01}.
Therefore the results that we present in the rest of the manuscript are of relevance 
to a wide class of complex fluids.

{\bf Note:}
Equation (\ref{eq:NanoGov-PDE}) may be alternatively derived using the variational or gradient dynamics 
formulation \cite[see][]{Mitlin1993,Thiele2012}. 
%.
In this thermodynamically motivated approach the governing equation is expressed as
\begin{equation} \label{eq:NanoGov-GD}
h_t + \nabla \cdot \left[ Q(h) \nabla \left( \frac{\delta F}{\delta h} \right) \right]   =  0 \; ,
\end{equation}
where $Q(h)$ is the mobility function, and $F$ is the free energy functional. 
Previous work \citep[see][]{Lin2013a} showed that, with consistent boundary conditions, 
the gradient dynamics formulation is equivalent to the long wave approximation
of the Leslie-Ericksen equations. 
In terms of our model, $Q(h)=h^3$ and
\begin{equation} \label{eq:FreeEnergyFunctional}
 F(h) = -\cC \left[ 1 + \frac{\left( \nabla h \right)^2}{2} \right] - \frac{\cK b^2}{2 h}\left[ \frac{b}{h}-2\right] + \cN \left[ \frac{m(h)^2}{2 h} - \int \frac{m(s)m'(s)}{s} \; ds \right]  \; .
\end{equation}

\subsection{Scalings and Physical Parameters} \label{sec:Scalings}

We now discuss the choice of scaling parameters mentioned in the beginning of 
$\S$~\ref{sec:GovEqn} and the material parameters in (\ref{eq:nondimensionalParas}) and
(\ref{eq:viscosities}) for our model of thin \gls{NLC} films (\ref{eq:EffDisjoingPressure})
and (\ref{eq:NanoGov-PDE}).  
In experimental studies of dewetting of \gls{NLC} films, the film thickness
is typically of the order 10 to 100 nanometers~\citep[see][]{Herminghaus1998,Vandenbrouck1999};
therefore we set the representative film height $\hat{H}$ (used to scale $z$) to 100~nm.
For temperatures
far from a phase transition, the lower bound of the forbidden range of thickness
is a few nanometers
(it corresponds, in the experiments of~\citet{Cazabat2011}, to a trilayer of molecules).
We define this lower bound to correspond to the equilibrium film thickness and 
set the equilibrium film thickness $\hat{b} = 1$ nm. 
Bulk elastic energy is expected to be comparable to surface
anchoring energy for film thicknesses in the 100 nanometers range; therefore we set 
$\hat{\beta} = 100$~nm.
Periodic undulations are observed on the free surface of films with a height on the order of 
$\hat{H}$, and in experiments by \citet{Cazabat2011} and \citet{Herminghaus1998}, the wavelength of such undulations  is observed to be in the range of
10-100 $\mu$m. Choosing the length scale in the $(x,y)$ domain
$\hat{L}=10$~$\mu$m, we fix the aspect ratio $\delta=10^{-2}$.
We choose the timescale 
of fluid flow $\hat{T}_F=\hat{L}/\hat{U}$ to match the timescale on which undulations develop in experiments,
typically seconds to minutes~\citep[see][]{Herminghaus1998}; thus
we set $\hat{T}_F =1$ s, corresponding to a velocity scale $\hat{U}=10$~$ \mu$m~s\textsuperscript{-1}.

The remaining parameters to be determined are: 
$\hat{\gamma}$, the surface tension;
$\hat{K}$, the bulk elastic constant;
$\theta_F$, free surface azimuthal anchoring angle;
$\theta_S$, substrate azimuthal anchoring angle;
$\hat{A}$, a force per unit volume, proportional to the Hamaker constant; 
and $\hat{\mu}$, a representative viscosity. 
Many experimental studies, such as~\citet{Ausserre2011,Cazabat2011,Effenterre2003,Poulard2005,Schulz2011,Vandenbrouck1999} use 
the \gls{nCB} series of liquid crystals (where n refers to the number of carbons in the aliphatic tail), 
in particular, \gls{5CB}. Therefore, we choose these
physical parameters to match \gls{5CB} data.
It is noted by~\citet{Cazabat2011} that, due to impurities in the \gls{NLC}, published measurements of surface tension 
are scattered. However, to estimate the order of the non-dimensional
parameters in our model, we note that surface tension appears typically to be of order $10^{-2}$ N~m\textsuperscript{-1}, see \citet{Rey2002,Cazabat2011}.
The elastic constant, $\hat{K}$, is of the order $10^{-11}-10^{-10}$~N~\citep[see][]{Poulard2005,Delabre2008}. 
For the \gls{nCB} series of NLCs, the molecules prefer to align normal to the free surface
(homeotropic anchoring) and
in practice, substrates are often chosen and treated such that molecules align parallel to the substrate e.g. SiO$_2$;
therefore we set  $\theta_F =0$ and  $\theta_S =\pi/2$. Accurate values for the characteristic
viscosity of \gls{5CB} are difficult to find; however, \citet{Mechkov2009} was able to fit solutions of a thin film 
equation to experimental results
(this work considers micron-scale drop thicknesses and the model used assumes strong free surface anchoring and 
neglects van der Waals interactions) with $\hat{\mu}$ of order $10^{-2}$~Pa~s, a value in line with reported data~\citep[e.g.][]{Inoue2002}. Taking account of this available data, we choose
 $\hat{K} = 1\times10^{-11}$~N,  $\hat{\gamma}=60\times10^{-3}$~N~m\textsuperscript{-1},
and $\hat{\mu}=60\times10^{-3}$~Pa~s.

In summary, the dimensional scaling factors are then fixed as  
\begin{equation} \label{eq:ScalingChoice}
  \hat{L} = 10~\mu\textrm{m} \; , \;\;
  \hat{H} = 100~\textrm{nm}  \; , \;\;
  \hat{T}_F = 1~\textrm{s} \; ,   \;\;
  \hat{T}_R = 60~\mu\textrm{s} \; ,   \;\;
  \hat{U}= 10~\mu\textrm{m~s\textsuperscript{-1}} \; .
\end{equation}
Note that the assumption in $\S$~\ref{sec:WeakAnchoring} that the
timescale of elastic reorientation across the layer ($\hat T_R$) is much slower 
than that of fluid flow ($\hat{T}_F$) is satisfied. 
In addition, with the value of the contact angle
to be discussed in $\S$~\ref{sec:ContactAngle}, the contact line velocity 
of a spreading NLC drop measured by~\citet{Poulard2005} (at the chosen contact angle) 
is of a similar order to our derived velocity scale $\hat{U}$.

Based on the scales discussed above, the dimensionless parameters we use (at least, for our initial investigations) are
\begin{equation} \label{eq:ParameterChoice}
  \delta  =  0.01 \; , \;\;
  \cC  =  0.0857 \; , \;\;
  \cN  = 1.67  \; , \;\;
  \cK= 0.0017 \hat{A}~\textrm{m\textsuperscript{3}~J\textsuperscript{-1}} \; , \;
  b = 0.01 \; , \;\; 
  \beta = 1 \; ,
\end{equation}
and $\hat{A}$, the volume energy density associated with van der Waals forces,
is discussed next. 

\subsubsection{Contact Angle} \label{sec:ContactAngle}
To define a value for $\cK$ (\ref{eq:nondimensionalParas}),
we relate the spreading parameter (a function 
of the contact angle) to the stored energy per 
unit area of the film 
using the Young-Laplace equation, see~\citet{Diez2007}. 
The spreading parameter $\hat{S}$ is given by
\begin{equation} \label{eq:SpreadingParameter}
  \hat{S} = \hat{\gamma} (1-\cos \theta_c) = \hat{E}(\hat{h}=\infty) =  -\int_{\hat{b}}^{\infty} \hat{\Pi}_\textrm{S} (\hat{h}) \, d\hat{h} \;, %= \left[\kappa  u(h^*) + K \frac{m^2(h^*)}{2h^*} + K \cG (h^*) \right] \;,
\end{equation}
where $\theta_c$ is the equilibrium contact angle, $\hat{E}(\hat{h})$ is the stored energy per unit area,
and $\hat{\Pi}_\textrm{S} (\hat{h}) = (\hat\mu \hat{U}/\delta \hat{L}) \Pi_\textrm{S} (h)$,
where $\Pi_\textrm{S}$ is defined by (\ref{eq:EffDisjoingPressure}): the
usual van der Waals' disjoining pressure plus the elastic ``structural'' component dependent on the 
function $m(h)$ and with prefactor $\cN$. This elastic term of the integral in (\ref{eq:SpreadingParameter}) 
can not be evaluated analytically, due to the complicated functional
form of $m(h)$ as defined in (\ref{eq:mh}).
However, direct numerical integration shows (for the chosen values of $b$, $\beta$, and $w$), 
that the value of (\ref{eq:SpreadingParameter}) is approximately the same if the 
approximation $g(h)=1$ in the definition of $m(h)$ is made (less than 1\% difference).
Setting $g(h)=1$, the integral in (\ref{eq:SpreadingParameter}) may be evaluated
analytically; but the result is not simple to work with. Therefore,
we asymptotically expand in $\varepsilon = b/\beta= 0.01 \ll1$ and keep
the leading order term. The contact angles for \gls{NLC} on Si (or SiO$_2$) substrates were measured by
\citet{Poulard2005} to be around $10^{-2}$~--~$10^{-1}$ radians;
thus assuming the contact angle is small, $\theta_c \ll 1$, (\ref{eq:SpreadingParameter})
may be expressed as
\begin{equation} \label{eq:kappa2}
  \hat{A}  = \frac{\hat{\gamma} \theta_c^2 }{\hat{b}}  \left( 1 +  \kappa \left[ \frac{ \pi}{4}  \varepsilon  + O( \varepsilon^4) \right] \right) \; , \quad
  \kappa =  \frac {\hat{K}  }{ \hat{\gamma} \hat{b} \theta_c^2 } \;. %\approx 46.4\;. 
\end{equation}
To express (\ref{eq:kappa2}) in terms of the nondimensional parameters (\ref{eq:ParameterChoice}),
the order $\varepsilon^4$ term in (\ref{eq:kappa2}) is neglected, and the approximated equation
is multiplied by $\delta/\hat{\mu} \hat{U}$ to yield
\begin{equation} \label{eq:kappa2nondim}
 \cK b \approx \cC \left( \frac{\theta_c}{\delta} \right)^2 + \frac {\cN}{\beta \pi }\; ,
\end{equation}
where $b=\hat{b}/\hat{H}$ is the dimensionless equilibrium thickness.
For the chosen values of $\hat{\gamma}$ and $\hat{K}$, and $\theta_c=0.06$,
$ {\kappa \pi \varepsilon/4} \approx 0.364$ in (\ref{eq:kappa2}),
therefore, it is reasonable to neglect the elastic contribution to the spreading parameter. 
This is in agreement with
liquid crystals of low molar mass, where the anchoring energy 
on the free surface is usually significantly 
smaller than the isotropic surface tension energy, see~\citet{Rey2000}.
Neglecting the elastic term in (\ref{eq:kappa2}) then,
$\hat{A} = \hat\gamma \theta_c^2 /\hat{b},$
and with $\theta_c=0.06$, we then obtain
$\cK = 36.0$.

\subsection{Linear Stability Analysis (LSA)} \label{sec:2DNanoLSA}
We now focus on the stability properties of a flat film,
and in particular on the influence of the average film thickness, $H_0$. 
For this purpose, we first carry out LSA of a flat film solution to our governing equation 
(\ref{eq:NanoGov-PDE}). Assuming a solution of the form $h(x,t) = H_0 + \epsilon e^{iqx+i\omega  t}$
where $0<\epsilon \ll 1$, substitution in (\ref{eq:NanoGov-PDE}) yields the dispersion relation
\begin{equation} \label{eq:NanoDispersionRelation}
  \omega = i  H_0^3 \left[ \cC q^2 - \Pieff'(H_0)  \right]q^2 \; ,
\end{equation}
where 
\begin{equation} \label{eq:DerivEffDisjoingPressure}
 \Pieff'(h) = \cK f'(h) - \cN \frac{M(h)}{h^3} \;; \quad M(h) = m^2(h) - hm(h)m'(h) ;
\end{equation}
 with $f(h)$ defined by (\ref{eq:DisjoingPressure}), $m(h)$ by (\ref{eq:mh}) and parameters $\cC$, $\cK$, and $\cN$ by
(\ref{eq:nondimensionalParas}); their values are given in \S~\ref{sec:Scalings}.

\begin{figure}
\centering
 \includegraphics[width=.65\textwidth]{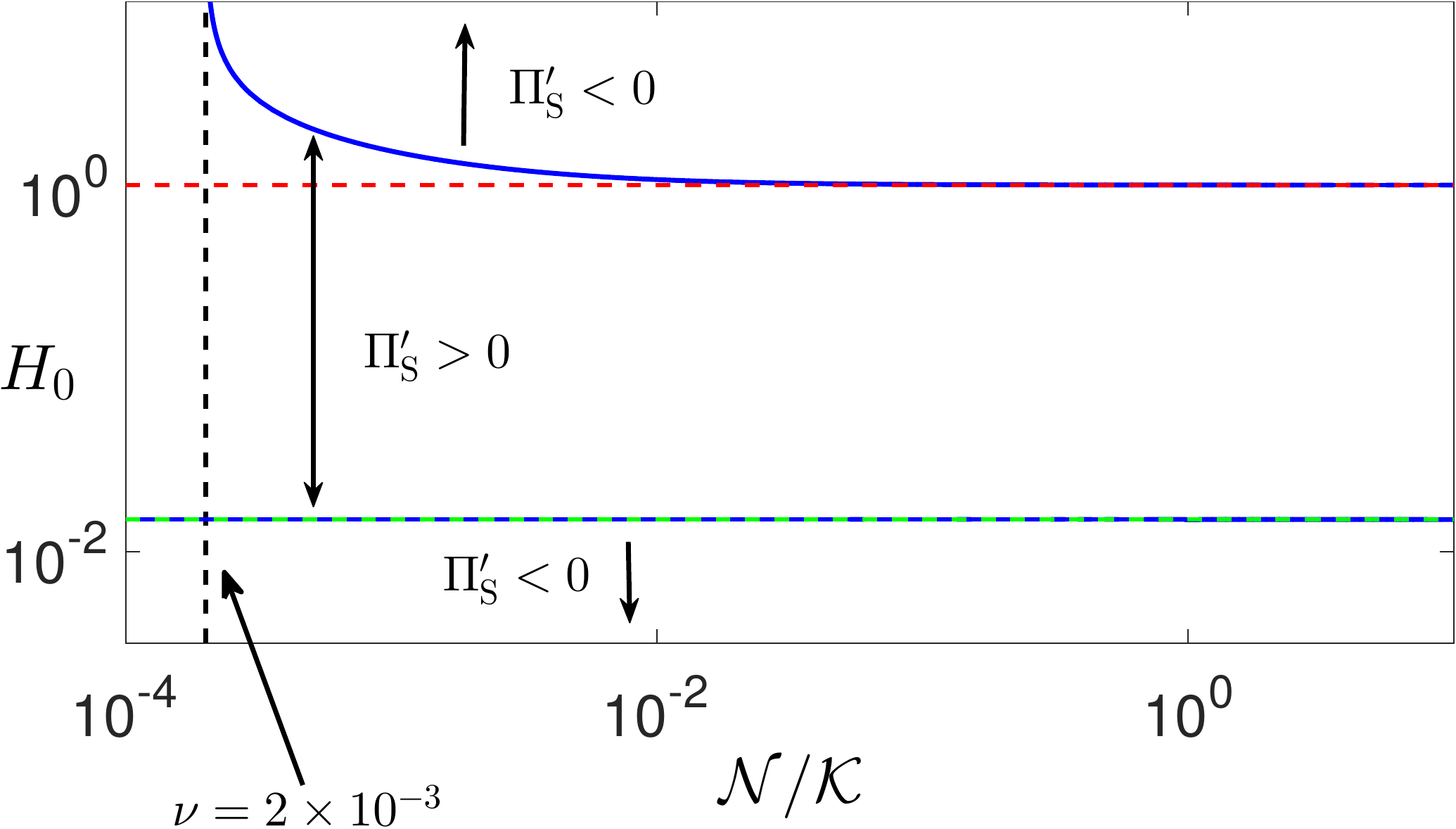}
  \caption[]{ The zeros of (\ref{eq:RootDerivEffDisjoingPressure}) (blue curves), 
              the zero of $M(H_0)$ (dashed red line); and the zero of $f'(H_0)$ (dashed cyan line, lying almost entirely on the lower blue curve).
              The region enclosed by the blue curves denotes film thicknesses that are linearly unstable, i.e. $\Pieff'(H_0)>0$.
              The  dashed black line corresponds to a critical $\cN/\cK$ value; below that value, the destabilising van der Waals force
              is stronger than the stabilising elastic response.
              }
  \label{fig:PhaseDiagram2}
\end{figure}

Assuming that $q$ is real, a film is always stable if $\Pieff'(H_0)<0$, and
is unstable otherwise.
Therefore, the values of $H_0$ for which the film is unstable are the roots of 
\begin{equation} \label{eq:RootDerivEffDisjoingPressure}
 \frac{\Pi_{\rm S}'(H_0)}{\cal K} = f'(H_0)- \frac{\cN}{\cK} \frac{M(H_0)}{H_0^3} \; ,
\end{equation}
which are calculated numerically and
 plotted (as two blue curves) in figure~\ref{fig:PhaseDiagram2} in the $(\cN/\cK,H_0)$-plane. 
In addition the roots of $f'(H_0)=0$ and $M(H_0)=0$ are also plotted.
Denoting the two roots of (\ref{eq:RootDerivEffDisjoingPressure}) as $H_{-}$ (the smaller root) 
and $H_{+}$ (the larger root), and noting that the region enclosed by blue curves
corresponds to $\Pieff'(H_0)>0$,
figure~\ref{fig:PhaseDiagram2} demonstrates that for $H_0\in(H_{-},H_{+})$, a flat film
is unstable; therefore we conjecture that the unstable range of thicknesses
in our model may be related to a range of so-called ``forbidden thicknesses'' noted experimentally by~\citet{Cazabat2011}.
It can be seen that, for the range of $\cN/\cK$ considered,  $H_{-}$ 
is closely approximated by the zero of the van der Waals term, i.e. $f'(H_{-})\approx 0$; thus $H_{-}\approx 3b/2$ (see (\ref{eq:DisjoingPressure})): the lower thickness threshold
between linear stability and instability is primarily determined by the van der Waals forces.

For $\cN/\cK > 0.01$, the largest root of (\ref{eq:RootDerivEffDisjoingPressure}), $H_{+}$, corresponds approximately to the root of $M(H_0)$,  suggesting that for sufficiently strong elasticity,
the upper threshold between linear instability and stability is determined by the antagonistic anchoring conditions (elastic response). 
For $H_0\gg b=0.01$, we may simplify $M(H_0)$ (see its definition in (\ref{eq:DerivEffDisjoingPressure})) by setting $g(H_0)=1$ in (\ref{eq:mh}), and $H_+ \approx \beta$ can be found analytically. 
This is confirmed numerically using the values from $\S$\ref{sec:Scalings}, which give $\cN/\cK = 0.0514$;  
the second root of (\ref{eq:RootDerivEffDisjoingPressure}) is then found to be
$ H_{+} \approx 1.0150 \approx \beta$ (see (\ref{eq:ParameterChoice})).

In the linearly unstable regime, only perturbations with wavenumber $q\in(0,q_c)$ are
unstable; here $q_c$ is the critical wavenumber
given by
$q_c  =  \sqrt{\Pieff'(H_0)/\cC}$.   
The most unstable mode is given by
$q_m = q_c\sqrt{2} = 2\pi/\lambda_m$,
with the growthrate 
\begin{equation}
\omega_m = {Q(H_0) \left[ \Pieff'(H_0)\right]^2\over 4 \cC}  = {H_0^3 \left[ \Pieff'(H_0)\right]^2\over 4 \cC}\, ,   
\label{eq:growthrate}
\end{equation}
(see (\ref{eq:NanoGov-GD})). We will use these quantities later in the text.
Note that in dimensionless form, $q_c$ and $q_m$ do not depend on the mobility function $Q(h)$ and the range(s) of unstable
thicknesses is/are determined by $\Pieff(h)$. Furthermore, the mobility function only affects the growth rate of the instability.

\subsection{Comparison to other Models}

There are several differences between the structural disjoining
pressure used in our model (\ref{eq:EffDisjoingPressure})
and others used in the literature. 
Unique to
our model is weak free surface anchoring, which relaxes the free surface
azimuthal anchoring as the film height evolves. This leads to
a critical thickness value, $\beta$, below which the elastic contribution 
(ignoring van der Waals forces) is linearly
destabilising for a flat film, and above which it is
linearly stabilising. Furthermore, and as discussed above,
figure~\ref{fig:PhaseDiagram2} shows that for $\cN \gg \cK$,
$H_+ \approx \beta $, i.e. the upper bound between linear
instability and stability is controlled by the elastic response.
In comparison, the model presented by \citet{Vandenbrouck1999}
sets the free surface azimuthal anchoring conditions to be a constant
for all film thickness (the value of which is determined by 
the average initial film thickness). This is equivalent to strong anchoring
($m(h)=1$ in our model), which leads to the elastic response being
always linearly stabilising; therefore, 
 in the theory of \citet{Vandenbrouck1999},
the upper threshold
between linear instability and stability, $H_+$, is determined
by the balancing of the destabilising van der Waals forces
and the stabilising elastic response, regardless of the values
of $\cN$ and $\cK$. 

Furthermore, we note that the structural
disjoining pressure term in our model may be
expressed as
\begin{equation}
\begin{split}
\Pieff'(h) h_x & = \cK f'(h)h_x - \frac{\cN}{h^3} \left[  m^2(h) - hm(h)m'(h) \right]  h_x \\
& = \cK f'(h)h_x - \frac{\cN m^2(h)}{h^3}  h_x - {\nabla \cG (h)\over h} \;,
\label{eq:G}
\end{split}
\end{equation}
where $\nabla \cG(h)$ is the gradient of the free surface anchoring energy. 
The derivation of $\cG(h)$ is nontrivial and we refer the reader to our 
previous work for further details \citep[][]{Lam2014,Lin2013}. Since the second term on the right hand 
side of the last equation in (\ref{eq:G})
is always stabilising, instability arises from the 
gradients of the free surface anchoring energy (the last term in (\ref{eq:G})), present  
only for non-constant $m(h)$ (i.e., only for weak anchoring).
Thus our model
includes an alternative instability mechanism due to elastic forces not captured by the model in \citet{Vandenbrouck1999}.

Excluding the $h^{-3}$ term from the van der Waals contribution
to the disjoining pressure (\ref{eq:DisjoingPressure}), the structural
disjoining pressure in our model (\ref{eq:EffDisjoingPressure})
is qualitatively similar to the total disjoining pressure derived by \citet{Ziherl2000}. 
This term is, however, essential for the purpose of carrying out 
numerical simulations.
In the work by \citet{Ziherl2000}, the structural component to the disjoining pressure
was derived by first considering the elastic energy due to fluctuations
in the director field from some initial state. A critical
thickness was defined such that, below this threshold, the initial 
director field state is uniform, and above, the director field is in the \gls{HAN} state.
Therefore our model has similar underlying physical mechanisms
driving instability: 
1) transition between a uniform director field for thin films to the \gls{HAN} state for thick films;
2) interplay between bulk elastic energy and surface anchoring energy as the film thickness changes; and 
3) inclusion of van der Waals forces.

However, there are also the following key differences: 
1) the transition
between the planar state and the \gls{HAN} state is continuous
in our model, whereas it is discontinuous in \citet{Ziherl2000};
2) \citet{Ziherl2000} assumed that the anchoring at the free surface was stronger than that at the substrate, therefore the uniform director field
state for thin films was homeotropic in their model, while it is planar in ours;
3) to obtain similar linearly stable and unstable film thickness values with their model would require setting
$\beta\approx 0.2$; and
4) only the total disjoining pressures are similar between the two models, the individual components (elastic contribution
and van der Waals term minus the $h^{-3}$ term), are qualitatively different.

\subsection{Numerical Scheme}

In the following we will present a number of numerical simulations of our
model in order to study its behaviour in detail.
To solve the governing equation (\ref{eq:NanoGov-PDE}) numerically,
we choose a spatial domain $x\in[0,x_L]$, and enforce periodic  
boundary conditions, 
\begin{equation} \label{eq:nano_bc_x}
   h_x(0,t)=h_x(x_L,t)= h_{xxx}(0,t)= h_{xxx}(x_L,t)=0\; ,
\end{equation}
where the right boundary $x_L$ will be fixed depending on the initial condition.
The numerical method employed is a fully-implicit Crank-Nicolson type discretization scheme,
with adaptive time stepping. A Newton iterative method
is implemented at each time step to deal with nonlinear terms.  More details about the numerical 
scheme  can be found e.g. in~\cite{siam}. For all numerical simulations
shown in this paper, the grid size is fixed at the equilibrium film thickness, i.e.
$\Delta x = b = 0.01$ and, unless stated otherwise, the parameters in governing
equation (\ref{eq:NanoGov-PDE}) are as given in \S~\ref{sec:Scalings}.

\section{Films exposed to global perturbations} \label{sec:RandomPert}

In this section, we present simulation results for
films of thickness $H_0$ with superimposed perturbations that are global
in character (cover the whole domain) and are intended to model infinitesimal random 
perturbations that are expected to be present in physical experiments.   
We focus on investigating film thicknesses in the linearly unstable regime, 
$H_0\in(H_-,H_+)\approx(0.015,1.015)$, and the morphology of dewetted films as  
a function of $H_0$. 
We start by discussing results for films perturbed by a single wavelength, on a domain 
equal in size to the imposed perturbation wavelength, and then proceed to consider random perturbations
and large domains.

\subsection{Simulations: Single Wavelength Perturbation}

\begin{figure}
\centering
  \subfigure[$H_0=0.05$]{
   \includegraphics[width=0.30\textwidth]{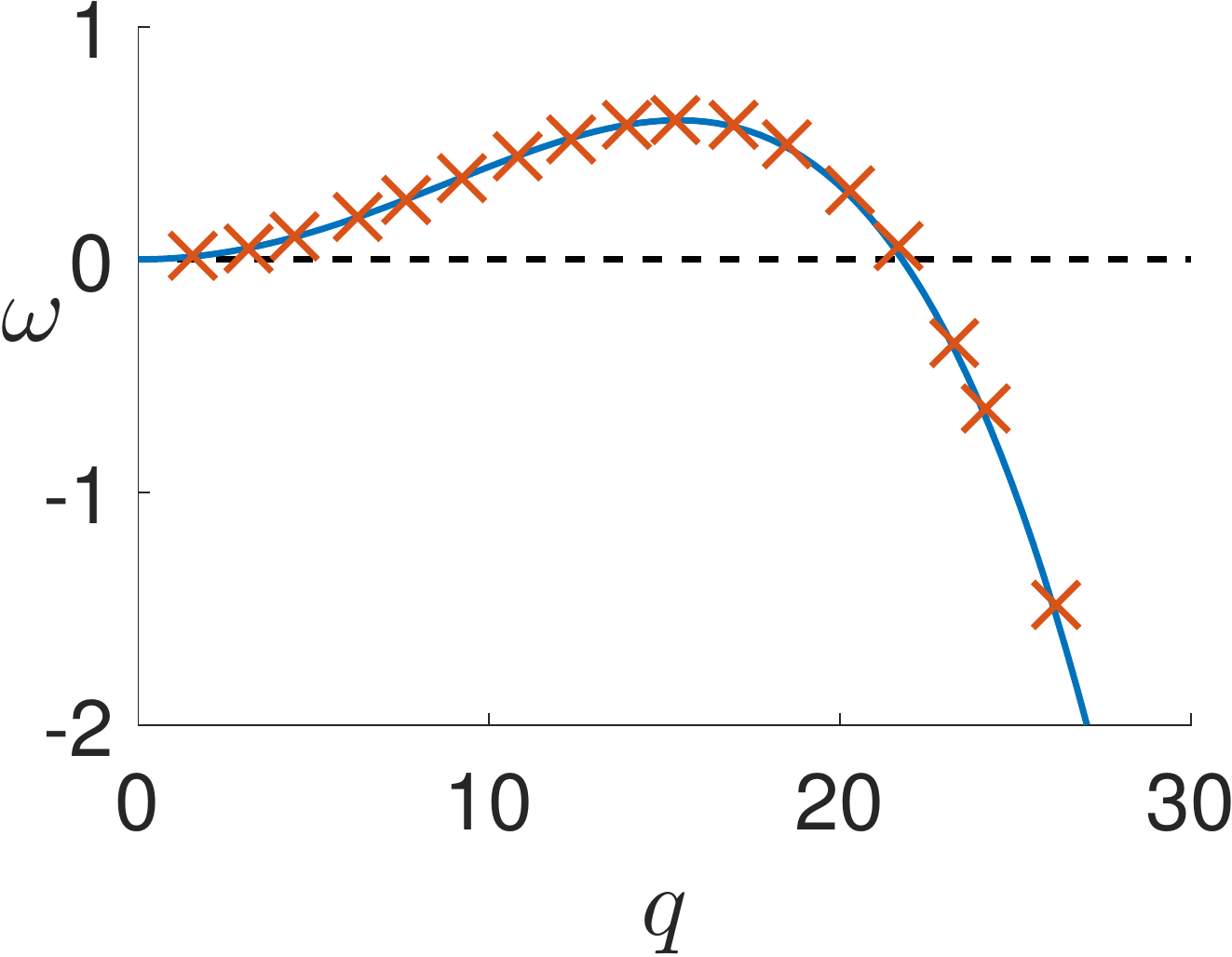}
  \label{fig:NanoDispersionRelation1} }
  \subfigure[$H_0=0.3$]{
   \includegraphics[width=0.30\textwidth]{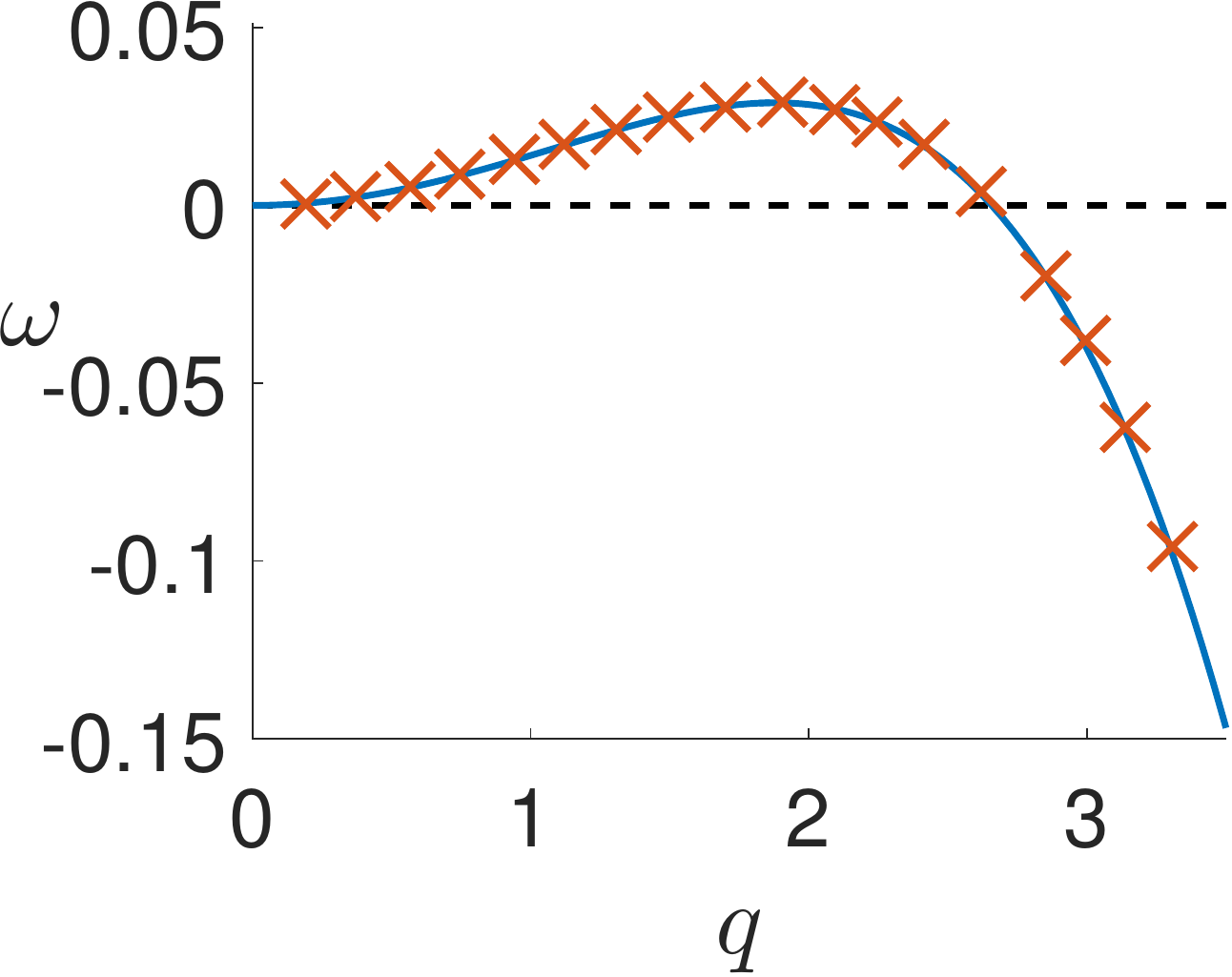}
  \label{fig:NanoDispersionRelation2} }
  \subfigure[$H_0=0.8$]{
   \includegraphics[width=0.30\textwidth]{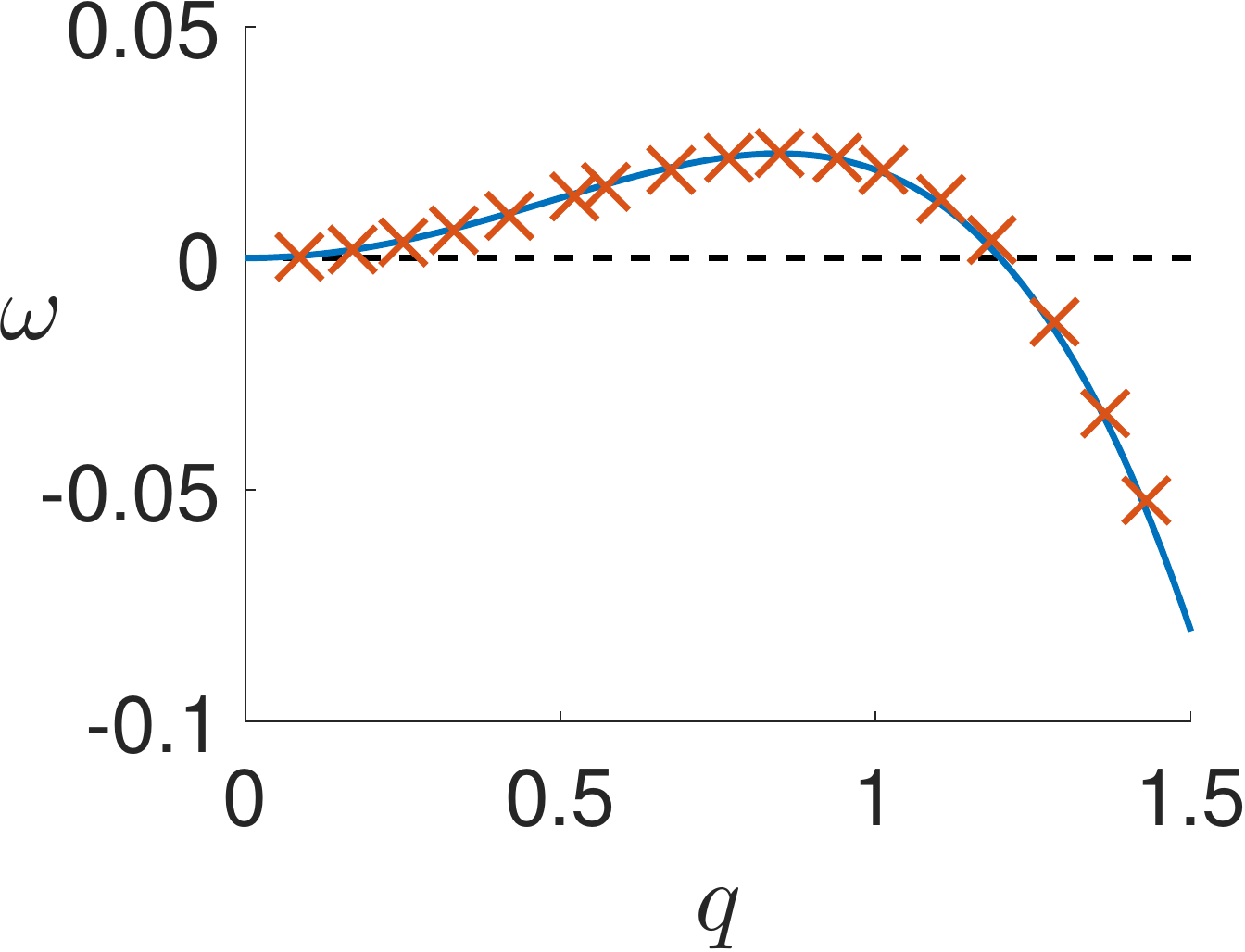}
  \label{fig:NanoDispersionRelation3} }
\caption{Comparison of the dispersion relationship (solid blue curves) (\ref{eq:NanoDispersionRelation}), 
and the growth rates extracted from numerical simulations ({\color{red}$\times$}) for $H_0=0.05,\,0.3$ and $0.6$. 
Other parameters are given in $\S$\ref{sec:Scalings}.} 
\label{fig:NanoDispersionRelation}
\end{figure}

To begin, we validate the numerical scheme by comparing the growth rates
extracted from numerical simulations to the expression (\ref{eq:NanoDispersionRelation}) 
derived from LSA. The initial condition is of the form
\begin{equation} \label{eq:NanoICSingle}
  h(x,t=0;H_0,q) =  H_0 \left[ 1 + 0.01\cos ( q x ) \right] \; , \quad x\in[0,x_L] \;, \quad x_L =  \frac{2 \pi}{ q}   \;,
\end{equation}
where $H_0$ and $q$ will be used to parameterise (\ref{eq:NanoICSingle}). 
In figure~\ref{fig:NanoDispersionRelation}, the growthrates extracted from
simulations are compared to those given by the dispersion relation. 
We see that the numerical simulations and LSA agree extremely well over the 
range of unstable film thicknesses and $q$ values considered.

To proceed, we investigate in detail how the resulting film morphologies depend on the initial film
thickness, $H_0$. Our simulations reveal that droplets may be classified according to their heights, 
with distinct droplet sizes emerging at different times over the course of a simulation. 
We therefore plot free surface height on a log scale in order to properly visualise smaller drops; also
we show the free surface profile at four different times to illustrate the timescales on 
which dewetting and droplet formation occur.
\begin{figure}
\centering
  \subfigure[$H_0=0.2$, $\omega_m^{-1}\approx 34.22$]{
   \includegraphics[width=0.35\textwidth]{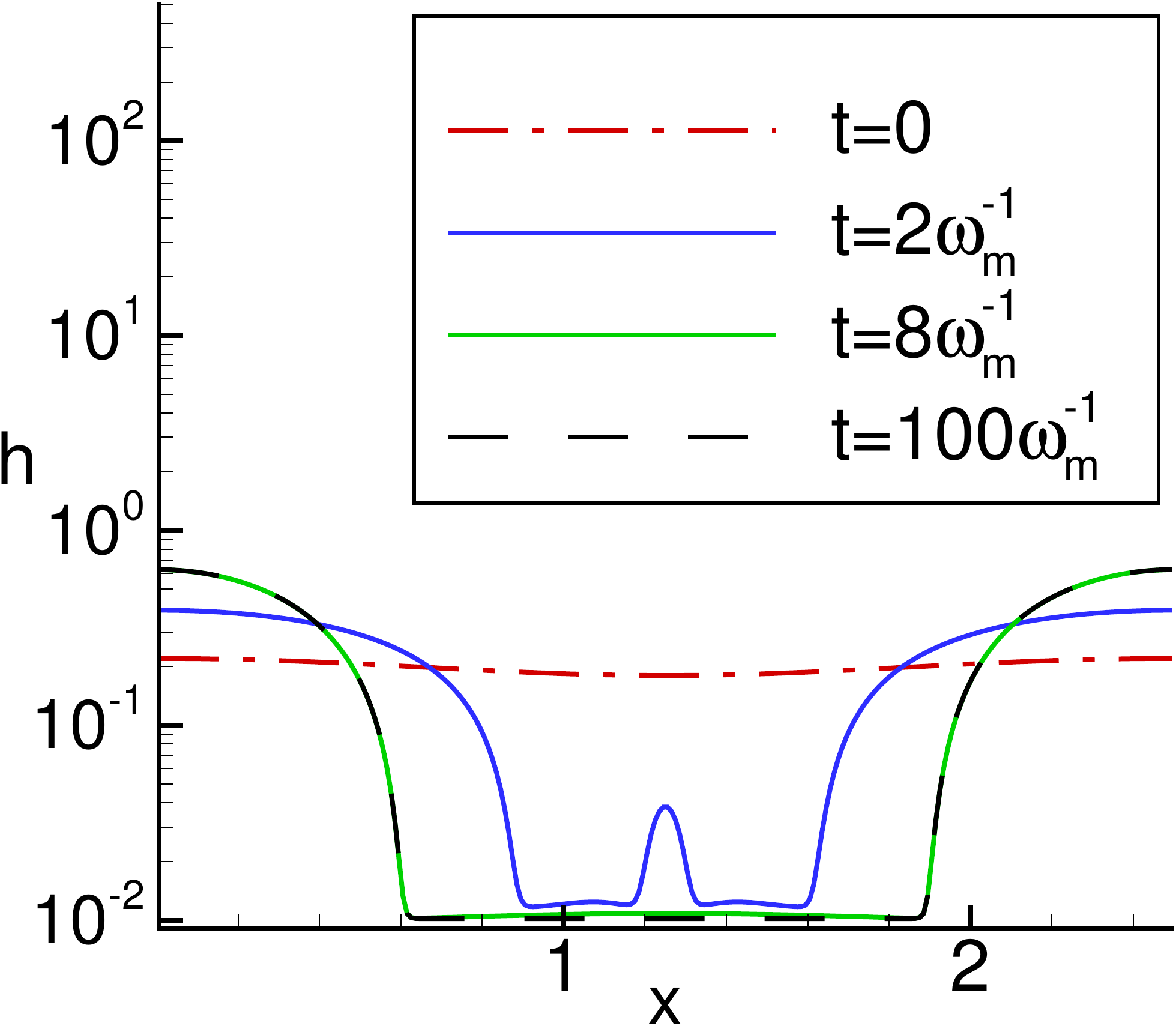}
  \label{fig:NLCvdw_002_Single} }
    \quad \quad
  \subfigure[$H_0=0.3$, $\omega_m^{-1}\approx 34.64$]{
   \includegraphics[width=0.35\textwidth]{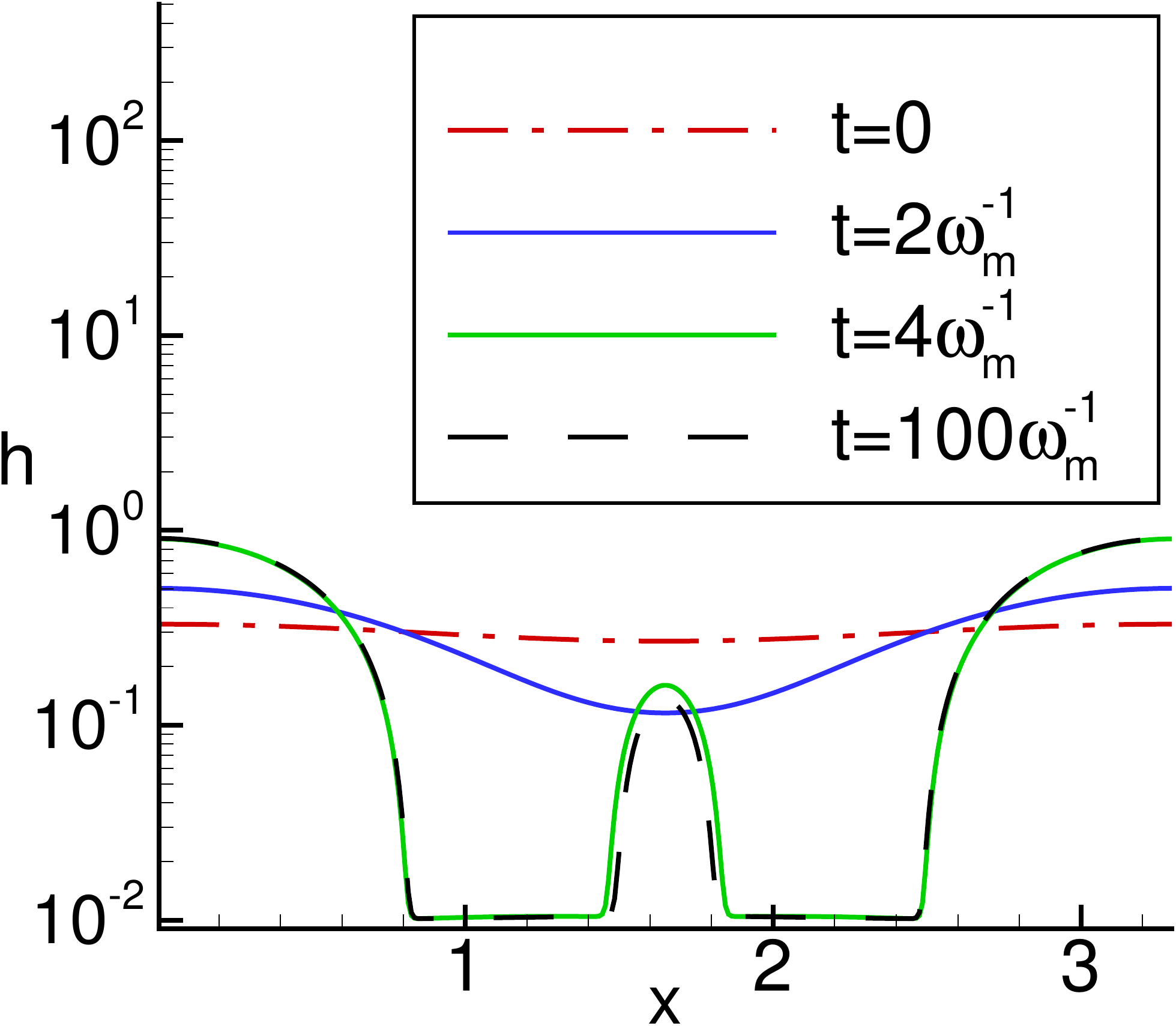}
  \label{fig:NLCvdw_003_Single} }
  \\
  \subfigure[$H_0=0.6$, $\omega_m^{-1}\approx 19.21$]{
   \includegraphics[width=0.35\textwidth]{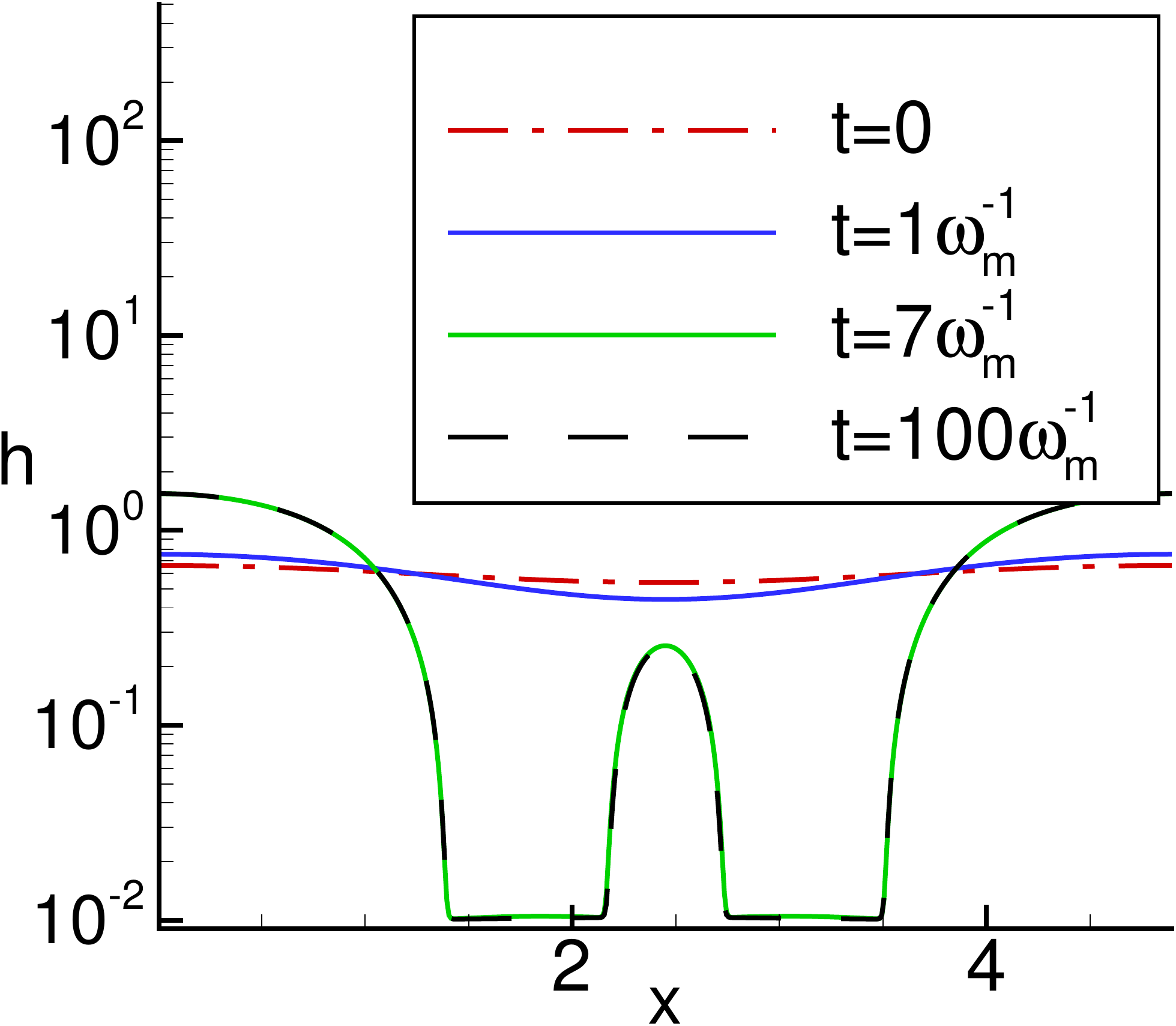}
  \label{fig:NLCvdw_006_Single} }
    \quad \quad
  \subfigure[$H_0=0.95$, $\omega_m^{-1}\approx 423.4$]{
   \includegraphics[width=0.35\textwidth]{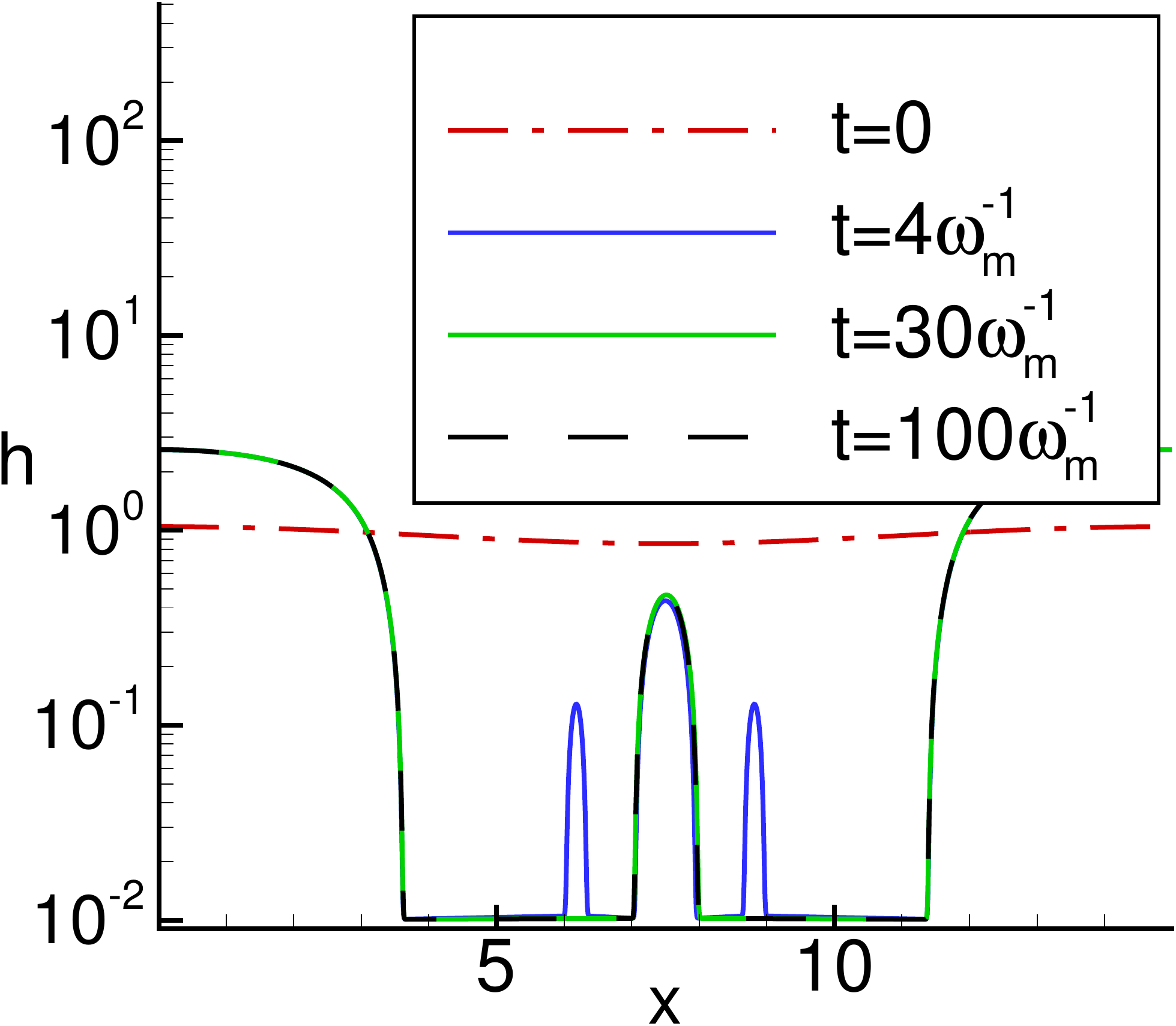}
  \label{fig:NLCvdw_010_Single} }
\caption{Free surface evolution for various initial average film thicknesses, $H_0$. The initial condition
is given by (\ref{eq:NanoICSingle}) with $x_L=\lambda_m$. 
Other parameters are given in $\S$\ref{sec:Scalings}.} 
\label{fig:NLCvdw_Single}
\end{figure}

Figure~\ref{fig:NLCvdw_Single}, where  $q=q_m$ (thus $x_L=\lambda_m)$ in (\ref{eq:NanoICSingle}),
shows the free surface evolution of the \gls{NLC}
film for several different values of the initial average film thickness, $H_0$. Initially, the perturbation
grows, eventually forming large drops at the domain boundaries, as expected from LSA, 
and an additional secondary drop at the centre.
On timescales much longer than the breakup time, comparable to  $\omega_m^{-1}$, different dynamics emerge, 
however, depending on the film thickness $H_0$.    
For thinner films, shown in figure~\ref{fig:NLCvdw_002_Single}, the secondary drop persists for a short
time only; while for thicker films, see figures~\ref{fig:NLCvdw_003_Single}
and \ref{fig:NLCvdw_006_Single}, the secondary drops exist
on a long time scale. For  films near the critical thickness of linear instability, 
see figure~\ref{fig:NLCvdw_010_Single}, we observe an additional set of tertiary drops
forming on either side of the secondary drop.  These tertiary drops
form at the same time as the primary drops, and exist on a timescale that is longer 
than that of dewetting, but shorter than that on which the secondary drop persists.
 We note that the times considered here are not asymptotically long, 
as considered by~\cite{glasner_pre03}.  
\begin{figure}
\centering
 \includegraphics[width=0.7\textwidth]{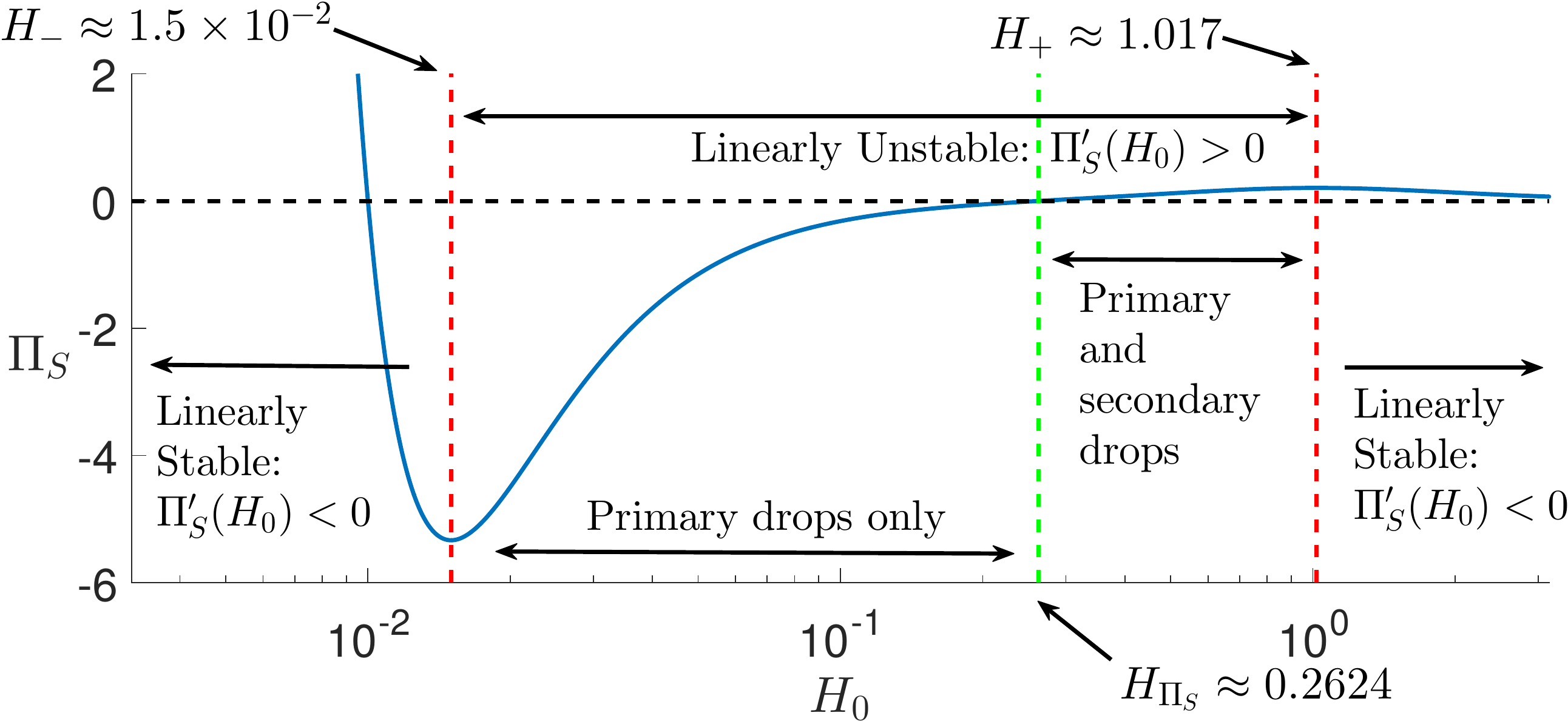}
  \caption[]{
  A summary of the final states obtained in simulations and their relationship to the structural disjoining pressure. The region 
enclosed by the dashed red lines is unstable. The    dashed green line, at $H = H_\Pieff$, corresponds to
the zero of the structural disjoining pressure.  These line patterns will be used consistently in all the following figures.
}
  \label{fig:DisjoiningPressureDiagram}
\end{figure}

Figure~\ref{fig:DisjoiningPressureDiagram}
shows schematically our findings regarding film stability and 
existence of secondary drops on time scales that are long compared to $\omega_m^{-1}$.   
The region enclosed by the  dashed red lines denotes the linearly unstable regime (determined by (\ref{eq:DerivEffDisjoingPressure})), and the
 dashed green line corresponds to $H_\Pieff\approx 0.2624$, the zero of the structural disjoining pressure $\Pieff$ (\ref{eq:EffDisjoingPressure}).
The secondary drops are found only for film thicknesses with positive disjoining pressure, 
as we discuss further below in $\S$~\ref{sec:NanoRandomSims}.
To the best of our knowledge, this finding is novel, and is one of the
predictions of the present work.  It would be of interest to verify it experimentally, not only for
\gls{NLC} films, but also for other films where positive disjoining pressures may be found, 
for example, for polymer films where positive disjoining pressure is due to fluid/substrate interaction; 
or for ferrofluid films in electric 
or magnetic fields~\citep[see][]{Seric2014}.   

\subsection{Simulations: Multiple Random Perturbations} \label{sec:NanoRandomSims}
To  test the connection between the formation of secondary drops 
and the sign of the disjoining pressure, we carried out simulations in
large domains with imposed perturbations of random amplitudes.
The initial condition is defined as
 \begin{equation} \label{eq:NanoICRandom}
   h(x,t=0;H_0,x_L) =  H_0 \left[ 1+ \ep \chi(x;x_L) \right]  \;  , \quad
   \chi(x;x_L) = \sum_{n=1}^{300} a_n \cos\left( \frac{n x}{x_L} \right) \;  , \quad
    x\in[0,x_L] \; ,
 \end{equation}
where $a_n \in [-1,1]$ are uniformly distributed, $\ep$ is chosen such that
$| \ep \chi(x;x_L) | \le 0.01$ (i.e. the total perturbation size is bounded
by $0.01$)
%{\color{red} *** Is this a roundabout way of saying $\ep=0.01$? If so, why not just give the value used? ***}
, and $H_0$ and $x_L$ will be used to parameterise (\ref{eq:NanoICRandom}).

 % -----------------------------------------
 % - Note: png used instead of pdf due to  -
 % -       Tecplot error displaying legend -
 % -       text. May be to due to larger   -
 % -       aspect ratio                    -
 % -----------------------------------------
\begin{figure}
\centering
  \subfigure[$H_0=0.2$, $\omega_m^{-1}\approx 34.22$]{
   \includegraphics[width=0.95\textwidth]{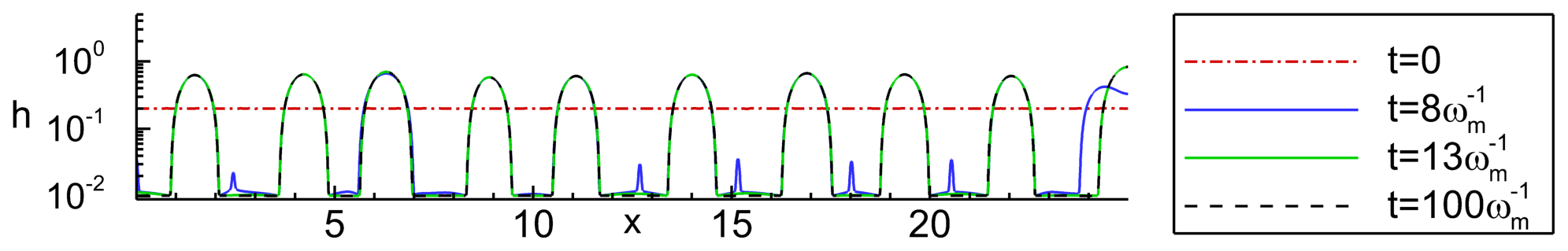}
  \label{fig:NLCvdw_002_Random} }
  \subfigure[$H_0=0.3$, $\omega_m^{-1}\approx 34.63$]{
   \includegraphics[width=0.95\textwidth]{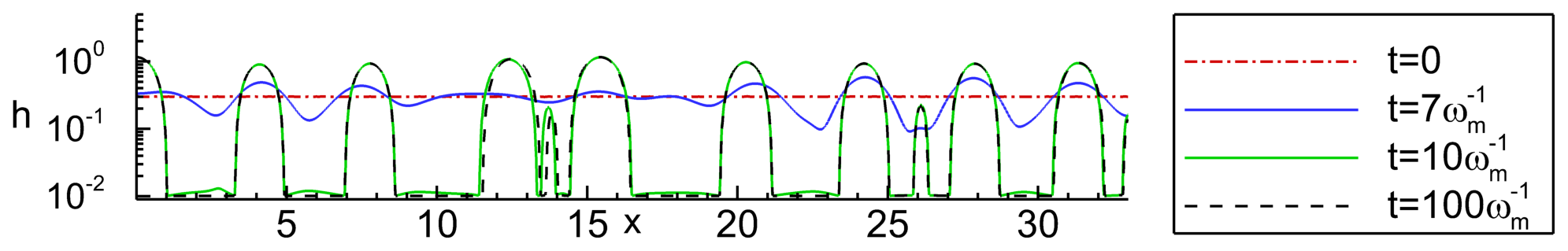}
  \label{fig:NLCvdw_003_Random} }
  \\
  \subfigure[$H_0=0.6$, $\omega_m^{-1}\approx 19.21$]{
   \includegraphics[width=0.95\textwidth]{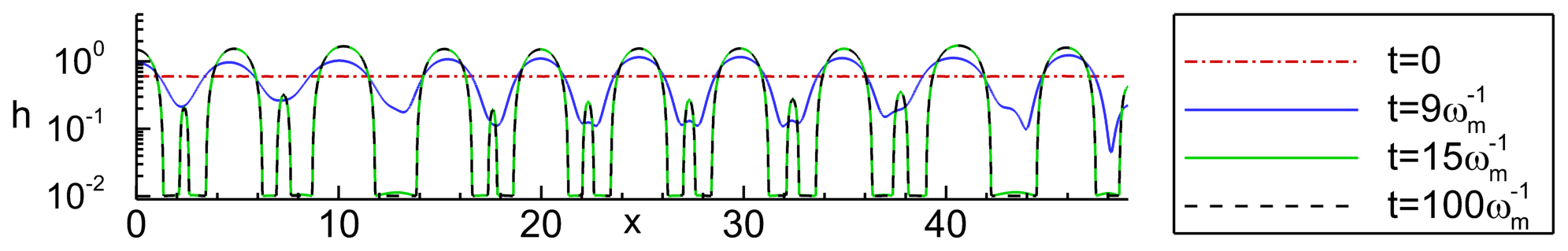}
  \label{fig:NLCvdw_006_Random} }
  \subfigure[$H_0=0.95$, $\omega_m^{-1}\approx 423.4$]{
   \includegraphics[width=0.95\textwidth]{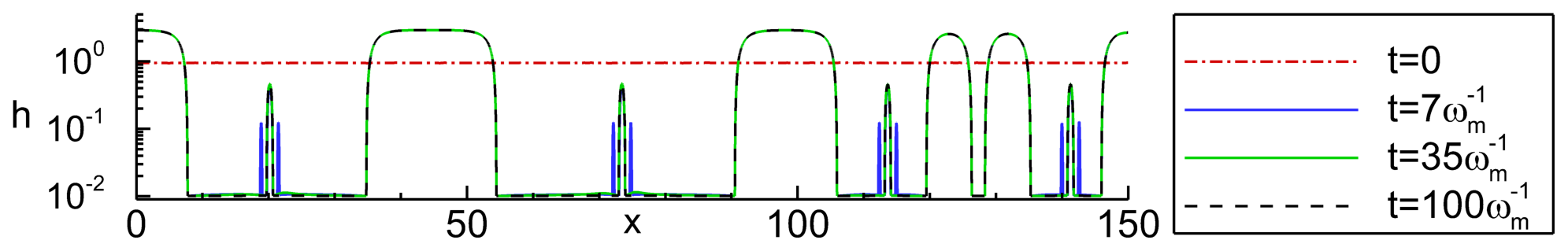}
  \label{fig:NLCvdw_010_Random} }
\caption{Free surface evolution for various initial average film thicknesses, $H_0$. The initial condition
is given by (\ref{eq:NanoICRandom}) with $x_L=10\lambda_m$. Other parameters are given in $\S$\ref{sec:Scalings}.} 
\label{fig:NLCvdw_Random}
\end{figure}
%
% Figure~\ref{fig:NLCvdw_Random} shows the free surface evolution for a range of initial thicknesses $H_0$, and
% for the same parameters as in Figure~\ref{fig:NLCvdw_Single}. 
 Figure~\ref{fig:NLCvdw_Random} shows the free surface evolution for initial condition
 (\ref{eq:NanoICRandom}) with $x_L=10\lambda_m$, and the same parameters and same range of initial 
 thicknesses, $H_0$, as in figure~\ref{fig:NLCvdw_Single}. 
The main observation is that a minimum film 
thickness is still required for formation of long-lived secondary drops.

\begin{figure}
\centering
  \subfigure[$H_0=0.2$]{
   \includegraphics[width=0.35\textwidth]{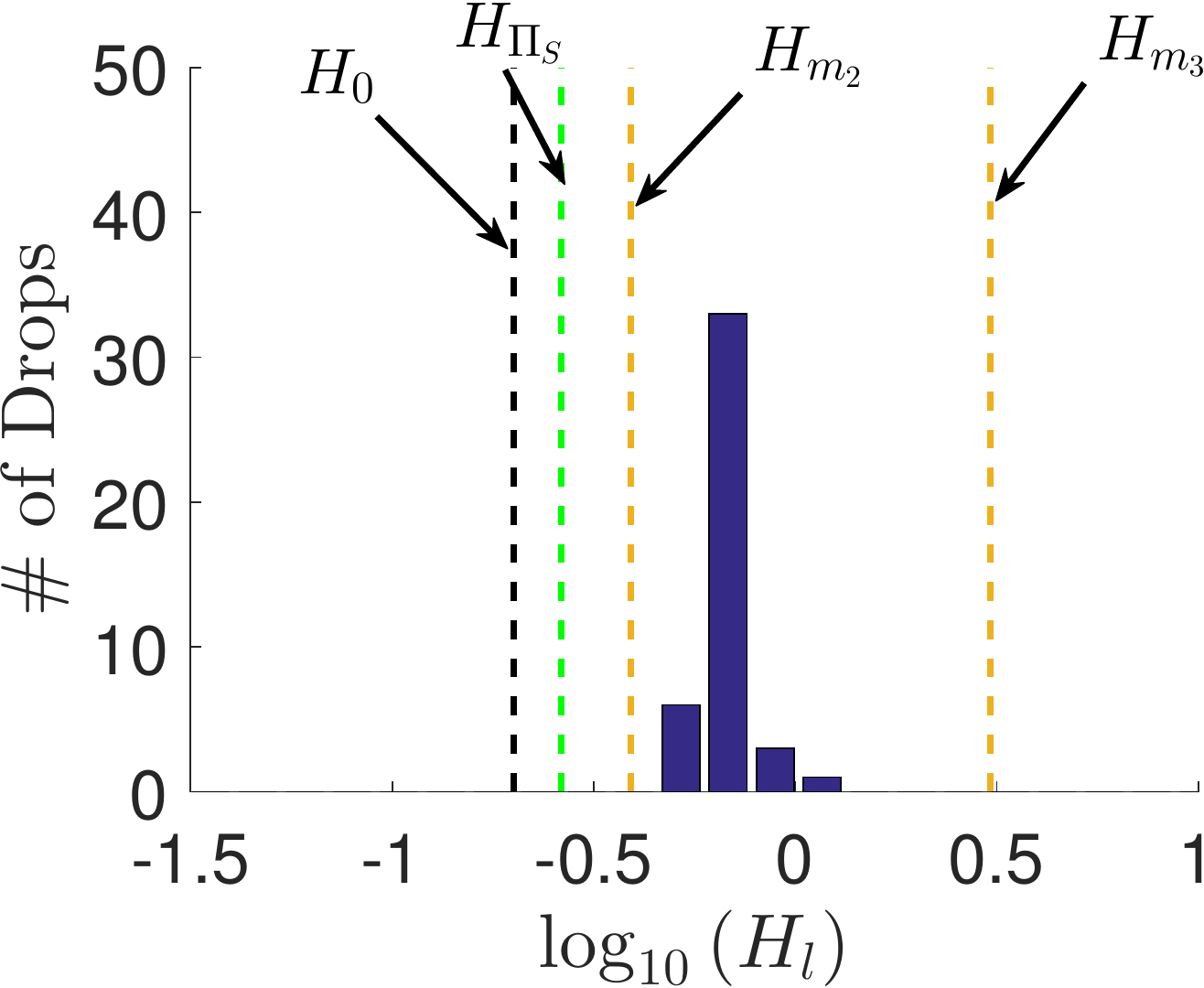}
  \label{fig:NLCvdw_002_DropHeight} }
    \quad \quad
  \subfigure[$H_0=0.3$]{
   \includegraphics[width=0.35\textwidth]{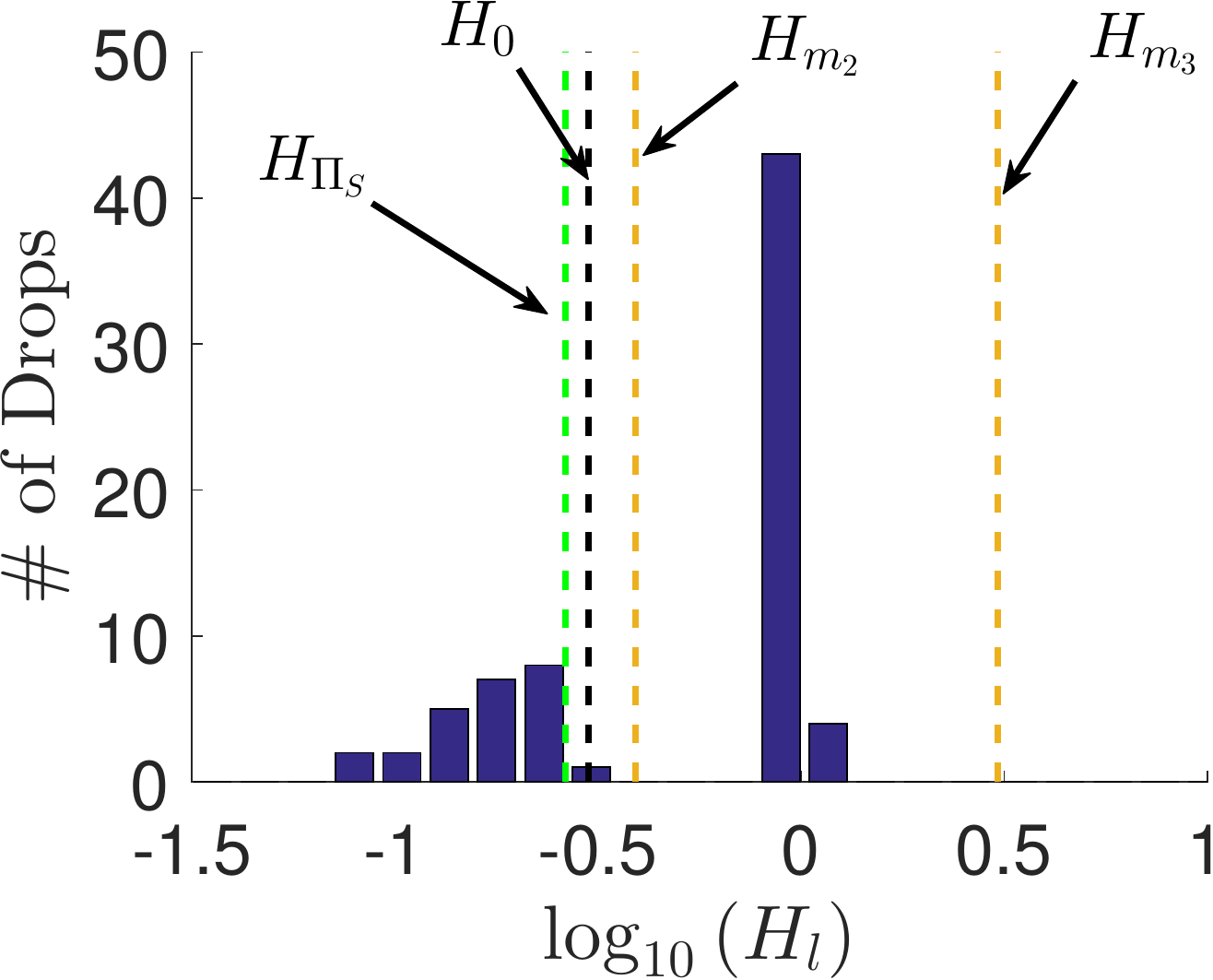}
  \label{fig:NLCvdw_003_DropHeight} }
  \\
  \subfigure[$H_0=0.6$]{
   \includegraphics[width=0.35\textwidth]{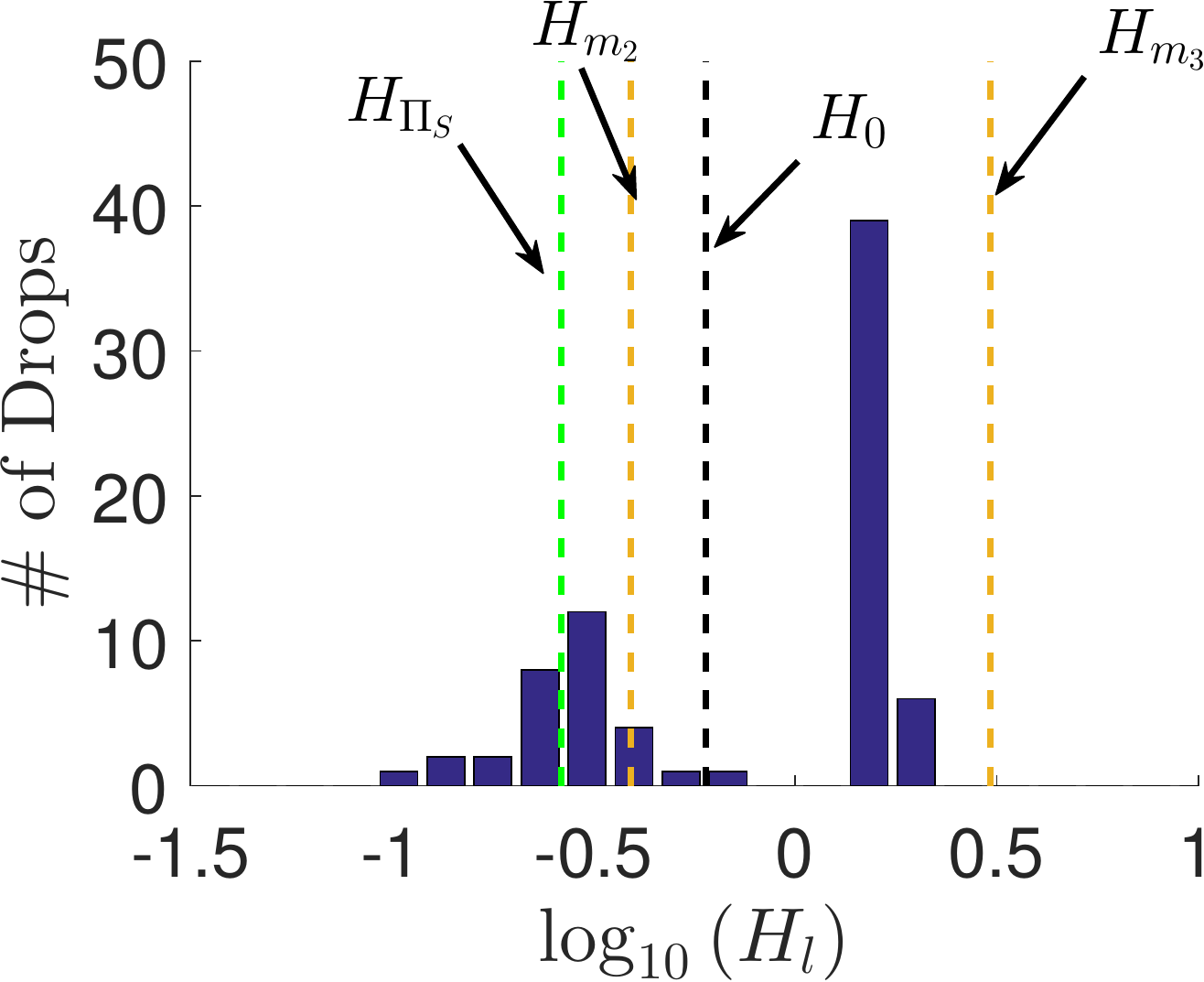}
  \label{fig:NLCvdw_006_DropHeight} }
    \quad \quad
  \subfigure[$H_0=0.95$]{
   \includegraphics[width=0.35\textwidth]{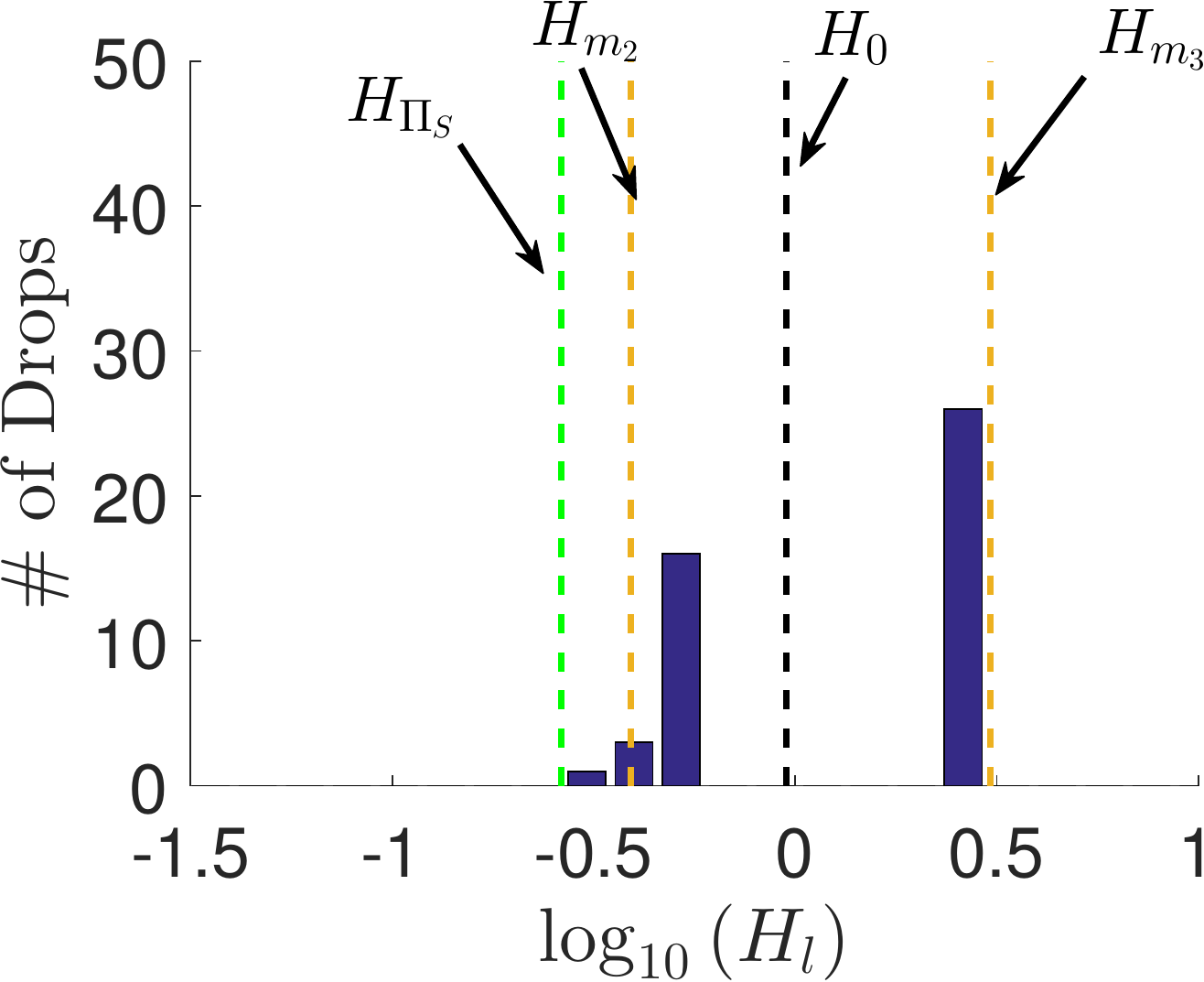}
  \label{fig:NLCvdw_010_DropHeight} }
\caption{Histogram of initial individual drop heights extracted from the corresponding
simulations in figure~\ref{fig:NLCvdw_Random}.  We see
that when $\Pieff(H_0)>0$, panels (b), (c), and (d),
the distribution is bimodal (primary and secondary drops), whereas for $\Pieff(H_0)<0$, panel (a),
the distribution is unimodal (primary drops only).
The black dashed line 
denotes the initial film thickness $H_0$ and dashed gold lines
denote the range of metastable thicknesses, to be discussed in $\S$~\ref{sec:MetastableAnaylsis}.
 } 
\label{fig:NLCvdw_DropHeight}
\end{figure}

To discuss this finding in more detail, we also consider the drop sizes, 
defined by the heights of drop centres, $H_l$.  
Figure~\ref{fig:NLCvdw_DropHeight} shows a histogram of the log of the drop heights extracted from
simulations corresponding to the same $H_0$ values as used in figure~\ref{fig:NLCvdw_Random}, but carried 
out on a larger domain ($x_L = 50 \lambda_m$) to improve statistics.
We see that for $H_0<H_\Pieff$,  figure~\ref{fig:NLCvdw_002_DropHeight},
the distribution is unimodal; and for $H_0>H_\Pieff$, 
figures~\ref{fig:NLCvdw_003_DropHeight}-\ref{fig:NLCvdw_010_DropHeight},
the distribution is bimodal (recall that $H_\Pieff$ is the zero of the structural disjoining pressure, $\Pieff (H_\Pieff)=0$).  
In addition, the  dashed black line denotes the initial film thickness $H_0$,
and it may be seen that $H_0$ provides an adequate threshold value between (larger) primary
drops and (smaller) secondary  drops (if they exist). The relevance of the  dashed
gold lines will be discussed shortly in $\S$~\ref{sec:RandomPrevWork}, but for now, note that
as $H_0\rightarrow H_+$ (see figure~\ref{fig:NLCvdw_010_DropHeight}), the height of primary drops 
converges to the rightmost gold line.

\begin{figure}
\centering
  \subfigure[]{
   \includegraphics[width=0.40\textwidth]{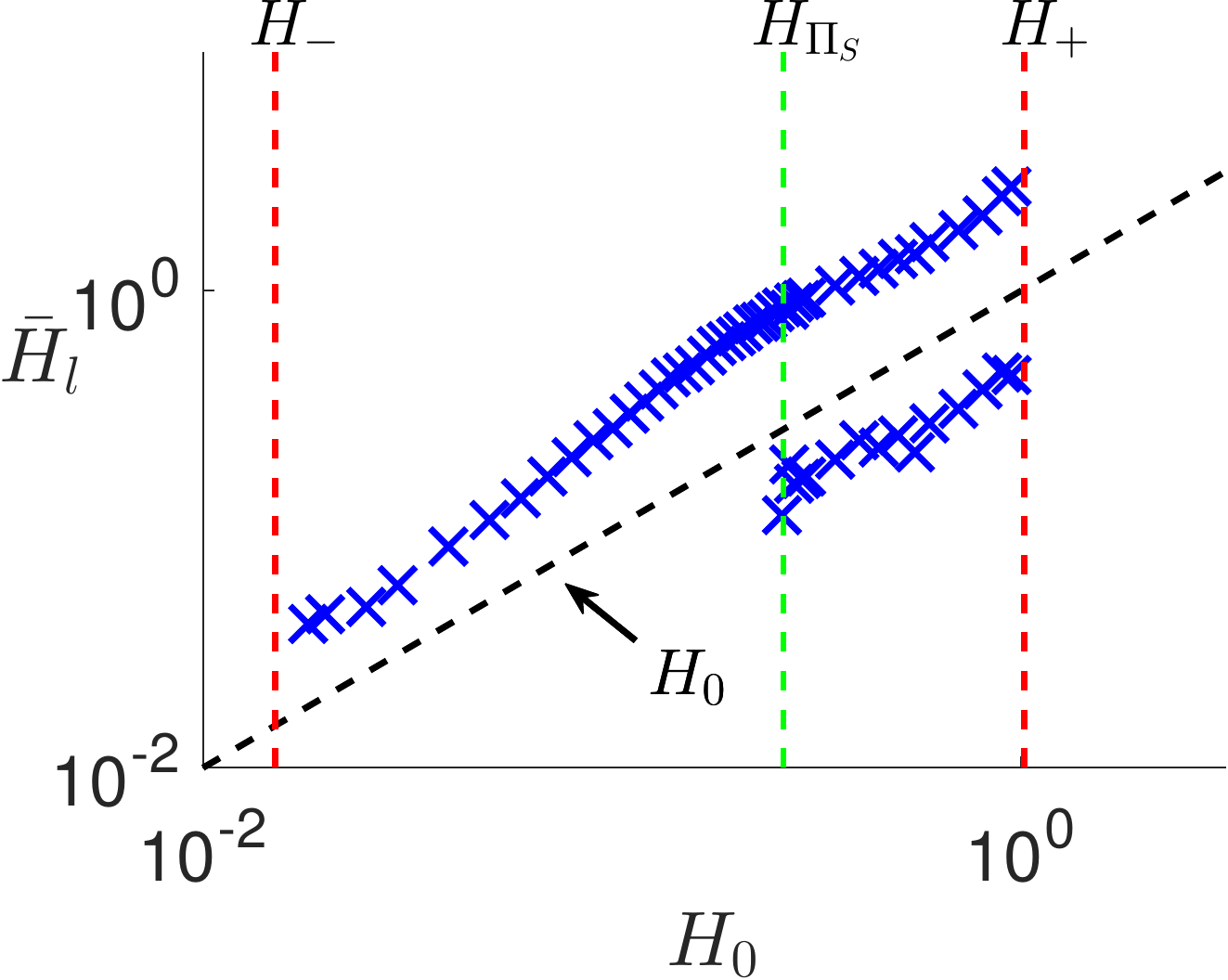}
  \label{fig:DropHeightsModeVsThickness1} }
    \quad \quad
  \subfigure[]{
   \includegraphics[width=0.40\textwidth]{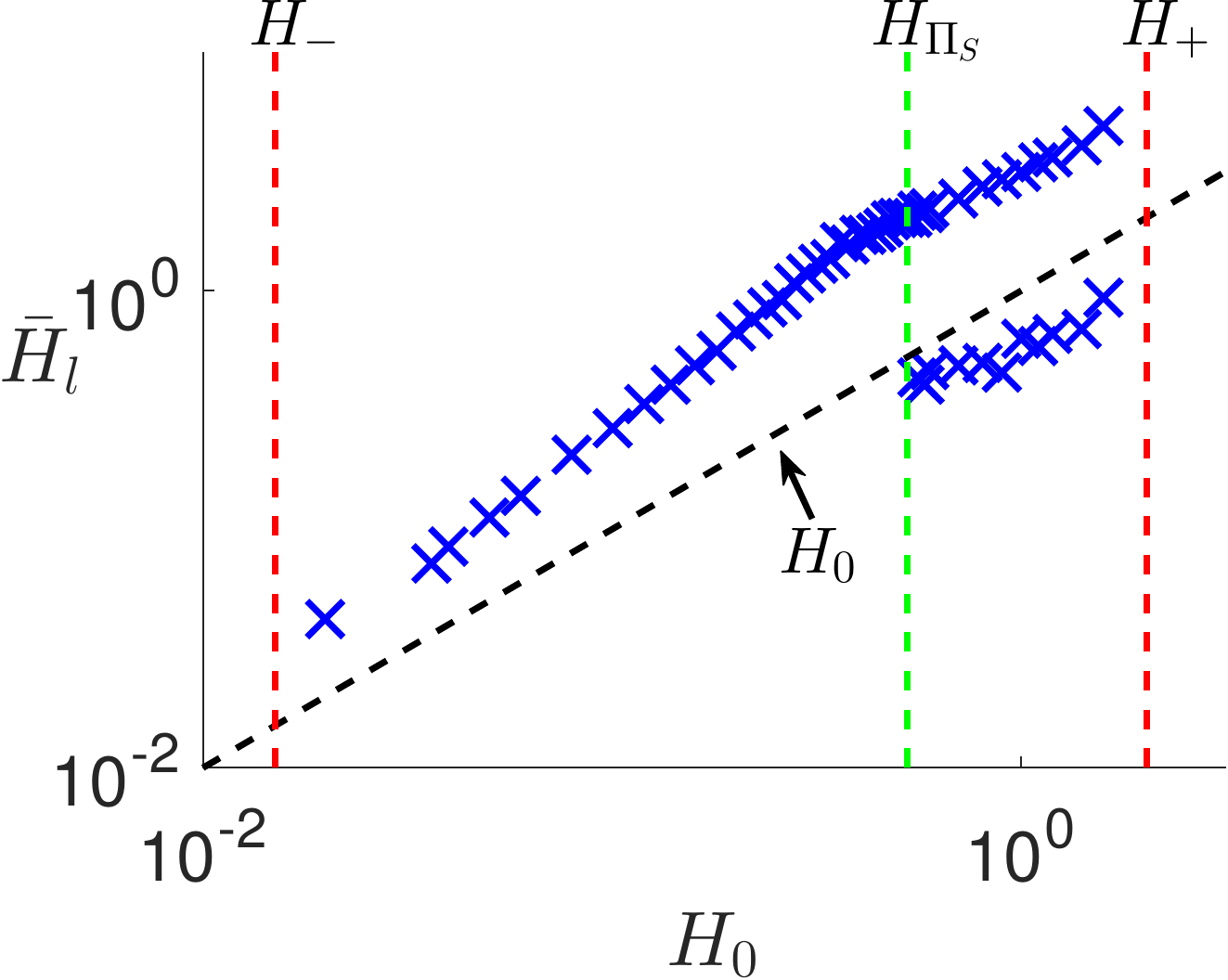}
  \label{fig:DropHeightsModeVsThickness2} }
\caption{Mode(s) of drop heights (peak(s) of histogram data in figure~\ref{fig:NLCvdw_DropHeight}) for 
(a) $\beta=1$, and (b) $\beta =2$. When
$\Pieff(H_0)$ changes from negative to positive values (green dashed
line), the distribution of the height of drop centres transitions from unimodal (primary drops only) to a bimodal
distribution (primary and secondary drops). Furthermore, the initial film thickness, $H_0$ (black dashed line), proves an adequate threshold between the heights of the primary and secondary drops.
 Note that for $\beta=1,~2$, $H_{\Pieff}=0.2624,~0.5275$, respectively.
}
\label{fig:DropHeightsModeVsThickness}
\end{figure}

To further illustrate the generality of our finding that secondary drops occur only if $\Pieff(H_0)>0$,
we next consider the mode(s) of the drop heights, $\bar{H}_l$,
the value(s) of which is/are given by the peak(s) of the distribution of $H_l$ in the histograms in figure~\ref{fig:NLCvdw_DropHeight},
for two values of $\beta$  (the parameter defining the thickness where the bulk elastic energy is 
comparable to surface anchoring energy, see (\ref{eq:EffDisjoingPressure})). 
Figure~\ref{fig:DropHeightsModeVsThickness}
plots $\bar{H}_l$ as a function of $H_0$:  the part (a) corresponds to the parameters used 
in figure~\ref{fig:NLCvdw_DropHeight} ($\beta=1$) and 
the part (b) shows the data for simulations with $\beta=2$. We clearly see that, in both cases, 
the secondary droplets appear only for film heights such that $\Pieff(H_0)>0$,
and that $H_0$ (black dashed line)
can be used to differentiate  primary and secondary drops.  

\subsection{Relation to Previous Work} \label{sec:RandomPrevWork}

\begin{figure}
\centering
  \subfigure[]{
   \includegraphics[width=0.31\textwidth]{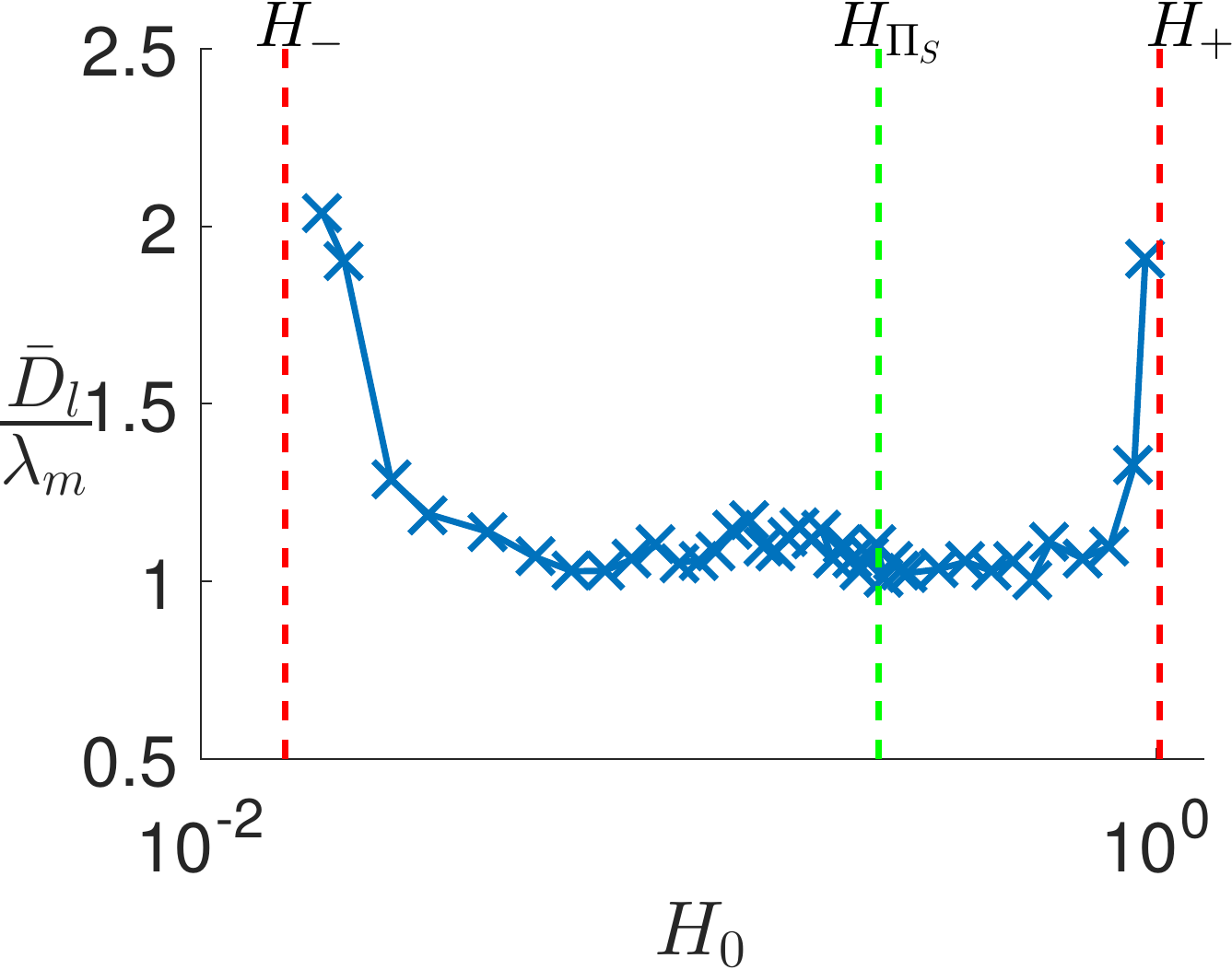}
  \label{fig:RandomMeanSpacing} }
    \subfigure[]{
   \includegraphics[width=0.31\textwidth]{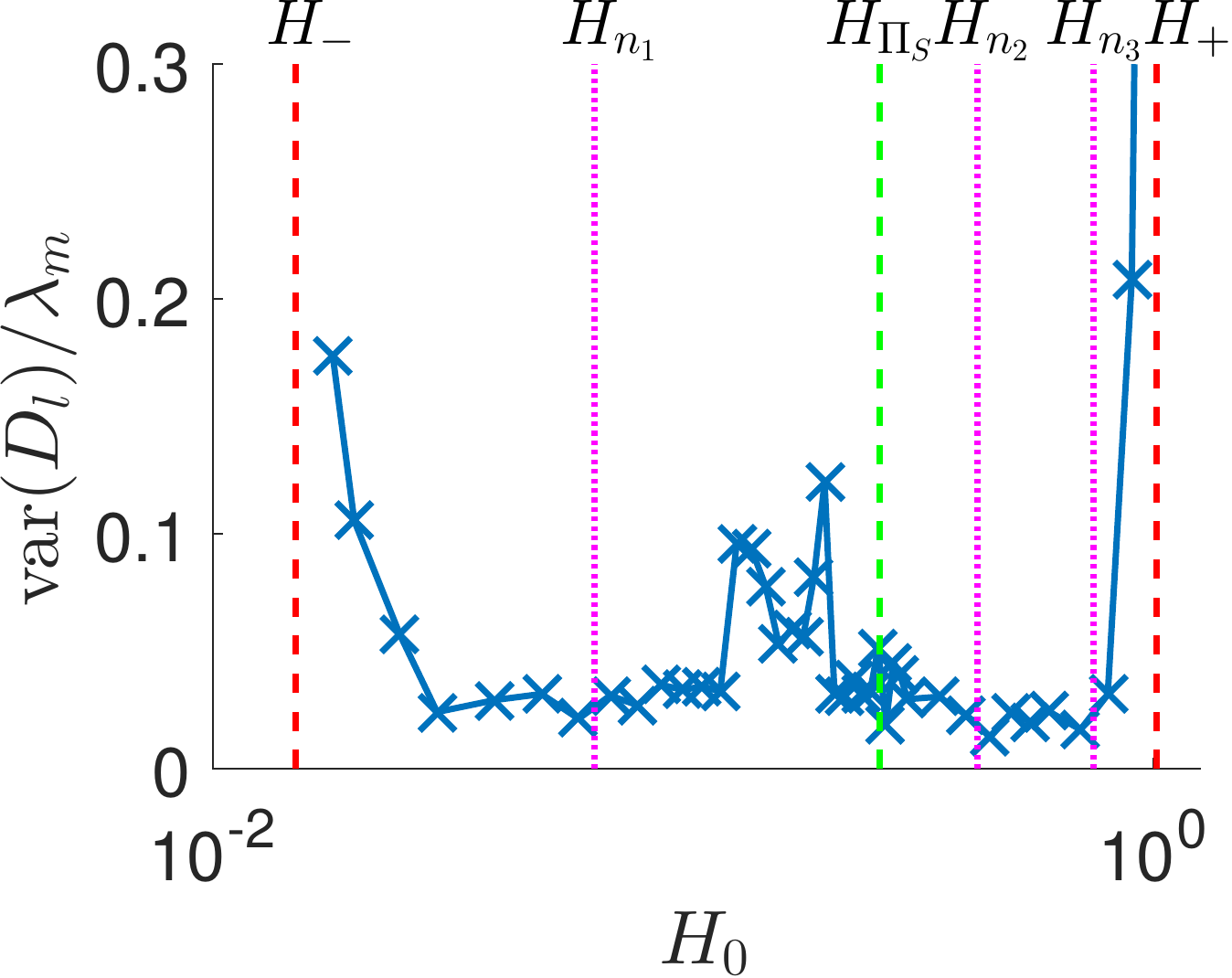}
   \label{fig:RandomVarSpacing}}   
  \subfigure[]{
   \includegraphics[width=0.31\textwidth]{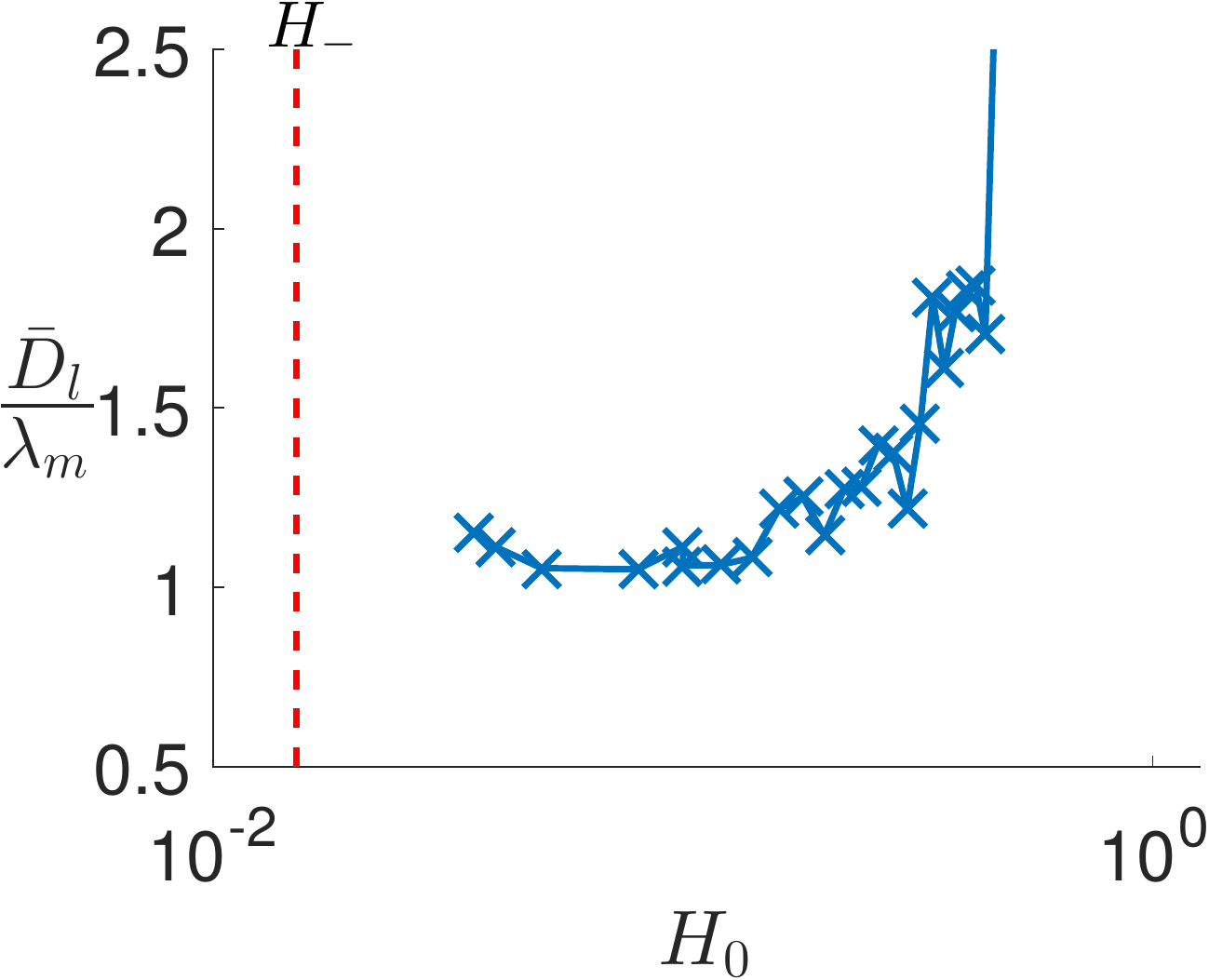}
  \label{fig:RandomMeanSpacingVdW} }
\caption{ (a) Mean spacing between primary drop ($H_l> H_0$) centres, $\bar{D}_l$, normalised by
$\lambda_m$; and (b) (variance in ${D}_l)/\lambda_m$, for a randomly perturbed NLC film of thickness $H_0$.
The dotted  magenta lines denote thickness threshold values extracted from simulations of a film exposed 
to localised perturbations. The derivation and meaning of these lines will further discussed in $\S$~\ref{sec:LocalizedPert}
and compared; for now, we note the increase in the variance in the region bounded by $H_{n_1}$ and $H_{n_2}$, and the region bounded by $H_{n_1}$ and $H^+$.
(c) Distance between drop centres (no secondary drops) for a Newtonian film ($\cN=0$). Note that
here there is no upper bound between linear instability and stability (i.e. $H^+$ is not defined)
and all films thicker than $H_-$ (dashed red line) are unstable.}
\label{fig:RandomSpacing}
\end{figure}

Before closing this section we comment on the relation of our results to those previously reported in the literature on instabilities of 
polymer films.  To begin, we extract the distances, $D_l$, between neighbouring primary drop centres (ignoring secondary drops with $H_l<H_0$),
and normalise $D_l$ by $\lambda_m$. Figure~\ref{fig:RandomSpacing} shows the mean distance, $\bar D_l$, 
for both the present model (figure~\ref{fig:RandomMeanSpacing}) and for the Newtonian 
equivalent (figure~\ref{fig:RandomMeanSpacingVdW}). 
For $H_0$ not near the boundaries ($H_{\mp}$) of the linearly unstable regime  
(for Newtonian films, $H_+ \rightarrow \infty$ within
the present model that ignores gravity), $\bar D_l \approx 1.15 \lambda_m$,  consistent with LSA. Near the boundaries, however, $\bar D_l$
increases dramatically and is no longer related to $\lambda_m$. For very thin films (near $H_-$), the increase of $\bar D_l$
 is due to coarsening occurring on a time scale similar to that of
dewetting; therefore, drops merge before the entire film completely dewets. For thicker films (near $H_+$) 
the increase in the distance between drop centres appears to be due to a different mechanism that  appears rather general, 
since it is operational for both NLC and Newtonian films.     Note that for the same range of thicknesses,  \citet{Seemann2001b} suggest that thermal 
(homogeneous) nucleation may be relevant;  our results suggest that even without thermal effects (not included in our model), similar behaviour may emerge.

Regarding the formation of secondary drops, we note that 
 previous works, such as \citet{Sharma2004} or~\citet{Thiele2001}, do not report their
existence.  A possible reason for this may be the difference in scales.  
To see this, consider the ratio $\rho=H_+/H_-$.
For the problem considered here, $\rho = O(10^2)$, while in \citet{Thiele2001} and \citet{Sharma2004}, $\rho = O(1)$.   
Therefore, in the present problem it is possible to distinguish clearly features that may be invisible 
for smaller $\rho$. 

Next, we discuss the shape of the primary drops.
Figure~\ref{fig:NLCvdw_Random} shows that for  $H_0\rightarrow H_{+}$, 
the base of these drops widens.   Such behaviour is found  consistently for  simulations 
with initial film thickness in the  linearly unstable regime.  In the example shown 
in figure~\ref{fig:NLCvdw_010_Random},
the primary drops at $x\approx 0,45,98,150$
are approximately flat for ten or more length units, then transition in height
to the equilibrium film; therefore, ignoring the
secondary drops, figure~\ref{fig:NLCvdw_010_DropHeight} may be interpreted as showing coexistence of two flat films, 
resembling \gls{NLC} films observed on 
silicon (or silicon oxide) substrates, see \citet{Cazabat2011}.
Similarly to our results, as $H_0 \rightarrow H^+$,
\citet{Sharma2004} observe the height profiles transitioning
from ``hills and valleys'' (e.g. figure \ref{fig:NLCvdw_006_Random}) 
to ``pancakes'' (wide drops), see figure~\ref{fig:NLCvdw_010_Random}.

\citet{Sharma2004} also show that an analytical result may be obtained for the minimum (equilibrium film thickness)
and the maximum (height of drop centres) of steady state solutions.
We will discuss an extension 
of these results to our setup  later in $\S$~\ref{sec:MetastableAnaylsis}; for the present 
purposes it is sufficient to mention the prediction that, for sufficiently long times, the 
thickness value  denoted by $H_{m_3}$ in figure~\ref{fig:NLCvdw_DropHeight}
should be reached.   We observe such behaviour only for $H_0 \lessapprox H_{m_3}$ for the 
times considered.
Once again, it is quite possible that the details of the influence of the film thickness could not 
be captured in~\citet{Sharma2004} due to insufficient separation of scales.

\section{Films exposed to a Localised Perturbation}\label{sec:LocalizedPert}

In this section we focus on instabilities induced by a localised perturbation, with the main motivation 
of comparing results with those of the preceding section obtained by global perturbations.   Note that in physical experiments, such localised perturbations 
(also referred to as nucleation centres) may be a results of impurities, substrate defects, or some 
localised modification of the film structure.  

We use the following initial condition
 \begin{equation} \label{eq:NanoICLocal}
   h(x,t=0) =  H_0 \left[ 1 - d \exp \left( -\left[ \frac{x}{0.2 W} \right]^2 \right)  \right] \;,
\end{equation}
i.e. a film of thickness $H_0$ with a localised perturbation at the left boundary, $x=0$, where 
a line of symmetry is assumed.  
The parameters $W$ and $d$ control the width and depth of the perturbation, respectively.
We first present and discuss results for film heights in the linearly
unstable regime, and then focus on  the linearly stable regime in $\S$~\ref{sec:LocalLinearlyStable}.

\subsection{Linearly Unstable Regime} \label{sec:LocalUnstable}

Here we fix $d=0.1$ and $W=\lambda_m$ in (\ref{eq:NanoICLocal}).
The domain length is initially set to $5\lambda_m$, and the right boundary is dynamically expanded
as a simulation progresses.   Our simulations should therefore be representative 
of a localised perturbation imposed on a film of infinite extent.

\subsubsection{Computational Results} \label{sec:LocalSimsUnstable}
\begin{figure}
\centering
  \subfigure[$H_0=0.05$, $\omega_m^{-1}\approx 1.681$]{
   \includegraphics[width=0.9\textwidth]{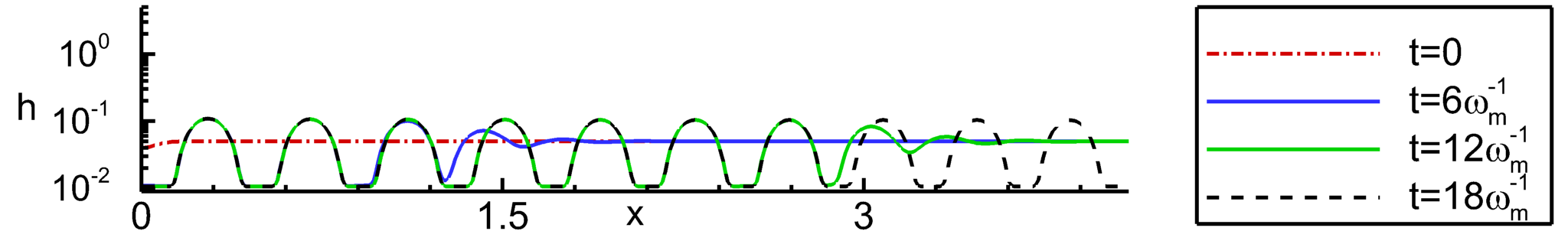}
  \label{fig:NLCvdw_002_Local} }
  \subfigure[$H_0=0.2$, $\omega_m^{-1}\approx 34.22$]{
   \includegraphics[width=0.9\textwidth]{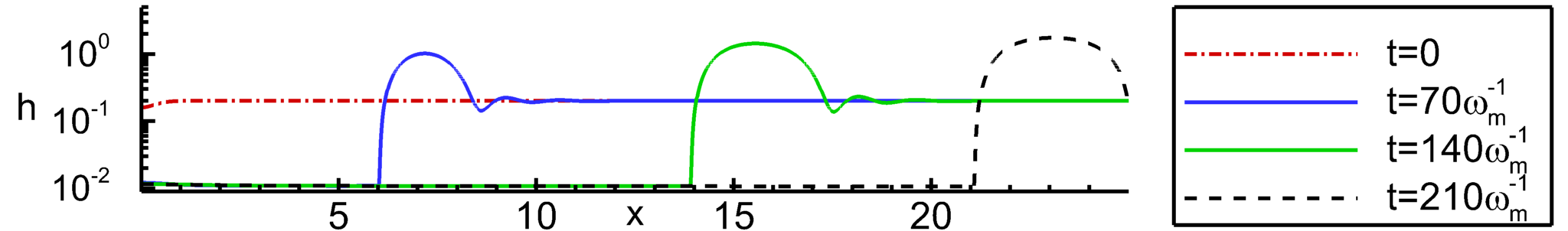}
  \label{fig:NLCvdw_002_Local} }
  \subfigure[$H_0=0.3$, $\omega_m^{-1}\approx 10.27$]{
   \includegraphics[width=0.9\textwidth]{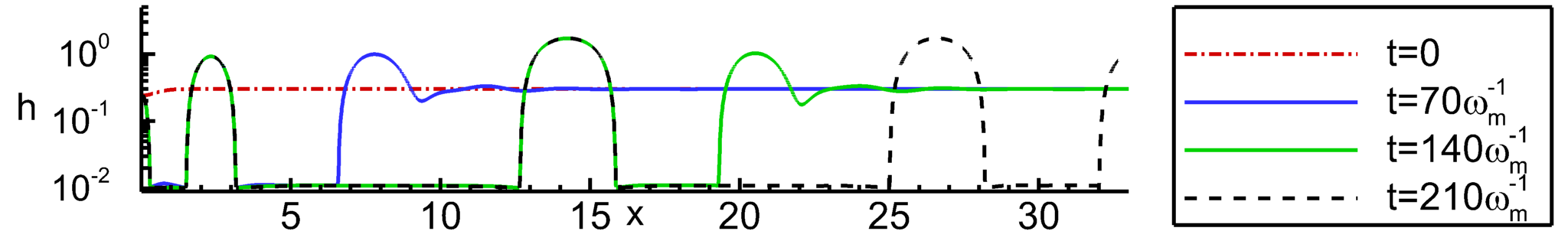}
  \label{fig:NLCvdw_003_Local} }

  \subfigure[$H_0=0.6$, $\omega_m^{-1}\approx 423.4$]{
   \includegraphics[width=0.9\textwidth]{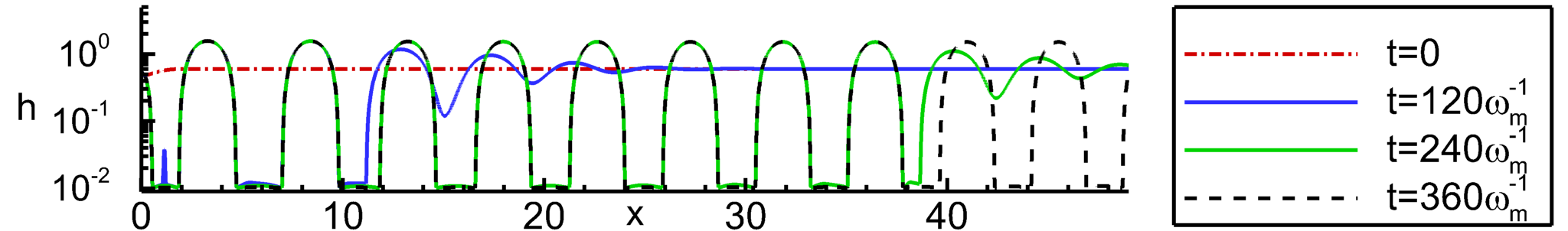}
  \label{fig:NLCvdw_006_Local} }
  \subfigure[$H_0=0.95$, $\omega_m^{-1}\approx 423.4$]{
   \includegraphics[width=0.9\textwidth]{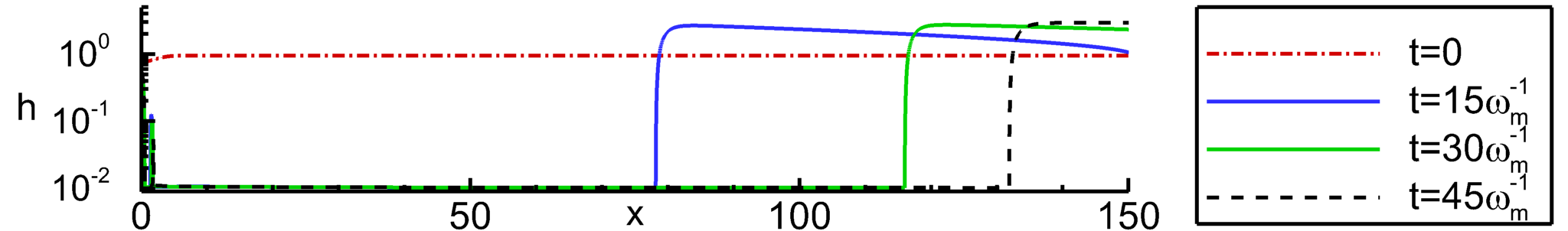}
  \label{fig:NLCvdw_010_Local} }
\caption{The free surface evolution for various initial average film thicknesses, $H_0$. The initial condition
is given by (\ref{eq:NanoICLocal}).  Note that the film profiles shown in (b) and (e) form drops, but for times longer than shown here.  The domain size shown in each panel is $10\lambda_m$; the simulation domains extend beyond the right boundary.  The other parameters are given in $\S$\ref{sec:Scalings}.  
}
\label{fig:NLCvdw_Local}
\end{figure}
Figure~\ref{fig:NLCvdw_Local} shows
simulation results obtained with initial condition (\ref{eq:NanoICLocal}) and using the same parameter sets as in 
figures~\ref{fig:NLCvdw_Single} and~\ref{fig:NLCvdw_Random},
as well as an additional thickness value, $H_0=0.05$.
The main observation is that the localised perturbation propagates
to the right and successively touches down forming drops; however, comparing the results of localised perturbation simulations to those 
obtained using random infinitesimal perturbations
we note two main differences: first, no secondary drops are seen for localised perturbations; and 
second, the distance between drops is close to $\lambda_m$ only for some values of $H_0$; for example, in
Fig.~\ref{fig:NLCvdw_Local} the agreement with LSA is found only  for $H_0 = 0.05$ and $H_0 = 0.6$, panels (b) and (d), respectively. 

\begin{figure}
\centering
\subfigure[]{
   \includegraphics[width=0.31\textwidth]{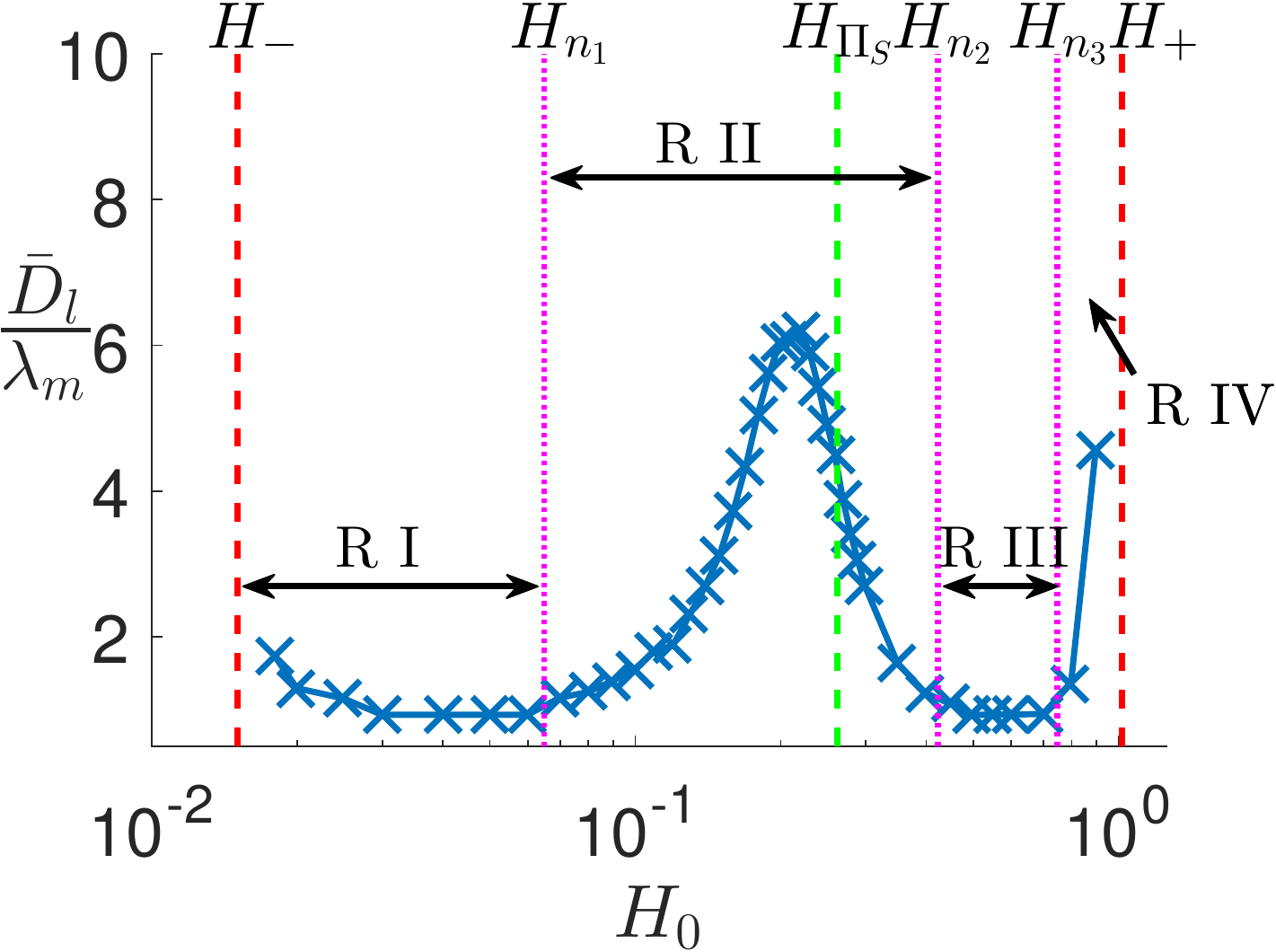}
  \label{fig:LocalSpacing} }
\subfigure[]{
   \includegraphics[width=0.31\textwidth]{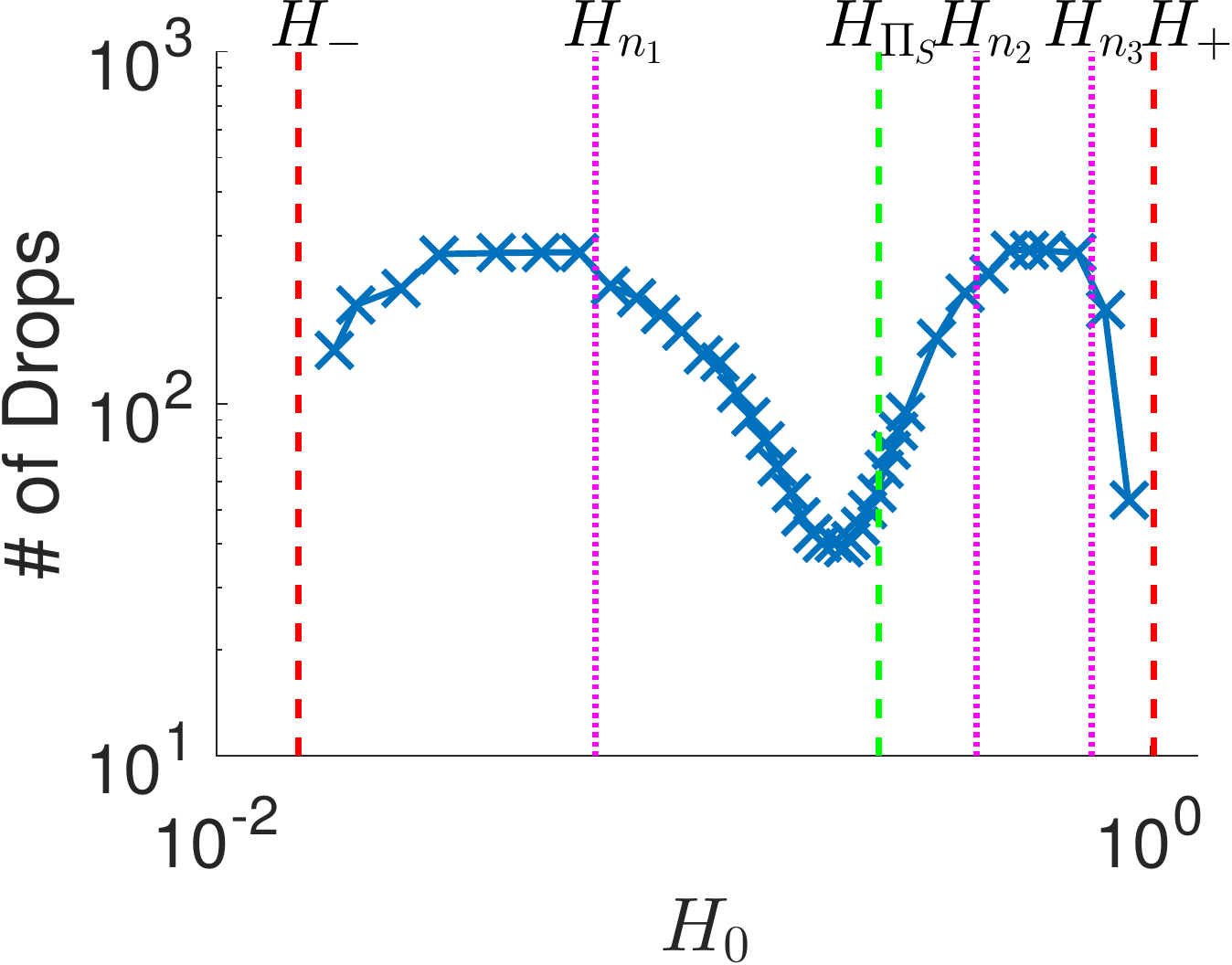}
  \label{fig:LocalCount} }
  \subfigure[]{
   \includegraphics[width=0.31\textwidth]{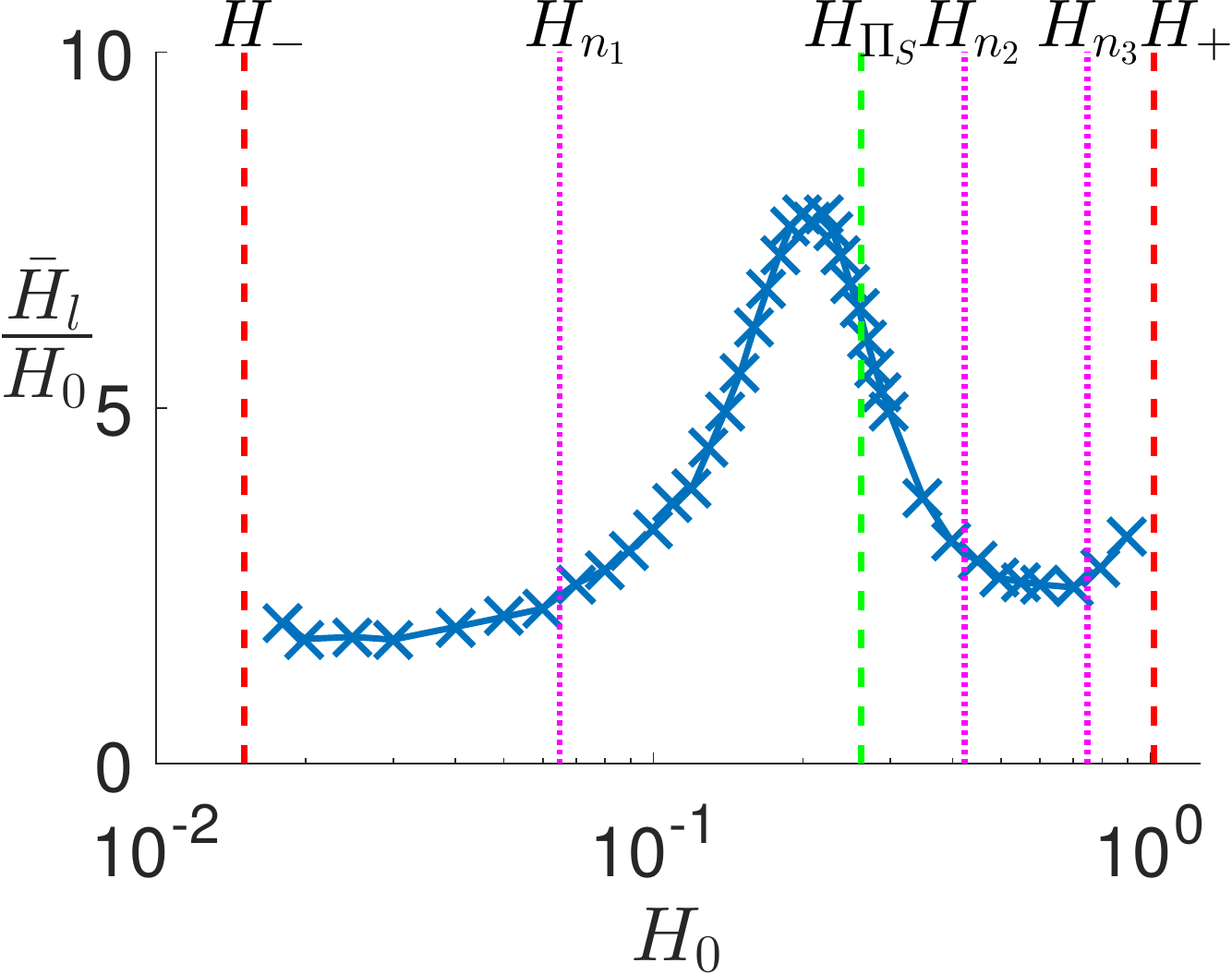}
  \label{fig:LocalHeight} }  
  \caption{
  (a) The mean distance between
drop centres, $\bar D_l$, normalised by $\lambda_m$, at $t=350 \omega_m^{-1}$;  
(b) The corresponding number of drops; and  
(c) The mean height of drop centres, $\bar H_l$, normalised by $H_0$.
The magenta dotted lines correspond to the values of $H_0$ for which $\bar D_l/\lambda_m$ crosses the value $1.1$.
The other parameters are given in $\S$\ref{sec:Scalings}.}
\label{fig:LocalSimData}
\end{figure}
To investigate further the dewetting process we calculate
the mean distance between drop centres, $\bar D_l$. 
Figure~\ref{fig:LocalSpacing} shows the results; to improve statistics, 
the simulations are carried out in large domains expanded in time to 
$320 \lambda_m$.    We observe 
the existence of four different
regimes characterised by the magnitude of  $\bar D_l/\lambda_m$, separated
by three distinct film heights (marked by the  dotted magenta lines showing 
where $\bar D_l/\lambda_m\approx1.1$, to characterise departure from LSA predictions)
and denoted $H_{n_1}$, $H_{n_2}$, and $H_{n_3}$.
We see that in the regions marked R I and R III, the distance between
drop centres is well approximated by $\lambda_m$; 
however, there exist regions R II and R IV, where there is no correlation
between $\bar D_l$ and $\lambda_m$.

Regimes for which $\bar D_l \not\approx \lambda_m$ have been reported before 
(\citet{Thiele2001,Diez2007}) for different thin film models;
however, the main findings differ from those presented here.  In~\citet{Thiele2001} and~\citet{Diez2007} it is shown 
that there exists a $\bar D_l  \not\approx \lambda_m$  regime near the upper threshold $H_+$ between
linear instability and stability (see the R IV regime in figure~\ref{fig:LocalSpacing}).
The work by~\citet{Thiele2001} furthermore discusses a corresponding regime close to $H_-$; in the present formulation, this 
is a subset of the R I region very close to $H_-$ where, similarly to the results for a film exposed to
random perturbations (see figure~\ref{fig:RandomMeanSpacing})
the difference between $\bar D_l$ and 
$\lambda_m$ is due to coarsening occurring on the considered time scale.  Neither of the previous 
works, however, discusses the existence of a (separate) R II regime.  

Before exploring the instability development due to a local perturbation more rigorously, we first note a possible connection to our 
previous results for randomly perturbed films. Recalling the results for the mean distance between drop centres, $\bar{D}_l$,
for films exposed to random perturbations (see figure~\ref{fig:RandomMeanSpacing}), while $\bar{D}_l/\lambda_m \approx 1.15$
mostly everywhere, there appears to be small peak in $\bar{D}_l/\lambda_m$ in the thickness range corresponding to the
R II regime. To highlight this connection further, in figure~\ref{fig:RandomVarSpacing}
the variance in $D_l/\lambda_m$ is plotted as a function of the initial film thickness $H_0$. 
We see that the variance increases in the R II and R IV regime, suggesting that there is a connection between the results for randomly and locally perturbed films.
A possible explanation of this connection is that on large domains, the film does not dewet everywhere at the same time even
if exposed to random perturbations; therefore, at the locations where the film dewets initially, the solution may appear as
due to a localised perturbation, with retracting fronts modifying the distance between the drops that form eventually.  
The influence of these fronts may result in a large variance, explaining  the results captured by 
figures~\ref{fig:RandomMeanSpacing} and~\ref{fig:RandomVarSpacing}.

We also note a possible connection to the theoretical and experimental results presented by \citet[]{Diez2007};
\citet[]{Seemann2001b}; and \citet[]{Thiele2001}. \citet[]{Seemann2001b} suggest that 
as $H\rightarrow H_+$ (or $H_-$), where by definition $\Pieff'(H_+)=0$ 
($\Pieff'(H)$ corresponds to $\phi''(h)$ in the referenced work),
the barrier to nucleation vanishes, thus nucleation by thermal activation is possible. While our model
does not include thermal effects, our results are similar to those observed by 
 \citet[]{Seemann2001b}. Therefore, the R IV regime 
in our results, and the nucleation reported by \citet{Diez2007} and \citet{Thiele2001} (again no thermal effects considered),
may be analogous to the thermal nucleation regime.    

% We also note a possible connection to the theoretical and experimental results presented by \citet[]{Diez2007};
% \citet[]{Jacobs2008}; and \citet[]{Thiele2001}. 
% \citet[]{Jacobs2008} suggest that 
% as $H\rightarrow H_+$ (or $H_-$), where by definition $\Pieff'(H_+)=0$ 
% ($\Pieff'(H)$ corresponds to $\phi''(h)$ in the referenced work),
% the barrier to nucleation vanishes, thus nucleation by thermal activation is possible. While our model
% does not include thermal effects, our results are similar to those observed by 
%  \citet[]{Jacobs2008}. Therefore, the R IV regime 
% in our results, and the nucleation reported by \citet{Diez2007} and \citet{Thiele2001} (again no thermal effects considered),
% may be analogous to the thermal nucleation regime. As mentioned before,
% the regime analogous to R II  was not reported in these other works. Thermal nucleation is not
% expected to play a role in the R II regime; therefore, the nucleation must be analogous to the heterogeneous type.
% \citet[]{Jacobs2008} observe heterogeneous nucleation
% in experiments for film thicknesses in the linearly unstable regime. If this observation is connected to our R II regime,
% the peak in the variance of $\bar{D}_l/\lambda_m$ observed in figure~\ref{fig:RandomVarSpacing} shows that nucleation
% may still have an effect even for random infinitesimally small perturbations, and suggests a reason
% why \citet[]{Jacobs2008} were unable to reduce the number of holes below a certain point 
% by improving preparation conditions. 

\begin{figure}
\centering
   \includegraphics[width=0.42\textwidth]{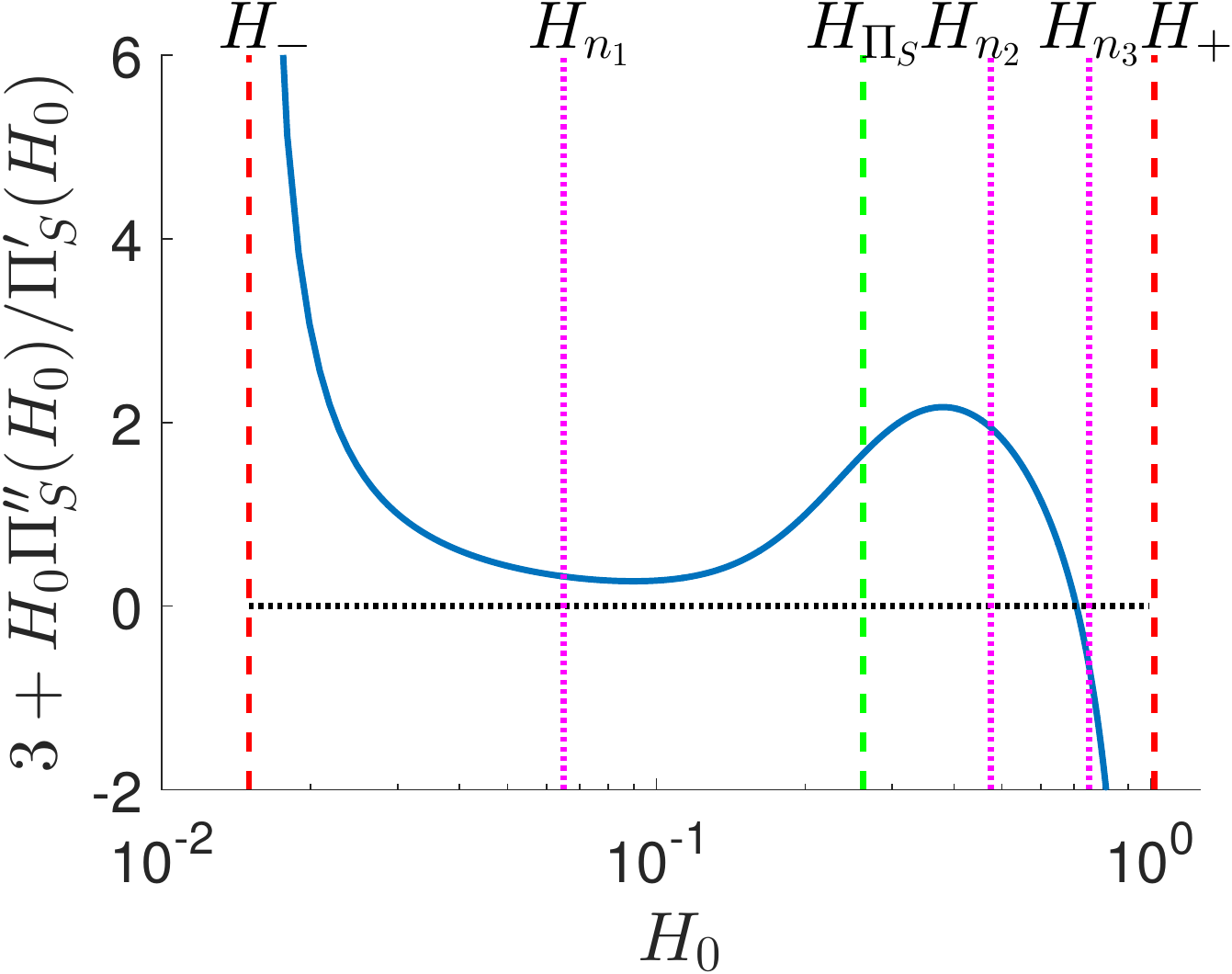}
\caption{The inner square bracketed term in (\ref{eq:NanoGovPDERescale3}).
}
\label{fig:term}
\end{figure}

{\bf Note: }  Under appropriate scaling, solutions of
the form $h(x,t) = H_0 [ 1 + \ep h_1(x,t) +\ep^2 h_2(x,t) + O(\ep^3)]$, where $\ep \ll 1$,
are to leading order independent of the model parameters.
The scalings are as follows:
 \begin{equation} \label{eq:Rescalings}
  h = H_0 \Psi \;, \quad  x= \sqrt{ \frac{\cC}{\Pieff'(H_0)} } \zeta \; , \quad \textrm{and} \quad t = \frac{\cC}{Q(h) \left[\Pieff'(H_0)\right]^2} \tau \;, % = \frac{\cC}{H_0^3 \left[\Pieff'(H_0)\right]^2} \tau \;.
 \end{equation}
 where for generality we keep the general mobility function, $Q(h)$.  
The rescaled leading order equation is 
 \begin{equation} \label{eq:NanoGovPDERescale}
  h_{1,\tau} + h_{1,\zeta\zeta\zeta\zeta}+ h_{1,\zeta\zeta}=0,
 \end{equation}
and the mode of maximum growth and its corresponding growth rate are
$\check{q}_m = {1/\sqrt{2}},\;~\check{\omega}_m = {1/4}$, 
respectively (here the check notation denotes the values obtained from the rescaled model).
We see that as long as $h_1(x,t)=O(1)$ (i.e., as long as linear theory is applicable), 
the evolution for different choices of $H_0$ differs only by spatial and temporal scales.
Recall that the initial conditions (\ref{eq:NanoICSingle}), (\ref{eq:NanoICRandom}),
and (\ref{eq:NanoICLocal}) were chosen to scale with $\lambda_m$ and $H_0$, so
the corresponding initial conditions in the rescaled variables are identical.
The governing equation for $h_2(x,t)$ is
  \begin{equation} \label{eq:NanoGovPDERescale3}
  h_{2,\tau} + h_{2,\zeta\zeta\zeta\zeta}+ h_{2,\zeta\zeta}=
  - \left[ h_1  \frac{Q'(H_0)}{Q(H_0)} \left(h_{1,\zeta\zeta\zeta} + \left[ 1  + \frac{ Q(H_0) \Pieff''(H_0)}{Q'(H_0) \Pieff'(H_0)} \right] h_{1,\zeta} \right) \right]_\zeta \; ,
 \end{equation}
for general $Q(h)$, and for $Q(h)=h^n$,
  \begin{equation} \label{eq:NanoGovPDERescale2}
  h_{2,\tau} + h_{2,\zeta\zeta\zeta\zeta}+ h_{2,\zeta\zeta}=
  - \left[ \frac{h_1}{H_0}\left( n h_{1,\zeta\zeta\zeta} + \left[ n + \frac{H_0 \Pieff''(H_0)}{\Pieff'(H_0)} \right] h_{1,\zeta} \right) \right]_\zeta \; .
 \end{equation}
For any unstable film thickness $H_0$, the second order equations only differ by the forcing term on the right hand side, which is independent of the inverse Capillary number, $\cC$,
suggesting that the structural disjoining pressure determines the transition between different dewetting morphologies. 
Figure~\ref{fig:term} plots the term enclosed by inner square brackets in (\ref{eq:NanoGovPDERescale2})
for $n=3$ and other parameters given in \S~\ref{sec:Scalings}.
Note that this term is an increasing function of $H_0$ in the most of the R II regime (between $H_{n_1}$ and $H_{n_2}$), in contrast to its behaviour elsewhere.   
While we do not have clear explanation of the connection of this observation to the film breakup properties that we observe in the R II regime, we expect 
that it may be useful in future analysis of different stability regimes.

\subsubsection{Marginal Stability Criterion: Outline} \label{sec:MSC}

Marginal stability criterion (MSC) theory  analyses front propagation into an unstable
state using Fourier transforms and asymptotic theory. The analysis 
provides an expression for the velocity at which a front
propagates into the unstable state asymptotically in time, and we will mainly focus our discussion
on computing this velocity.  
In addition, the method specifies other possible wavelengths that may govern film breakup, and allows us
to extract predictions for the expected time between the formation of successive drops \citep[see][]{Powers1998}. 
To begin, we give a brief overview of MSC theory and its main results in the present context; 
the reader is referred to the comprehensive review by \citet{Saarloos2003}
for further details. 
Then in the following section we discuss the correlation of MSC predictions with our computational results. 

We assumed that the solution of the rescaled leading order governing equation (\ref{eq:NanoGovPDERescale})
has a spatial Fourier series representation, i.e. 
 \begin{equation}  \label{eq:FTSolution}
    h_1(\zeta,\tau ) = \int_{-\infty}^{\infty} \breve{h}_1(\check q,\tau) e^{\check q \zeta i } \; d\check q \; .
 \end{equation}
Furthermore, with the ansatz 
 \begin{equation} 
  \breve{h}_1(\check q,\tau ) = \tilde{h}_1(\check q)e^{-  \check\omega( \check q) \tau i } \;,
 \end{equation}
substituting (\ref{eq:FTSolution}) into  (\ref{eq:NanoGovPDERescale})
yields
 \begin{equation} 
\left[  -i \check\omega( \check q) + \check q^4 - \check q^2 \right] 
 \int_{-\infty}^{\infty} \tilde{h}_1(\check q)e^{\check q \zeta i - \check\omega( \check q) \tau i } \; d\check q = 0\; ,
 \end{equation}
where $\tilde{h}_1(\check q)$ is the Fourier transform of the initial condition,
which is assumed to be an analytic function. 
Satisfying (\ref{eq:NanoGovPDERescale}) for arbitrary $ \tilde{h}_1(\check q)$ gives the
dispersion relation, 
\begin{equation} \label{eq:RescaledDispersionRelationship}
  \check\omega( \check q) = -i (\check q^4-\check q^2)  \;.
\end{equation}
In the reference frame travelling with velocity $\check V$,
the solution can be expressed as
\begin{equation} \label{eq:SDIntegral}
  h_1(s,\tau) = \int_{-\infty}^{\infty} \tilde{h}_1(\check q) e^{i \check q s} e^{-i[\check \omega(\check q)-\check V \check q] \tau} \;d \check q \; ,
\end{equation}
where $s=\zeta -\check V \tau$. If the above integral converges, $ h_1(s,\tau)$ satisfies
the rescaled governing equation (\ref{eq:NanoGovPDERescale}), and  $\check V$ is defined as the  ``linear spreading speed''
by \citet{Saarloos2003}. 

Following~\citet{Saarloos2003}, we now consider the limit $\tau \rightarrow \infty$ and use a saddle-point approximation (steepest
descent method) to approximate the integral in (\ref{eq:SDIntegral}). 
The saddle-point $\check q^*$ satisfies 
\begin{equation} \label{eq:MSC1}
  \check V  =  \left. \frac{d\check \omega(\check q)}{d \check q} \right|_{\check q^*} \;  .
\end{equation}
To eliminate exponent terms of the integrand in (\ref{eq:SDIntegral}) that lead to 
decay or growth in $\tau$, we set
\begin{equation} \label{eq:MSC2}
  \check V =  \frac{\check \omega_i(\check q)}{\check{q}_i}\; ,
\end{equation}
where $\check V \in \mathbb{R}$ (by definition), the $i$ subscript denotes the imaginary part of variables,
and an $r$ subscript will be used to denote the real part.
Recall $\check \omega(\check q)$ is a polynomial in $\check q$ and thus differentiable,  so that (\ref{eq:MSC1})
may be expressed as
 \begin{equation} \label{eq:MSC1a}
  \check V =  \textrm{Re} \left[ \left.\frac{d\check \omega}{d\check q} \right|_{\check q^*} \right]  =  
  \left.  \frac{1}{2}\left[ \pad{\check \omega_r}{\check{q}_r} +\pad{\check \omega_i}{\check{q}_i}  \right] \right|_{\check q^*} \;,
 \end{equation}
 and
  \begin{equation} \label{eq:MSC2a}
  \textrm{Im} \left[ \left.\frac{d\check \omega}{d\check q} \right|_{\check q^*} \right] = 
  \left. \frac{1}{2} \left[ \pad{\check \omega_i}{\check{q}_r} - \pad{\check \omega_r}{\check{q}_i} \right] \right|_{\check q^*} \;   =
  0 \;.
 \end{equation}
Furthermore, by definition,
$\check{\omega}_r$ and $\check{\omega}_i$ satisfy the Cauchy-Riemann equations, simplifying (\ref{eq:MSC1a}) and (\ref{eq:MSC2a}) to
 \begin{equation} \label{eq:MSC2b}
  \check V(\check q^*)  =  \left. \pad{\check \omega_r}{\check{q}_r} \right|_{\check q^*}  \;, \quad \textrm{and} \quad 
  \textrm{Im} \left[ \left.\frac{d\check \omega}{d\check q} \right|_{\check q^*} \right] =  \left. \pad{\check \omega_i}{\check{q}_r}  \right|_{\check q^*} = 0 \;,
 \end{equation}
respectively. Satisfying the second condition in (\ref{eq:MSC2b}) gives
an expression for $\check{q}_r$ as a function of $\check{q}_i$ and  
taking the derivative of (\ref{eq:MSC2}) with respect to $\check{q}_i$ yields 
 \begin{equation} \label{eq:MSC3a}
  \std{\check V}{\check{q}_i}  = \frac{1}{\check{q}_i} \left[ \pad{\check \omega_r}{\check{q}_r} + \std{\check{q}_i}{\check{q}_r} \pad{\check \omega_r}{\check{q}_i}  -  \frac{\check \omega_i(\check q)}{\check{q}_i} \right]\;.
 \end{equation}
Evaluating the above equation at the saddle point, and substituting (\ref{eq:MSC2}) and (\ref{eq:MSC2b}), we find 
 \begin{equation} \label{eq:MSC3a}
 \left. \std{\check V}{\check{q}_i} \right|_{\check q^*}  = 0 \;,
 \end{equation}
and thus $\check{q}_i^*$ is a critical point of $\check{V}$. We note that this critical point
is a minimum of $\check V$; however  this information is not required to determine $\check V$ itself
(see~\citet{Saarloos2003} for further details).
Equations (\ref{eq:MSC1}), (\ref{eq:MSC2b}), and (\ref{eq:MSC3a}) provide
the necessary conditions to determine $\check{q}_r^*$, $\check{q}_i^*$, and $\check V$. 
Therefore, with the rescaled dispersion relationship defined by (\ref{eq:RescaledDispersionRelationship}),
  \begin{equation} 
 % \check q^*  = \sqrt{ \frac{3 +\sqrt{7} }{2}} + \sqrt{\frac{1}{3}\left[ \sqrt{7}-1\right]} i \approx 1.188 + 0.370 i \;, 
  \check q^*  = \sqrt{ \frac{3 +\sqrt{7} }{4}} + \sqrt{\frac{1}{6}\left[ \sqrt{7}-1\right]} i \approx 0.791 + 0.262 i \;, 
 \end{equation}%
 and
 \begin{equation} \label{eq:MSCc}
   \check V= \frac{1}{9} \sqrt{1+\sqrt{7}}\left( 5 + \sqrt{7} \right) \approx 1.622 \;  \quad \textrm{and} \quad
   V \approx 1.622  \frac{Q(H_0) \left[ \Pieff'(H_0) \right]^\frac{3}{2} }{\sqrt{\cC}}\;,
 \end{equation}
where the second equation in (\ref{eq:MSCc}) is given in the unscaled variables.
We note in passing that MSC also provides the condition for
when a front propagating to the right will spread with the linear
spreading speed asymptotically.
If the initial condition decays faster
than $e^{-\check q^*\zeta}$ 
as $\zeta \rightarrow \infty$, then the instability will spread
with velocity $V$ given by (\ref{eq:MSCc}), and the solution
is referred to as a pulled front. For an initial
condition that does not decay fast enough, the instability
will propagate into the unstable state with a speed
greater than $V$. The solution is then referred to as a pushed 
front and deriving an expression for the speed at which it spreads is
nontrivial.  For the present problem, the initial condition (\ref{eq:NanoICLocal})
is a Gaussian perturbation, thus decays sufficiently
fast and a pulled front is expected.

Regarding the time scale for successive drop formation, we note that if a localised perturbation
leads to a coherent pattern behind the front, it is periodic in time, i.e., 
$H(s)=H(\zeta-\check V \tau)=H(\zeta +\check \lambda - \check V \tau)=H(\zeta-\check V(\tau+\check t_{\textrm{MSC}}))$, where
$\check t_{\textrm{MSC}}$ is the time between successive drops detaching
from behind the front at distances $\check \lambda$ apart.  
Satisfying $H(\zeta+\check \lambda - \check V \tau )=H(\zeta-\check V(\tau+\check t_{\textrm{MSC}}))$ yields
 \begin{equation} \label{eq:NanoBreakup}
   \check t_{\textrm{MSC}} = \frac{\check \lambda} { \check V} = \frac{2 \pi} { \check q \check V} \approx 5.478  \;, \quad \textrm{or} \quad 
   t_{\textrm{MSC}} = 5.478 \frac{\cC}{Q(H_0) \left[\Pieff'(H_0) \right]^2 }  \;,
 \end{equation}
 where, for future reference, $\check q = \check q_m$ was used in the first expression to obtain the second one.
In the next section we proceed with correlating the presented MSC results with the computational ones.

{\bf Note 1:}  The speed of the propagating front, (\ref{eq:MSCc}), 
may alternatively be obtained by using pinch point analysis, with a key
difference of how the complex integral path is chosen in 
the Fourier representation (\ref{eq:SDIntegral}).   Our 
previous results  \citep[see][]{Lam2014}
for the flow of an NLC film down an inclined plane
relied on the results of such analysis to derive the velocity 
of the wave packet boundaries ($V$ in the MSC analysis). 
In that work, analytical
and numerical simulations confirmed the existence of three
stability regimes behind the front travelling down the inclined plane:
stable, convectively unstable, and absolutely unstable.
%In our governing equation (\ref{eq:NanoGov-PDE}), there is no advection term, therefore,
%there is only the unstable regime, i.e. by definition the flat film thickness is linearly
%unstable in MSC, see \citet[]{Lam2014} for a detailed discussion on how an advection term
%can stabilise a linearly unstable film.
These same instabilities were discussed by \citet{Saarloos2003}
within a different physical context.   
 The two unstable regimes were further subdivided  into two cases: coherent pattern formation (periodic array
of drops) and incoherent pattern formation; however, a general analytical
theory for predicting these cases is not yet available.    We will discuss our computational results within this 
context briefly in the next section. 
% In our governing equation (\ref{eq:NanoGov-PDE}), there is no advection term, therefore,
% there is no distinction between the convectively unstable and absolutely unstable regimes, i.e.
% there are only the unstable and stable regimes. However, we see
% that the R I and R III regimes correspond to the coherent pattern formation case (see figure~\ref{fig:NLCvdw_006_Local}), while
% R II, typically characterized by multimodal drop distributions, is a mixture of the two cases (see figure~\ref{fig:NLCvdw_003_Local}).

{\bf Note 2:} Having explicit results for the time scales on which localised  and random perturbations
lead to dewetting and drop formation, given by $t_{MSC}$ and $\omega_m^{-1}$, respectively, allows one
to discuss which instability is relevant and actually destabilises a film.  Considering the ratio of these
time scales, however, shows it to be independent of the film thickness.  This finding 
at first sight appears at odds with the commonly accepted paradigm that for thin films spinodal instability is 
relevant, while for thick ones nucleation is dominant, see, e.g.~\citet{Seemann2001b}.  One explanation is that the
nucleation instability may be operational for thin films as well (assuming that nucleating centres are available), 
but this is not obvious due to the same resulting wavelengths (for example, in the R II regime, see Fig.~\ref{fig:LocalSimData}). 
However, more careful discussion of this important issue requires simulations in 3D, and we therefore 
defer it to future work. 

\subsubsection{Marginal Stability Criterion: Correlation with computational results}
We now return to simulations of our model, and consider them in light of the MSC predictions. 
Using the analytical result (\ref{eq:MSCc}) for the linear spreading speed, $V$, of the propagating 
perturbation, simulation results may be plotted in the reference frame moving with this speed.
Furthermore, to emphasise the difference between the dewetting dynamics for
different regimes, the travelling wave variable, $s=x-Vt$, is scaled
by $\lambda_m$.

\begin{figure}
\centering
  \subfigure[]{
   \includegraphics[width=0.43\textwidth]{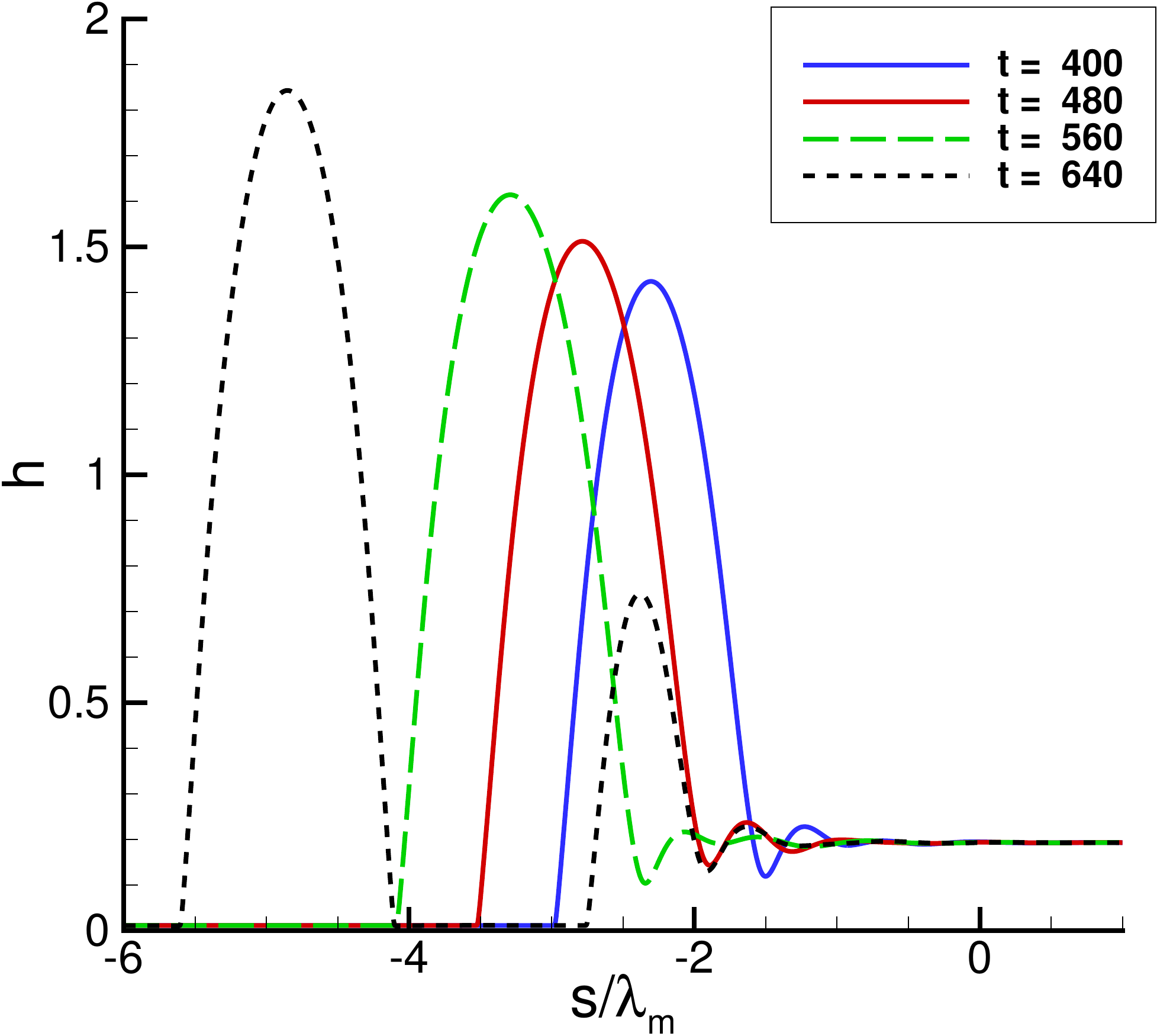}
  \label{fig:RII_TravelingFrame} }
  %}
%   %
    \quad \quad
  \subfigure[]{
   \includegraphics[width=0.43\textwidth]{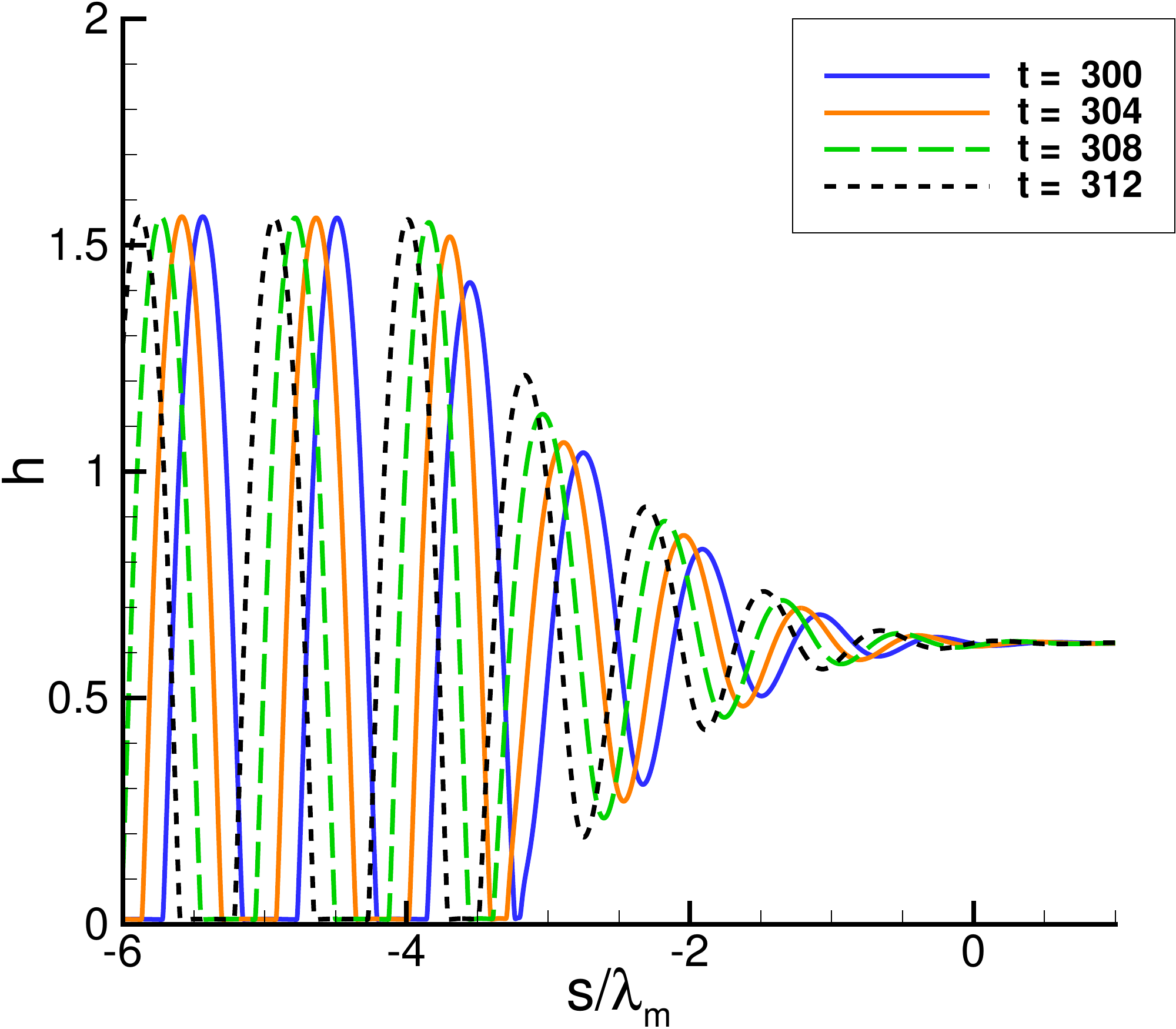}
  \label{fig:RIII_TravelingFrame} }

\caption{Typical evolution of the film thickness in the reference frame
travelling with the linear spreading speed $V$ in (a) R II  regime and (b) R III regime.
Note that the $x$-axis has been scaled by $\lambda_m$ to highlight departure from the LSA
predictions in R II regime. 
}
\label{fig:NLCvdw_Local_Traveling}
\end{figure}

Figure~\ref{fig:NLCvdw_Local_Traveling} shows typical results in R II and R III regimes (the results
in R I are similar to those in R III).  
First, we see that the instabilities caused by the localised perturbation
are essentially bounded below $s=0$, demonstrating that
MSC correctly predicts the speed with
which instabilities propagate into the unstable flat region.
Furthermore, since the front speed is given by MSC, we are in the pulled front case as expected.
Second, the differences between the dewetting 
dynamics in various regimes are easily observed, and we discuss these differences next.  Movies 
({\it Movie 1~-~4} in the online supplementary material) are also available to better illustrate the instability evolution.  

In R III regime, see figure~\ref{fig:RIII_TravelingFrame} and {\it Movie 1},
in the reference frame travelling with MSC speed $V$, the solution is
a leftward-propagating travelling wave, with a stationary envelope (in the travelling reference frame).
Furthermore, as the travelling wave propagates to the left, 
its amplitude (close to the front)  grows monotonically and 
dewets the film periodically with wavelength $\lambda_m$, therefore leading to 
coherent pattern formation.  

 In R II regime, however, the instability evolution is more complicated.   
 Here,  the amplitude of the waves connecting the capillary ridge to the flat film 
(see e.g. $s/\lambda_m\in[-2,1]$ in figure~~\ref{fig:RII_TravelingFrame})
does not grow monotonically as it travels to the left, but oscillates
until some critical amplitude is reached.   In this regime (and similarly in R IV, not 
discussed for brevity), both coherent and incoherent pattern formation is 
possible; see {\it Movies 2 -4} for examples that illustrate either 
coherent pattern formation ({\it Movie 2}); a bimodal pattern where drops of 
two different sizes are produced ({\it Movie 3}); or a disordered state, 
where the evolution resembles the R III regime sporadically interrupted by a 
different type of dynamics ({\it Movie 4}).   Clearly, more work is needed 
to understand the details of pattern formation in R II (and R IV) regimes; we 
leave this for the future and for now focus on the main features of the results, without
delving further into various modes of breakup in R II regime.

\begin{figure}
\centering
  \subfigure[]{
   \includegraphics[width=0.43\textwidth]{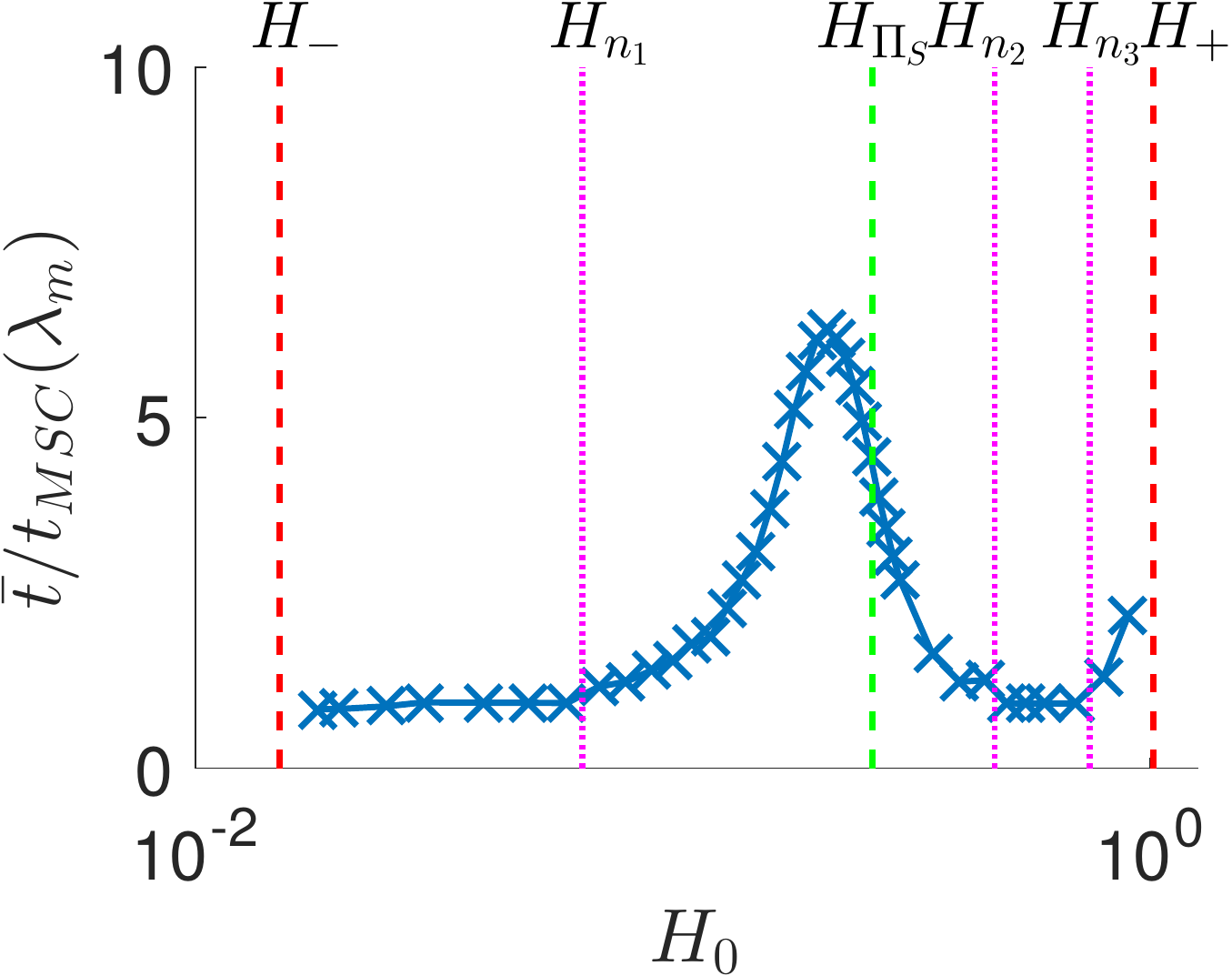}
  \label{fig:LocalMeanBreakup1} }
  %}
%   %
    \quad \quad
  \subfigure[]{
   \includegraphics[width=0.43\textwidth]{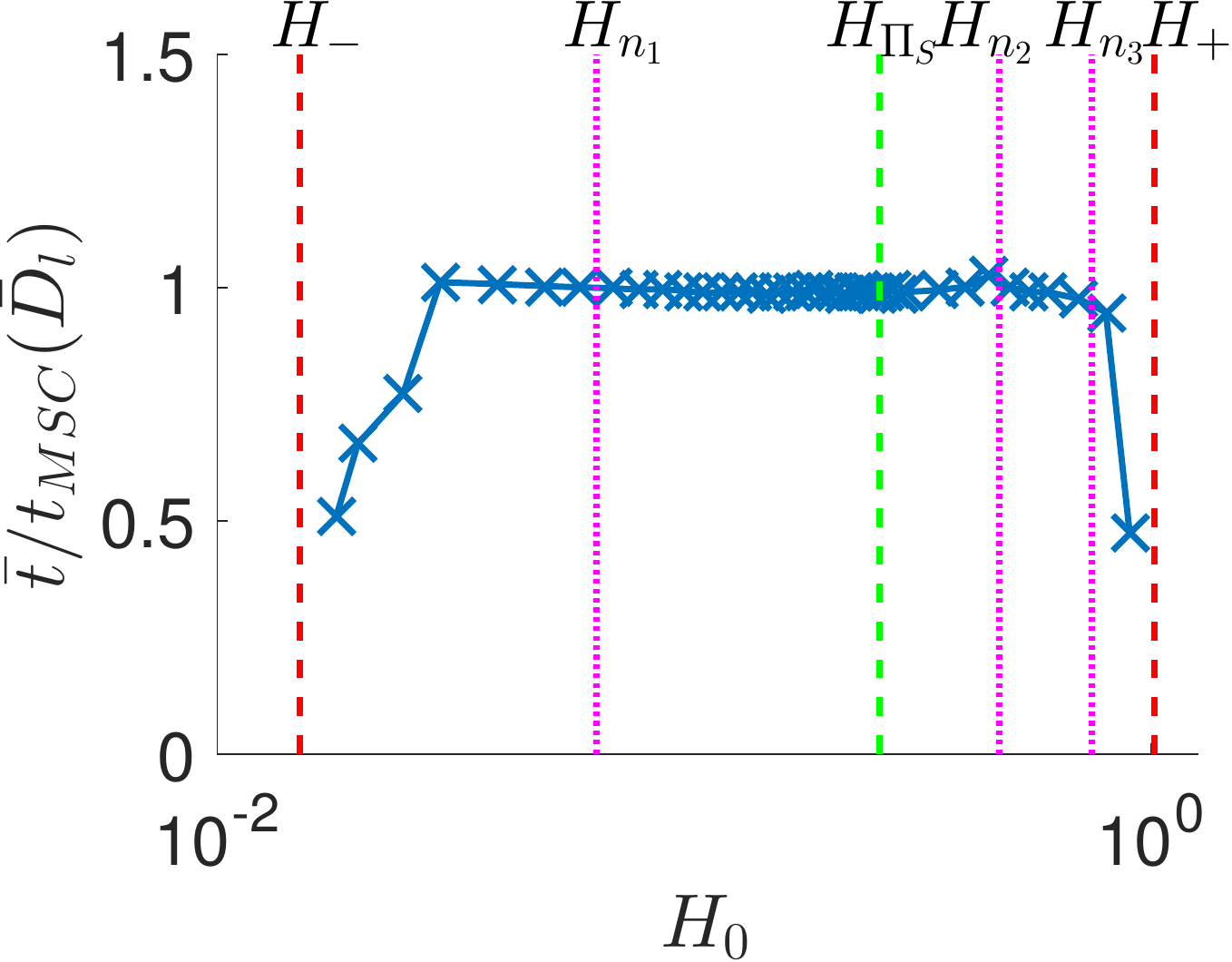}
  \label{fig:LocalMeanBreakup2} }

\caption{The mean time between drop formation extracted
from simulations, normalised by (\ref{eq:NanoBreakup}) with (a) $\lambda = \lambda_m$
and (b) $\lambda = \bar D_l$, i.e. the mean distance between drop centres obtained in simulations.
}
\label{fig:LocalMeanBreakup}
\end{figure}
To compare the film breakup times predicted by MSC (\ref{eq:NanoBreakup}) to the results of
simulations, we extract the mean time between formation of successive drops. 
The results are plotted in figure~\ref{fig:LocalMeanBreakup} as a function of the initial film thickness, $H_0$, 
normalised by (\ref{eq:NanoBreakup}) with (a) $\lambda = \lambda_m$
and (b) $\lambda = \bar D_l$ (the mean distance between drop centres extracted from simulations). 
Similarly to the results given in figure~\ref{fig:LocalSpacing}, figure~\ref{fig:LocalMeanBreakup1} shows that 
with $\lambda=\lambda_m$ the MSC prediction (\ref{eq:NanoBreakup}) gives good agreement 
with the numerics in regions R I and R III; however, there is no agreement in R II and R IV. Substituting
instead the mean distance between drop centres $\lambda=\bar D_l$
into (\ref{eq:NanoBreakup}), figure~\ref{fig:LocalMeanBreakup2} demonstrates that MSC
gives the correct result for all film thicknesses not near the threshold values between instability 
and stability (dashed red lines).   This result suggests that the MSC approach provides the 
correct result for breakup times once the instability wavelength is selected; however MSC does not 
provide information about this selected wavelength in the R II regime.  

We note that, while a nonlinear variant of MSC has been developed, it applies only to pushed 
fronts \citep[see][]{Saarloos2003} that
are not relevant here.  An alternative theory mentioned (but not thoroughly analysed)
by \citet{Saarloos2003} is the Eckhaus or Benjamin-Feir resonance,
based upon interaction between
unstable waves.   Further work is needed to analyse whether such interaction is responsible for the
observed instability in R II and R IV regimes.

To summarise: for linearly unstable films subjected to a localised perturbation, we have extracted 
the mean spacing between drop centres  and the mean formation times of successive drops, as a
function of the initial film thickness, $H_0$. It has been shown that there exists a region, 
R II in figures~\ref{fig:LocalSpacing} and~\ref{fig:LocalMeanBreakup1}, such that the results in the nonlinear 
evolution regime differ significantly from the predictions of LSA.    Our understanding is that the effects leading 
to the existence of regimes R II and R IV are fully nonlinear. 

\subsection{Linearly Stable Regime} \label{sec:LocalLinearlyStable}
We now switch focus to locally perturbed films with initial film thicknesses in
the linearly stable regime, in particular for  $H_0 > H_{+}$.  First, we investigate the behaviour
of linearly stable films exposed to a finite amplitude disturbance (metastability).
Then, once dewetting is induced, we investigate the speed at which a localised perturbation propagates
into the stable film region.

\subsubsection{Metastability: Analytical findings} \label{sec:MetastableAnaylsis}

We first adapt the results of \citet{Thiele2001,Diez2007}, and derive the metastable regime. 
Note that this regime is referred to as the heterogeneous
nucleation regime by \cite{Seemann2001b}; however, in contrast to the results to be 
presented here, \cite{Seemann2001b} provide no upper bound to this regime.

A linearly stable film is metastable if there exists a thinner, linearly stable, film 
with lower energy, hence, given a perturbation of sufficient size, 
the perturbation can push the thicker film into the energetically preferable thinner
film state (dewetting). To derive an expression for the energy per unit length
of a film, we analyse steady state solutions
to the governing equation (\ref{eq:NanoGov-PDE}) 
that connect two flat films of different thicknesses. 

To begin, we set $h_t(x,t)=0$ in (\ref{eq:NanoGov-PDE}), divide by $Q(h)=h^3$,
and integrate with respect to $x$ to yield
 \begin{equation} \label{eq:SteadyState_Step1}
  \cC h_{xxx} + \Pieff'(h)h_x = d_0 \;,
 \end{equation}
where $d_0$ is the constant of integration.
Note that the metastability analysis from this point on
is independent of the form of the mobility function $Q(h)$.
Similarly to~\citet{Sharma2004}, we use the far-field conditions 
\begin{equation} \label{eq:SteadyState_BC1}
  h(x\rightarrow-\infty ,t)=H_{\textrm{min}} \;, \quad 
  h(x\rightarrow\infty ,t)=H_{\textrm{max}}  \;, \quad
  H_{\textrm{min}} < H_{\textrm{max}} \;,
\end{equation}
and
\begin{equation} \label{eq:SteadyState_BC2}
  h_x(x ,t)=h_{xx}(x ,t)=h_{xxx}(x ,t) = 0  \;, \quad x \rightarrow \pm \infty \;,
\end{equation}
where $H_{\textrm{min}}$ and $H_{\textrm{max}}$ are constants to be determined.
Applying boundary conditions (\ref{eq:SteadyState_BC2}) to (\ref{eq:SteadyState_Step1}), we find $d_0=0$.
In the work of~\citet{Sharma2004}, the solution that satisfies the above far-field conditions
is referred to as a pancake solution. We will refer to this solution as a front
solution since (\ref{eq:SteadyState_BC1}) and (\ref{eq:SteadyState_BC2}) are
analogous to the far-field conditions used for a travelling front solution in our previous work \citep[see][]{Lam2014}.
Note that the same nomenclature is used by~\citet{Thiele2001}; the derivation
of the front solution in that work is slightly different
to that presented here, however, the two methods
yield the same result.
 
Equation (\ref{eq:SteadyState_Step1}) is integrated in $x$ to obtain
 \begin{equation} \label{eq:SteadyState_Step2}
  \cC h_{xx} + \Pieff(h) = \Pi_0 \;,
 \end{equation}
where $\Pi_0$ is the constant of integration. To satisfy (\ref{eq:SteadyState_BC1}) and (\ref{eq:SteadyState_BC2}), 
we require $\Pieff(h)$ to have the same value, $\Pi_0$, for at least two different thicknesses;  i.e.
ignoring the transition region between the flat films of thickness $H_{\textrm{min}}$ and $H_{\textrm{max}}$, the structural
disjoining pressure is constant across the film.
We may then parameterize steady state solutions, satisfying boundary conditions (\ref{eq:SteadyState_BC1}) and (\ref{eq:SteadyState_BC2}), by $\Pi_0$. 

\begin{figure}
\centering
 \includegraphics[width=0.7\textwidth]{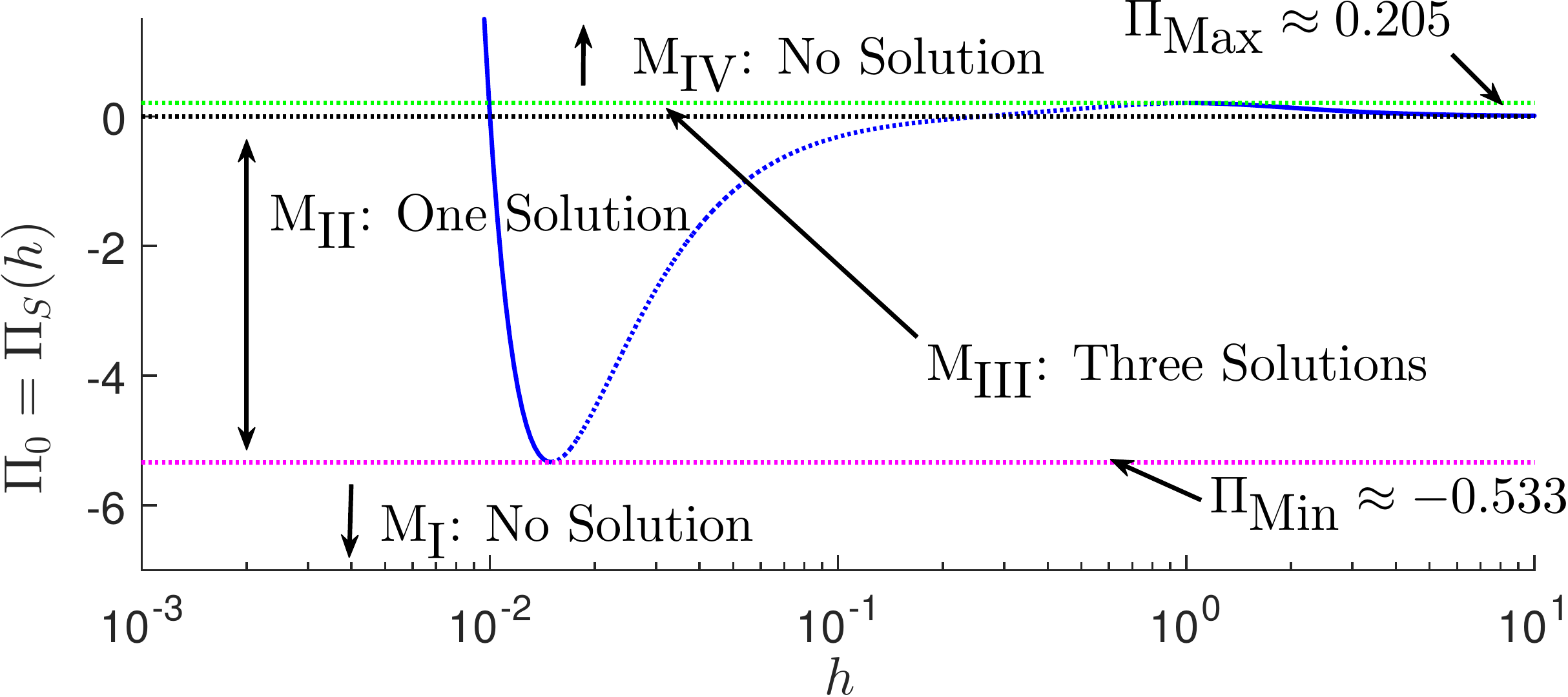}
  \caption[]{The solution space of (\ref{eq:SteadyState_Step2}) subject to conditions (\ref{eq:SteadyState_BC1}) and (\ref{eq:SteadyState_BC2}). Solid
blue curves denote linearly stable films, and the dotted blue curve denotes linearly unstable ones.}
  \label{fig:FrontPhaseDiagram}
\end{figure}
Recalling figure~\ref{fig:DisjoiningPressureDiagram}, there are, at most,
three different film thicknesses with the same disjoining pressure, hence
there are at most three roots to $\Pieff(h)=\Pi_0$,
which we denote as
$H_{m_1}$, $H_{m_2}$, and $H_{m_3}$, where $H_{m_1}<H_{m_2}<H_{m_3}$.
Furthermore, we define four regimes: 
M\textsubscript{I} (no roots), 
M\textsubscript{II} (two roots),
M\textsubscript{III} (three roots), and   
M\textsubscript{IV} (no roots), see 
figure~\ref{fig:FrontPhaseDiagram}.
Note that $H_{m_1}$ and $H_{m_3}$ correspond to the left and right solid blue curves, respectively
in Fig.~\ref{fig:FrontPhaseDiagram}
(thus represent linearly stable solutions), and $H_{m_2}$ is denoted by the central dotted blue curve 
(a linearly unstable solution).

In regions M\textsubscript{I} and M\textsubscript{IV} there are no
possible solutions to the steady state boundary value problem specified by 
(\ref{eq:SteadyState_BC1})--(\ref{eq:SteadyState_Step2}).  
 Solutions exist for $\Pi_0 \in ( \Pi_\textrm{Min} ,\Pi_\textrm{Max} )$,
where $\Pi_\textrm{Min}$ and $\Pi_\textrm{Max}$
are the local minimum and local maximum of $\Pieff(h)$ respectively,
the values of which are given in figure~\ref{fig:FrontPhaseDiagram} (for the chosen
parameter values).
In M\textsubscript{II}, there are  two roots of (\ref{eq:SteadyState_BC1})--(\ref{eq:SteadyState_Step2}), 
thus one solution joining these states, with
$H_\textrm{Min}=H_{m_1}$ and $H_\textrm{Max}=H_{m_2}$ i.e.
the solution in M\textsubscript{II} connects a thin linearly stable film to a thicker
unstable one.

\begin{figure}
\centering
  \subfigure[$\Pi_0<0$]{
   \includegraphics[width=0.38\textwidth]{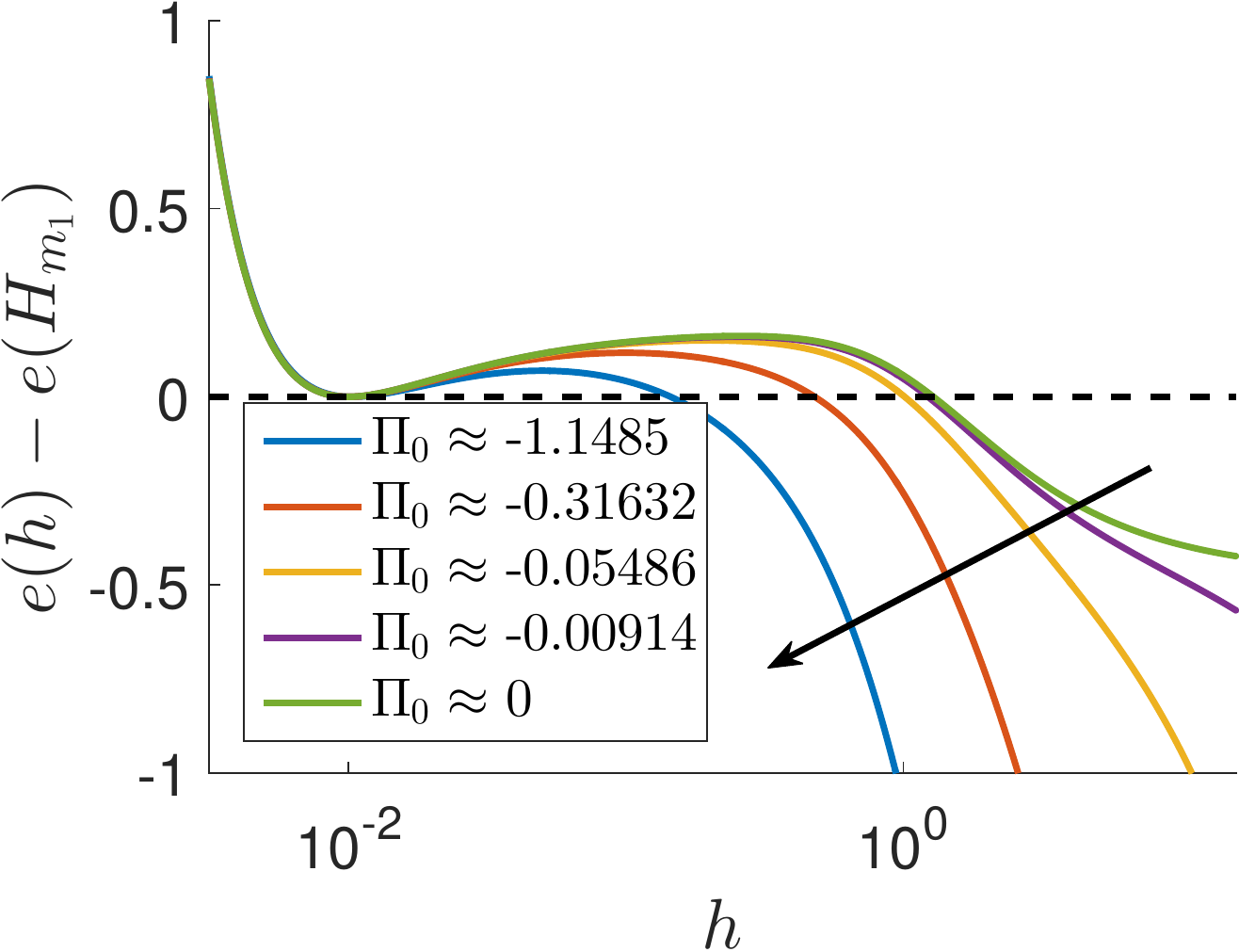}
  \label{fig:EnergyPerLength1} }
    \quad \quad
  \subfigure[$\Pi_0>0$]{
   \includegraphics[width=0.38\textwidth]{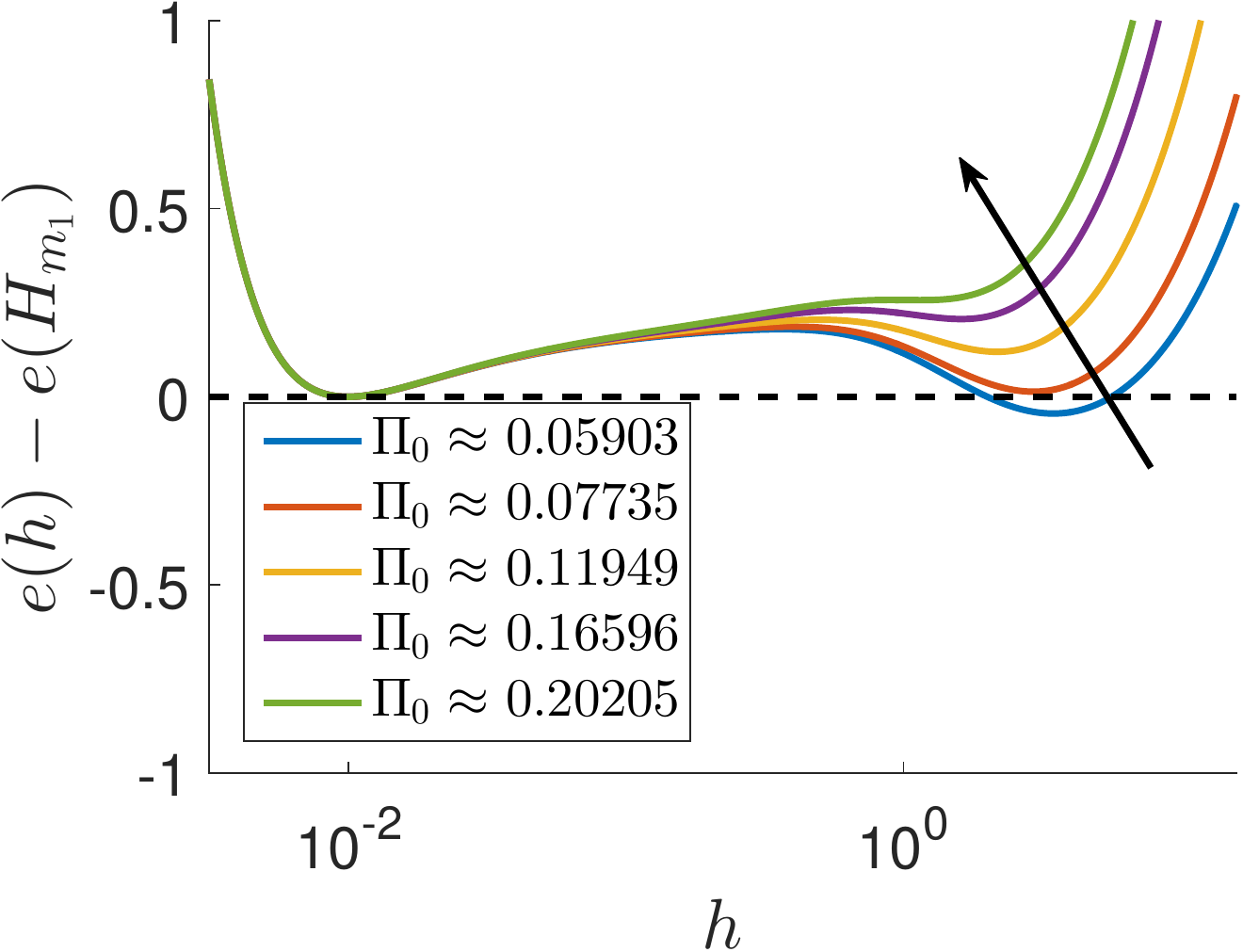}
  \label{fig:EnergyPerLength2} }

  \caption[]{ $e\big(h\big)-\big( H_{m_1} \big)$ where $e(h)$ is defined in (\ref{eq:SteadyState_Step3}) for an initial film profile with (a) negative structural disjoining
pressure, $\Pi_0<0$; and (b) positive structural disjoining pressure, $\Pi_0>0$. Arrows denote
  the direction of increasing $\Pi_0$. }
  \label{fig:EnergyPerLength}
\end{figure}

In M\textsubscript{III}, there are three roots; thus we can choose two
different $H_\textrm{Min}$ and $H_\textrm{Max}$, see figure~\ref{fig:EnergyPerLength}.
Since $H_{m_2}$ is a linearly unstable thickness, the corresponding steady state solutions
are also unstable. Ignoring the unstable solution, we see that there is a one parameter family of
stable steady solutions, which lies in M\textsubscript{III};
thus we set
\begin{equation} 
  \Pi_0 \in (0,\Pi_\textrm{Max})  \; ,\quad 
  H_\textrm{Min} = H_{m_1}  \; ,\quad \textrm{and} \quad
  H_\textrm{Max} = H_{m_3} \;.
\end{equation}
Integrating (\ref{eq:SteadyState_Step2}) with respect to $h$ yields (Euler--Lagrange equation)
 \begin{equation} \label{eq:SteadyState_Step3}
  \frac{\cC}{2} \left( h_x \right)^2 + e(h) = e_0   \;, \quad e(h)= r(h) + \Pi_0 h \;, 
 \end{equation}
where $e(h)$ is the energy per unit length of the film due to
the structural disjoining pressure.  Here, the constant of integration
 $e_0$ may be 
 interpreted as the total energy due to surface tension and structural disjoining pressure, and 
 \begin{equation} \label{eq:EnergyPerLength}
 r(h) = - \int \Pieff(h) \; dh = \frac{\cK b^2}{2 h}\left[ \frac{b}{h}- 2 \right]  - \cN \left[ \frac{m^2(h)}{2h} - \int \frac{m(s)m'(s)}{s} \, ds \right]   \; .
 \end{equation}
Recalling the free energy functional (\ref{eq:FreeEnergyFunctional}), we note that
$F(h) = - [ \cC( 1 + (h_x)^2/2 ] - r(h)$.
Setting $e_0=e(H_{m_1})$ we plot $e(h)-e_0$ in figure~\ref{fig:EnergyPerLength}
for various $\Pi_0$, and observe that only for $\Pi_0>0$  are there two local
minima for $e(h)-e(H_{m_1})$, see figure~\ref{fig:EnergyPerLength2}.
In addition, we note that the film thicknesses at the local minima ($H_{m_1}$ and $H_{m_3}$) 
correspond to linearly stable films, and the film thickness at the local
maximum ($H_{m_2}$) corresponds to a linearly unstable film.

Following~\citet{Thiele2001}, we seek a value of $\Pi_0$ such that the two local minima of
(\ref{eq:EnergyPerLength}) have the same energy.
Specifically, we seek a value of $\Pi_0$ that satisfies
 \begin{equation} \label{eq:ABSCritPoints}
    \pad{e\left(H_{\textrm{min}}\right)}{h} = \pad{e\left(H_{\textrm{max}}\right)}{h} = 0 \;, \quad
  \end{equation}
and 
 \begin{equation} \label{eq:ABSEqualEnergy}
    e \left( H_{\textrm{max}} \right) = e \left( H_{\textrm{min}} \right) \;.
  \end{equation}
Note that
 \begin{equation}
    \pad{e}{h} = 0 \,\, \Leftrightarrow \,\, \Pieff(h) = \Pi_0  \;,
  \end{equation}
therefore by inspection, (\ref{eq:ABSCritPoints}) and (\ref{eq:ABSEqualEnergy}) are equivalent to
satisfying the far-field conditions, (\ref{eq:SteadyState_BC1}) and (\ref{eq:SteadyState_BC2}), in (\ref{eq:SteadyState_Step2}) and (\ref{eq:SteadyState_Step3}),
respectively. 

We now apply these results to the specific problem in question.  
For the parameters chosen in $\S$~\ref{sec:Scalings}, 
$\Pi_0=\Pi_0^* \approx 0.07273$ satisfies (\ref{eq:ABSCritPoints}) and (\ref{eq:ABSEqualEnergy}),
and 
 \begin{equation} \label{eq:ABSSolution}
    H_\textrm{Min}^*=H_{m_1}^* \approx 0.99799\times 10^{-2} \;, \quad
    H_{m_2}^* \approx 0.39024 \;, \quad
    H_\textrm{Max}^*=H_{m_3}^* \approx 3.05160 \;,
  \end{equation}
where `*' superscripts denote the values obtained by using $\Pi_0^*$  in Equation~(\ref{eq:SteadyState_Step2}).
The metastability regime is thus given by
 \begin{equation}
    H_0 \in ( H_{m_1}^* , H_{-} ) \cup ( H_{+} , H_{m_3}^* ) \;.
  \end{equation}
The region $ ( H_{m_1}^* , H_{-} ) $ corresponds to a narrow
range of film thicknesses close to the equilibrium thickness, which we do not consider further.
Films thicker than $H_{m_3}^*$ are absolutely stable, thus no perturbations may induce dewetting.
The identified metastable region $(H_{+} , H_{m_3}^*)$ is considered by numerical 
simulations in the next section.   

{\bf Note 1:}
In \citet{Seemann2001b}, the metastable regime (heterogeneous nucleation in that work) 
is defined as $H_0>H_+$.   Extending our presented
metastability analysis to disjoining pressure models of \citet{Seemann2001b},
we find that for all positive disjoining pressures values ($\Pi_0$ in (\ref{eq:SteadyState_Step3})),
condition (\ref{eq:ABSEqualEnergy}) is never satisfied; specifically, we find that 
for all $\Pi_0>0$, $e \left( H_{\textrm{min}} \right)<e \left( H_{\textrm{max}} \right)$,
hence there is no upper bound for the metastable regime, consistent with~\citet{Seemann2001b}.  
Therefore, the existence of absolutely stable films is one of the distinguishing features
of the structural disjoining pressure considered here.

{\bf Note 2:}
Satisfying (\ref{eq:ABSCritPoints}) implies that $H_\textrm{Min}$ and $H_\textrm{Max}$ 
are critical points of $e(h)$; however, to satisfy condition (\ref{eq:ABSCritPoints}) with
$H_\textrm{Min} \ne H_\textrm{Max}$, there must exist at least three distinct critical points,
i.e., at least three different thicknesses must be characterised by the same disjoining pressure.
In our model, this corresponds to positive disjoining pressure. 
This finding is in agreement with \citet{Herminghaus1998,Poulard2006,Seemann2001b},
where it is stated that a disjoining pressure of the form shown in figure~\ref{fig:StructDisjoiningPressure}
will lead to metastability, a coexistence
of films of different thicknesses.

\subsubsection{Metastability: Numerical Simulations}

In this section, we investigate the evolution of localised perturbations in the metastable 
regime, $H_0 \in (H_{+}, H_{m_3}^*) \approx (1.015,3.0516)$.  
The initial condition is specified by  (\ref{eq:NanoICLocal}),
which is parameterized by the initial film height, $H_0$.  To explore the influence of 
the perturbation properties, we vary 
 $W$ and $d$, respectively the width and depth of the perturbation.
In particular, we discuss the quantity $d_\textrm{Min} ( H_0 ; W)$,
the minimum perturbation amplitude required to induce
dewetting. Since we are only interested if breakup occurs
for a particular set of $d$ and $W$,
it is not necessary to dynamically expand the right boundary in the $x$ domain;
therefore, to reduce simulation time, we fix $x\in[0,200]$.

\begin{figure}
\centering
   \includegraphics[width=0.42\textwidth]{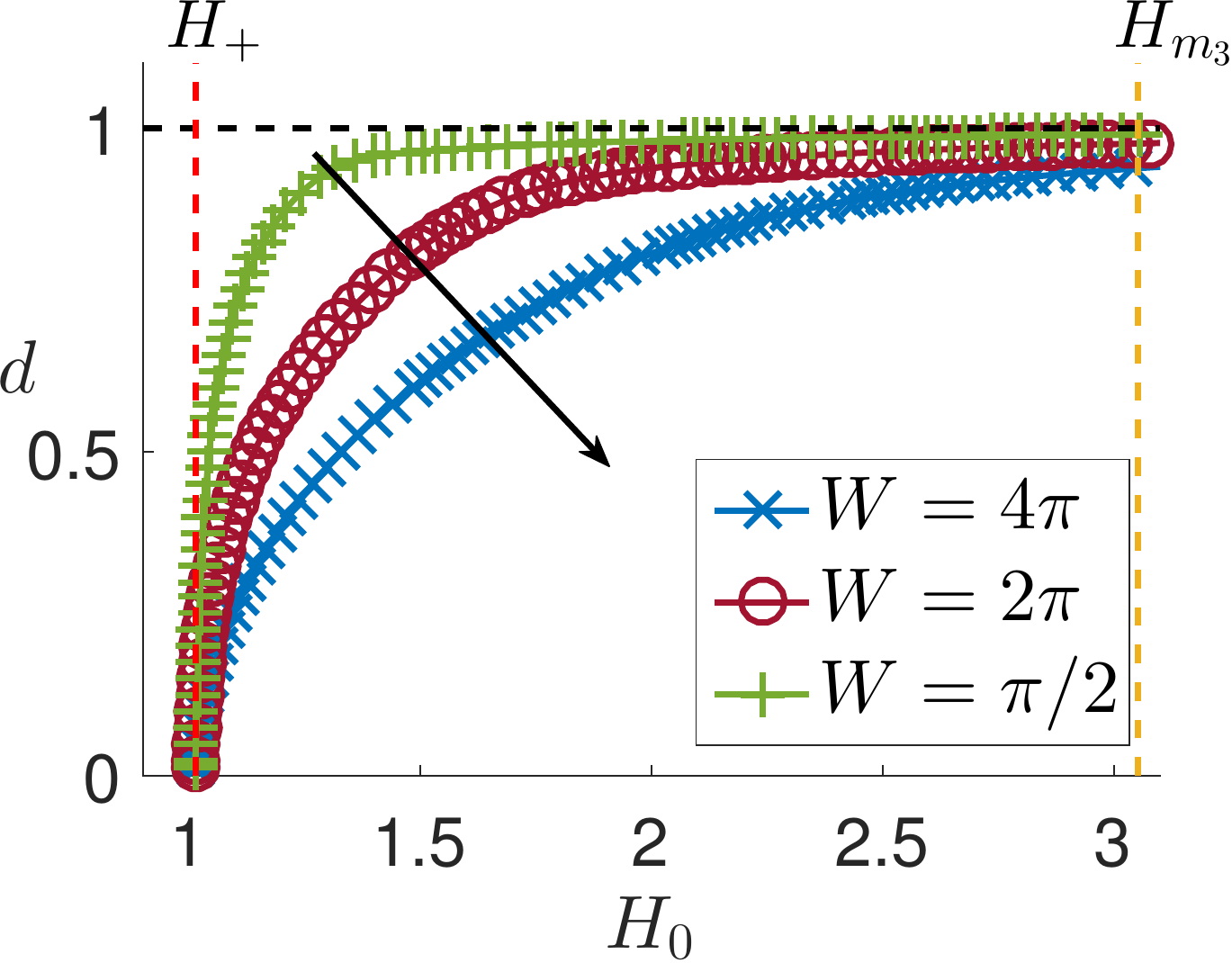}
\caption{ $d_\textrm{Min} ( H_0 ; W)$  as a function
of $H_0$ for various values of $W$. Black arrow denotes direction
of increasing $W$. }
\label{fig:MetaPhase}
\end{figure}

Figure~\ref{fig:MetaPhase} plots $d_\textrm{Min} ( H_0 ; W)$ 
as a function of $H_0$ for various values of $W$.  In
the region above $d_\textrm{Min} ( H_0 ; W)$ (curves with symbols)
and below the dotted black line ($d=1$), the
initial condition given by (\ref{eq:NanoICLocal})
induces dewetting. We see  that for all
values of $W$ considered, 
$d_\textrm{Min} ( H_0 ; W) \rightarrow 0$ as $H_0 \rightarrow
H_{+}$; i.e. as the film thickness approaches 
the threshold $H_+$ between linear stability and instability,
an infinitesimally small perturbation ($d\ll 1$)
 will induce dewetting, as expected. More importantly, figure~\ref{fig:MetaPhase} demonstrates
that as $H_0 \rightarrow H_{m_3}$, $d_\textrm{Min} ( H_0 ; W) \rightarrow 1$;
i.e. as the film thickness $H_0$ approaches the threshold $H_{m_3}$
between metastability and absolute stability, 
the initial condition (\ref{eq:NanoICLocal}) has to essentially rupture the film ($d\approx1$)
for dewetting to be observed. 
This demonstrates that the analysis in $\S$~\ref{sec:MetastableAnaylsis} provides
a correct upper bound for the metastable regime.
Note that $d_\textrm{Min} ( H_{m_3} ; W)$
is a monotonically decreasing function of the perturbation width $W$.  
This is expected, since the transition
region between a film of thickness $H_0$ and the minimum
of the initial condition (\ref{eq:NanoICLocal}) increases as 
$W$ increases, thus the transition region (which was 
assumed to be small and ignored in $\S$~\ref{sec:MetastableAnaylsis}) is no 
longer negligible. 

{\bf{Note 1:}} The presented results suggest that our model may transition between the regimes
called thermal and heterogenous nucleation, see \cite{Seemann2001b}.  As $H_0 \rightarrow H_+$
more and more holes are expected as the solution evolves, which may be a result of 
relatively small perturbations requiring more time to be experimentally measurable compared to the larger perturbations. In contrast, for $H_0$ significantly larger than $H_+$, 
only perturbations above the threshold value $d_\textrm{Min}$ will form holes, effectively enforcing an upper bound
on the time window within which hole nucleation is observed. Therefore, as $H_0 \rightarrow H_+$ from above, $d_\textrm{Min}\rightarrow 0$ and
the time window ($\propto d^{-1}_\textrm{Min}$) in which holes form transitions from a sharp window,
to a infinitely long time window, where more and more holes form in time.

\begin{figure}
\centering
\subfigure[]{
   \includegraphics[width=0.42\textwidth]{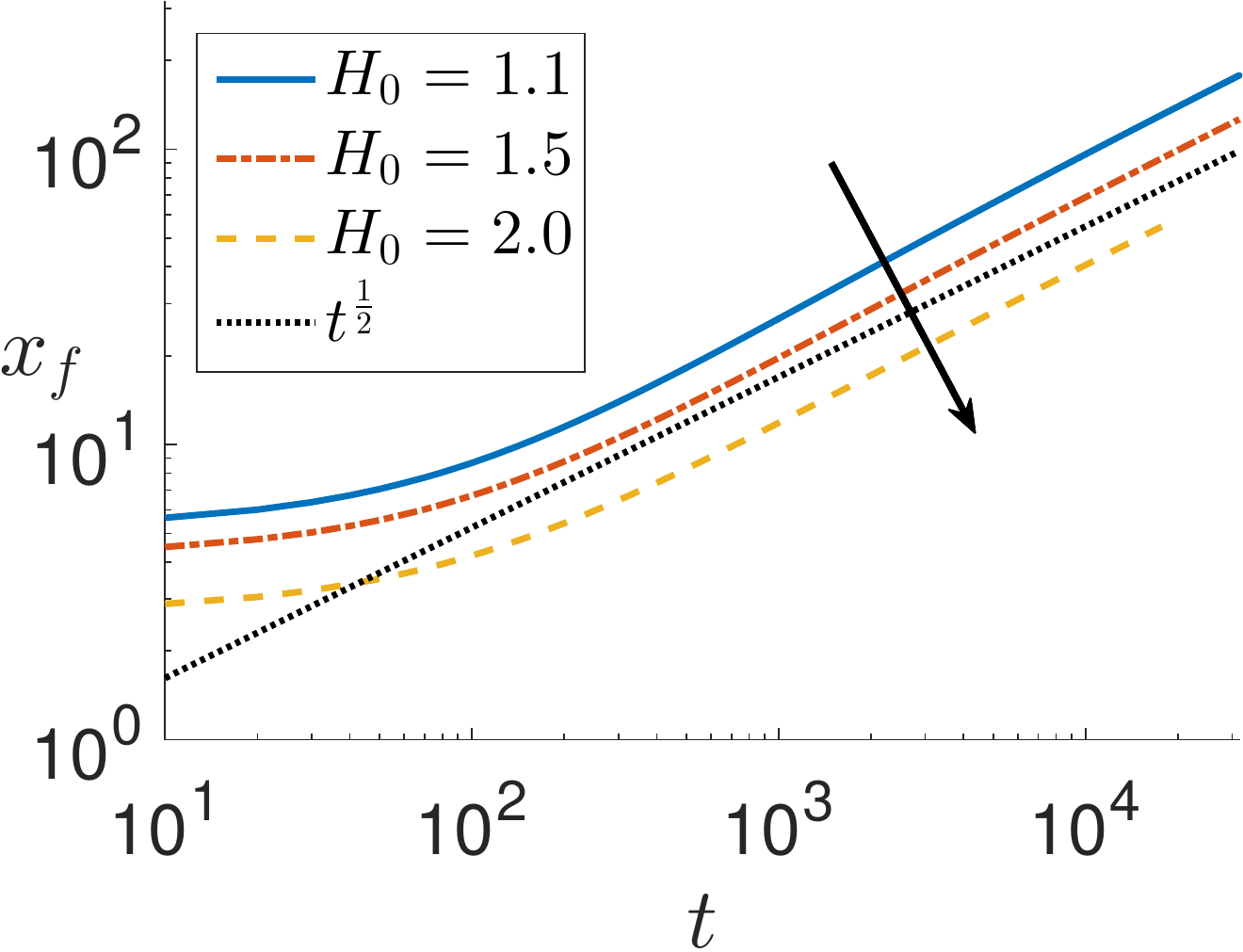}
  \label{fig:MetaFrontPosition} }
\subfigure[]{
   \includegraphics[width=0.42\textwidth]{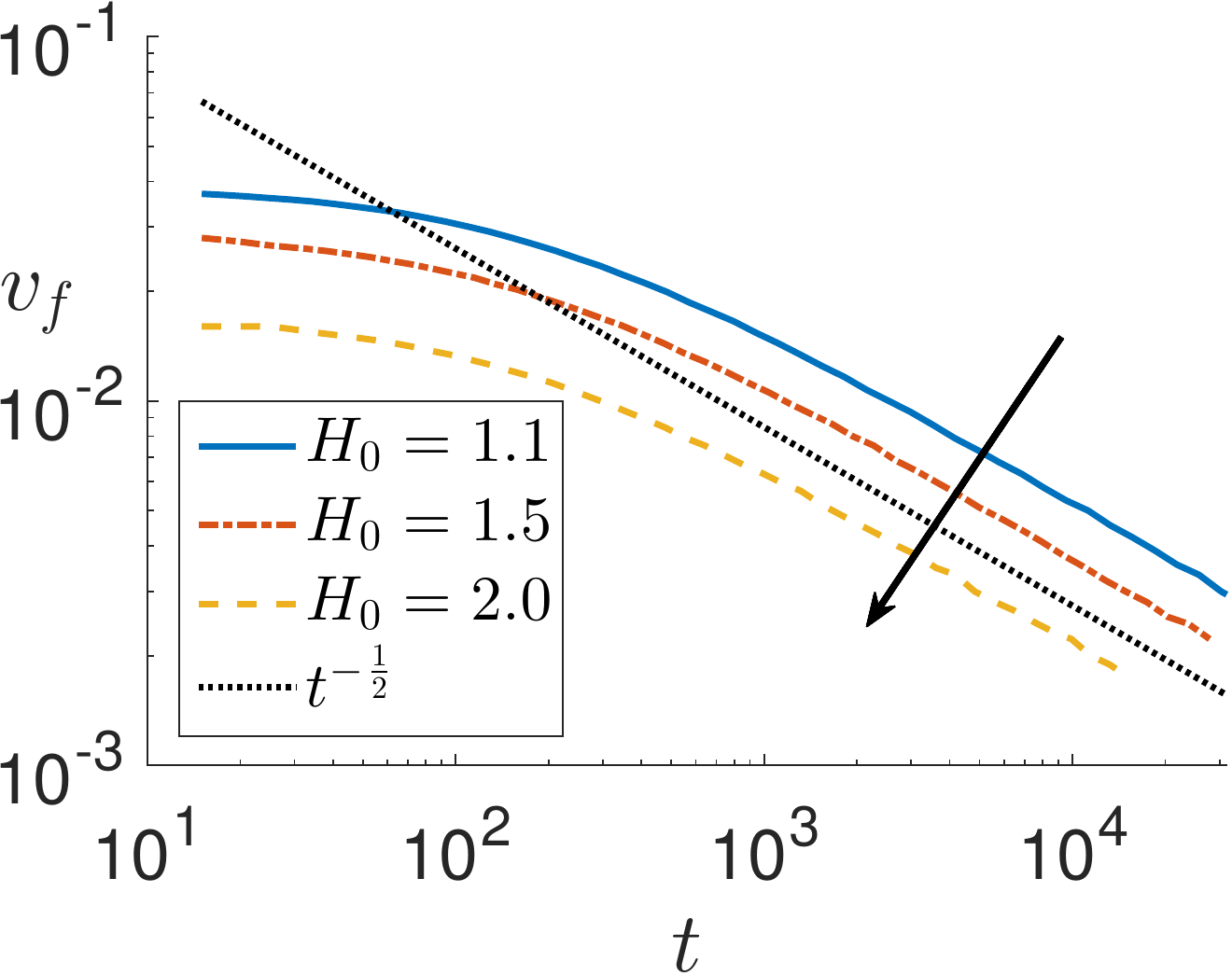}
  \label{fig:MetaFrontVelocity} }
\caption{(a) Front position and (b) front speed as a function of time
extracted for various initial film thicknesses, $H_0$, within the metastable regime. Black
arrows denote direction of increasing $H_0$.}
\label{fig:MetaFront}
\end{figure}

{\bf{Note 2:}}
We now briefly describe the film evolution once dewetting
is induced. 
Similarly to simulations in $\S$~\ref{sec:LocalSimsUnstable}, the right boundary in the $x$ domain
is dynamically expanded, and we set the initial domain to $[0,10]$.
Simulations show that when dewetting occurs, a retracting front
forms, and propagates without additional dewetting events into the flat film region.
Similarly to \citet{Muench2005}, we extract from simulation results the time dependence of the front position, $x_f (t)$,
and its speed, $v_f (t)$.  
Figure~\ref{fig:MetaFrontPosition} shows that at long times scales, $x_f(t)\propto t^{1/2}$.
Not surprisingly, this power law is different from the $t^{2/3}$,
power law found by \citet{Muench2005} for a thin film model in the slip dominated regime.
Figure~\ref{fig:MetaFrontVelocity} shows that at long times scales, $v_f(t)\propto t^{-1/2}$,
whereas in the linearly unstable regime, simulations and analysis show that propagation speed is a constant, see $\S$~\ref{sec:MSC}.

\section{Conclusions} \label{sec:conclusions}

\begin{figure}
\centering
   \includegraphics[width=0.95\textwidth]{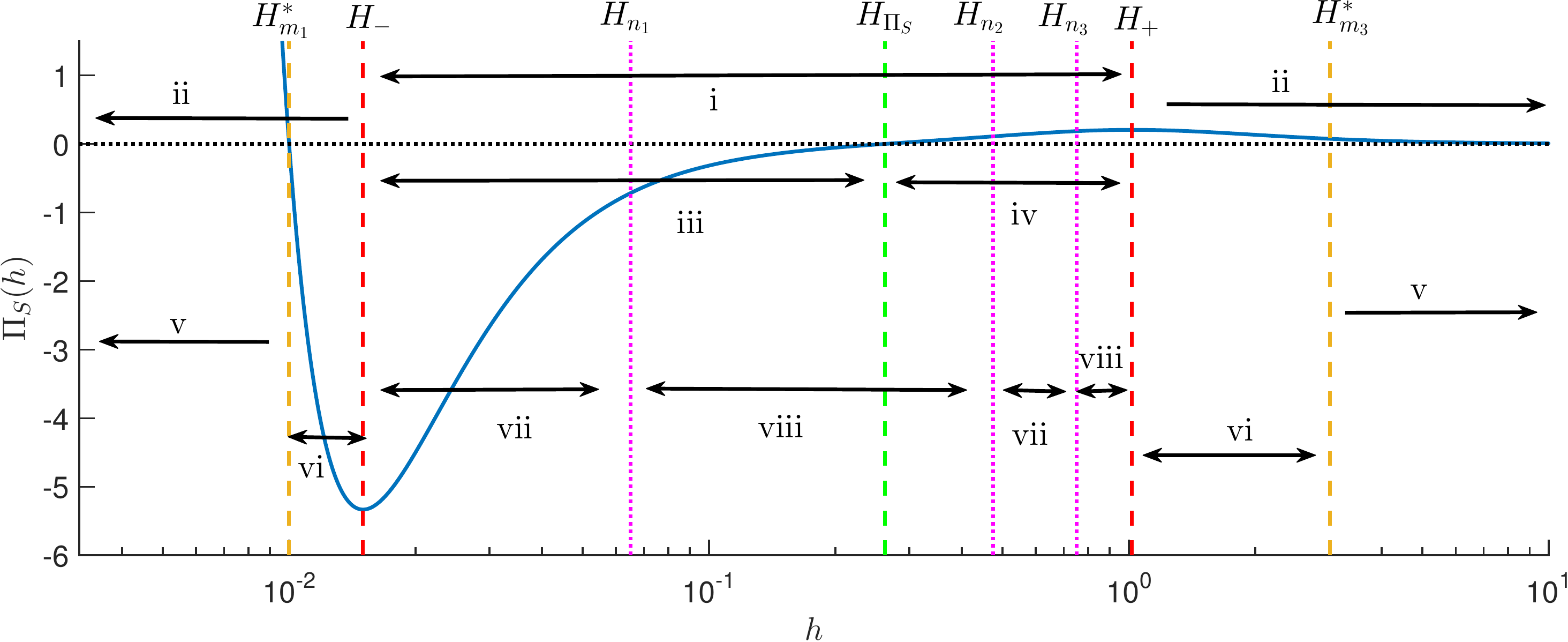}
\caption{  Summary of the various stability regimes discussed in this paper in relation to the structural/effective disjoining
pressure. The dashed lines denote analytically derived values and dotted lines denote values
extracted from numerical simulations. See the text for explanation of numbered regions. 
} 
\label{fig:CompleteDiagram}
\end{figure}

While this work has been motivated by  \gls{NLC} films, we have presented results that are relevant to a much 
wider class of problems involving thin films on substrates, such that the fluid-solid interaction involves a relatively complex form of effective 
disjoining pressure (sketched on figure~\ref{fig:CompleteDiagram}).  Such a form has been derived in the 
present work as a structural disjoining pressure that includes an elastic contribution for \gls{NLC} films; 
however similar functional forms have been extensively considered in the literature focusing on polymer films on Si/SiO$_2$ 
substrates, see, e.g.~\citet{Jacobs2008} and the references therein.

In the \gls{NLC} context, novel elements include dynamic relaxation of the free surface
polar anchoring, $\theta_F$, as a function of film thickness $h(x,t)$. In addition,
we have highlighted differences and similarities between the presented model, and others
discussed in the literature, e.g.~\citet{Cazabat2011,Effenterre2003,Schlagowski2002,Ziherl2000}.
Furthermore, our model exhibits a range of film thickness such that a film 
is linearly unstable, which is analogous to the so-called ``forbidden range'' discussed in the literature \citep[]{Cazabat2011}.

More generally, an extensive statistical analysis of our computational results, combined with novel analytical results 
based on the Marginal Stability Criterion, has allowed us to identify
a variety of instability mechanisms.  In addition to the linearly unstable regime, 
where infinitesimally small perturbations induce dewetting, there are regimes that require perturbations of sufficient magnitude to induce
dewetting  (metastability), and regimes that are stable with  respect to perturbations of any size (absolute stability).  
We have also identified the regimes that are linearly unstable, but susceptible 
 to nucleation dominated dynamics.   While we do not consider thermal effects, we find that for some film thicknesses (close to the 
 stability boundaries) our results are similar to other results attributed to a thermal nucleation mechanism, suggesting that consideration of 
 thermal effects may not in fact be needed to explain the dynamics.  Also, we have formulated analytical procedures for identifying most of the stability 
 regimes found computationally.   In particular, the application of marginal stability analysis has allowed us to reach new insight. 
 
Figure~\ref{fig:CompleteDiagram} summarises the various stability regimes uncovered, and correlates them with the form of structural/effective disjoining 
pressure. The following regions are identified: 
\begin{enumerate}[leftmargin=-.5in]
  \item Linearly unstable regime, $H_0\in(H_-,H_+)\approx(1.5b,\beta)$, where $b$ is the equilibrium
  thickness and $\beta$ is the film thickness at which elastic force are relevant (see $\S$~\ref{sec:Scalings} and $\S$~\ref{sec:2DNanoLSA}).
  \item Linearly stable regime, $H_0 \notin (H_-,H_+)$.
  \item Linearly unstable regime with negative disjoining pressure, $H_0\in(H_-,H_\Pieff)$, where $H_\Pieff$ is the thickness
	marking the sign change of $\Pieff$ ($\Pieff (H_\Pieff)=0$).   If randomly perturbed, the film will dewet and form 
	primary drops only (see $\S$~\ref{sec:RandomPert}).
  \item Linearly unstable regime with positive disjoining pressure, $H_0\in(H_\Pieff,H_+)$.  If randomly perturbed,  the film will dewet and form primary
	and secondary drops.
  \item Absolutely stable regime, $H_0 < H^*_{m_1}$ or $H_0 > H^*_{m_3}$. The film is stable with respect to any perturbation
  (see $\S$ \ref{sec:LocalLinearlyStable}). 
  \item Metastable regime, $H_0 \in ( H_{m_1}^* , H_{-} ) \cup ( H_{+} , H_{m_3}^* )$.
	A linearly stable film is unstable with respect to finite size perturbations.
  \item Spinodal regime, $H_0 \in ( H_{-} , H_{n_1} ) \cup ( H_{n_2} , H_{n_3} )\subset (H_-,~H_+)$. A linearly unstable film perturbed by either
  localised or random perturbations will result in the mean distance between drop centres, $\bar{D}_l$, as predicted 
  by LSA,  i.e. $\bar{D}_l\approx \lambda_m$  (see $\S$ \ref{sec:LocalUnstable}).
  \item Nucleation regime within linearly unstable regime, $H_0 \in ( H_{n_1} , H_{n_2} ) \cup ( H_{n_3} , H_{+} )\subset (H_-,~H_+)$. A linearly unstable
  film perturbed by a localised perturbation will lead to $\bar{D}\not\approx \lambda_m$,
  and for a randomly perturbed film the variance in the distribution of inter-drop distances is significantly larger than for the spinodal regime, although we still
  find that $\bar{D}\approx \lambda_m$.
\end{enumerate}

The regime often called `heterogenous nucleation' includes the regimes (vi) and (viii); the `spinodal' regime is defined under (vii). 

There are several potential directions for future work. Statistical analysis
has shown interesting connections between the observed phenomena and properties of the structural
disjoining pressure; however, no clear physical mechanism has been identified analytically.
Further investigation of the presented results is therefore required.  Also, generalising our results
to three dimensional flows is another area of interest, and is the subject of our ongoing research.
We hope that our computational results, statistical analysis, and theoretical results will inspire future experimental and analytical 
investigations, which may provide additional insight into the phenomena presented. 

\subsubsection*{Acknowledgements}

The authors gratefully acknowledge financial support from NSF DMS 1211713 and CBET 1604351.

%\subsubsection*{Supplementary Movies}

%Supplementary movies may be found in \textit{movies.zip}.

\newpage

\bibliographystyle{jfm}
\bibliography{LiquidCrystals,films}
\end{document}